\documentclass[12pt,preprint]{emulateapj}
\usepackage{apjfonts}
\usepackage{amsmath}
\usepackage{epsf}
\usepackage{color}
\shorttitle{CANDELS: The Rest-Frame UV Luminosity Function at $z =$ 4--8}
\shortauthors{Finkelstein et al.}

\newcommand{\sol}{$_{\odot}$}

\def\arcs{\hbox{$^{\prime\prime}$}}

\let\phi\varphi

\begin{document}
\slugcomment{Submitted to the Astrophysical Journal}
\title{The Evolution of the Galaxy Rest-Frame Ultraviolet Luminosity
  Function Over the First Two Billion Years}

\author{Steven L. Finkelstein\altaffilmark{1,a}, Russell E. Ryan
  Jr.\altaffilmark{2}, Casey Papovich\altaffilmark{3}, Mark
  Dickinson\altaffilmark{4}, Mimi Song\altaffilmark{1}, Rachel S. Somerville\altaffilmark{5}, Henry C. Ferguson\altaffilmark{2},
Brett Salmon\altaffilmark{3}, 
Mauro Giavalisco\altaffilmark{6}, Anton M.\ Koekemoer\altaffilmark{2}, 
Matthew L. N. Ashby\altaffilmark{7}, 
Peter Behroozi\altaffilmark{2}, 
Marco Castellano\altaffilmark{8},
James S. Dunlop \altaffilmark{9}, 
Sandy M. Faber\altaffilmark{10}, 
Giovanni G.\ Fazio\altaffilmark{7}, 
Adriano Fontana\altaffilmark{8},
Norman A. Grogin\altaffilmark{2}, 
Nimish Hathi\altaffilmark{11}, 
Jason Jaacks\altaffilmark{1}, 
Dale D. Kocevski\altaffilmark{12}, 
Rachael Livermore\altaffilmark{1}, 
Ross J. McLure\altaffilmark{9}, 
Emiliano Merlin\altaffilmark{8},
Bahram Mobasher\altaffilmark{13},
Jeffrey A. Newman\altaffilmark{14}, 
Marc Rafelski\altaffilmark{15}, 
Vithal Tilvi\altaffilmark{3}, 
S. P. Willner\altaffilmark{7}}

\altaffiltext{1}{Department of Astronomy, The University of Texas at Austin, Austin, TX 78712}
\altaffiltext{2}{Space Telescope Science Institute, Baltimore, MD 21218}
\altaffiltext{3}{George P. and Cynthia Woods Mitchell Institute for Fundamental Physics and Astronomy, Department of Physics and Astronomy, Texas A\&M University, College Station, TX 77843}
\altaffiltext{4}{National Optical Astronomy Observatory, Tucson, AZ 85719}
\altaffiltext{5}{Department of Physics \& Astronomy, Rutgers University, 136 Frelinghuysen Road, Piscataway, NJ 08854}
\altaffiltext{6}{Department of Astronomy, University of Massachusetts, Amherst, MA 01003}
\altaffiltext{7}{Harvard-Smithsonian Center for Astrophysics, 60 Garden Street, Cambridge, Massachusetts 02138}
\altaffiltext{8}{INAF -- Osservatorio Astronomico di Roma, via  Frascati 33, 00040 Monteporzio, Italy}
\altaffiltext{9}{Institute for Astronomy, University of Edinburgh, Royal Observatory, Edinburgh, UK}
\altaffiltext{10}{University of California Observatories/Lick Observatory, University of California, Santa Cruz, CA, 95064}
\altaffiltext{11}{Aix Marseille Universite, CNRS, LAM (Laboratoire d'Astrophysique de Marseille) UMR 7326, 13388, Marseille, France}
\altaffiltext{12}{Department of Physics and Astronomy, Colby College, Waterville, ME 04901}
\altaffiltext{13}{Department of Physics and Astronomy, University of California Riverside, Riverside, CA 92512}
\altaffiltext{14}{Department of Physics and Astronomy and Pitt-PACC, University of Pittsburgh, Pittsburgh, PA 15260}
\altaffiltext{15}{NASA Postdoctoral Program Fellow, Goddard Space Flight Center, Code 665, Greenbelt, MD 20771}
\altaffiltext{a}{stevenf@astro.as.utexas.edu}

\begin{abstract}
We present a robust measurement and analysis of the rest-frame ultraviolet (UV) luminosity
functions at $z =$ 4 to 8.  We use deep {\it Hubble Space Telescope}
imaging over the CANDELS/GOODS fields, the Hubble Ultra Deep Field and
the Hubble Frontier Field deep parallel observations near the Abell 2744 and
MACS J0416.1-2403 clusters.  The combination of these surveys provides an
effective volume of 0.6--1.2 $\times$10$^6$ Mpc$^3$ over this epoch,
allowing us to perform a robust search for faint (M$_{UV} = -$18) and
bright (M$_{UV} < -$21) high-redshift galaxies.
We select candidate galaxies using a well-tested photometric redshift
technique with careful screening of contaminants, finding a sample of 7446 candidate galaxies at 3.5 $< z <$ 8.5,
with $>$1000 galaxies at $z \approx $ 6 -- 8.
We measure both a stepwise luminosity
function for candidate galaxies in our redshift samples, as well as a Schechter
function, using a Markov Chain Monte Carlo analysis to measure robust
uncertainties.  At the faint end our UV luminosity functions agree with previous
studies, yet we find a higher abundance of UV-bright candidate galaxies
at $z \geq$ 6.  Our best-fit value of
the characteristic magnitude $M^*_\mathrm{UV}$ is consistent with $-$21 at $z
\geq$ 5, different than that inferred based on previous trends at
lower redshift, and brighter at $\sim$2$\sigma$ significance than
previous measures at $z =$ 6 and 7 \citep{bouwens07,bouwens11}.
At $z =$ 8, a single power-law provides an equally good fit to the UV
luminosity function, while at $z =$ 6 and 7, an exponential cutoff at
the bright end is
moderately preferred.  We compare our luminosity functions to
semi-analytical models, and find that the lack of evolution in $M^*_\mathrm{UV}$
is consistent with models where the impact of dust attenuation on the
bright end of the luminosity function decreases at higher redshift,
though a decreasing impact of feedback may also be possible.
We measure the evolution of the cosmic star-formation rate (SFR) density by integrating our observed luminosity functions
to M$_{UV} = -$17, correcting for dust
attenuation, and find that the SFR density declines proportionally to
(1$+z$)$^{-4.3 \pm 0.5}$ at $z >$ 4, consistent with observations at
$z \geq$ 9.  Our observed luminosity functions are
consistent with a reionization history that starts at $z \gtrsim$ 10,
completes at $z >$ 6, and reaches a midpoint (x$_{HII} =$ 0.5)
at 6.7 $< z <$ 9.4.  Finally, using
a constant cumulative number density selection and an empirically derived rising
star-formation history, our observations predict that the abundance of bright $z =$ 9
galaxies is likely higher than previous constraints,
though consistent with recent estimates of bright $z \sim$ 10 galaxies.
\end{abstract}

\keywords{early universe --- galaxies: evolution --- galaxies: formation --- galaxies: high-redshift --- ultraviolet: galaxies}

\section{Introduction}
The past half-decade has seen a remarkable increase in our
understanding of galaxy evolution over the first billion years after the
Big Bang, primarily due to the updated near-infrared capabilities of
the {\it Hubble Space Telescope}.  Robust
galaxy samples at $z >$ 6 now include more than 1000 objects
\citep[e.g.,][]{bouwens10a,oesch10,finkelstein10,mclure10,bunker10,finkelstein12a,yan12,oesch12}
with a few candidate galaxies having likely redshifts as high as 10
\citep[e.g.,][]{bouwens11c,ellis13,mclure13,oesch13,coe13,oesch14,bouwens15}.
These galaxies are selected photometrically, primarily based on
a sharp break at rest-frame 1216 \AA\ due to absorption by intervening
neutral hydrogen in the intergalactic medium (IGM).  

Studies of galaxies at $z >$ 6 have revealed a number of interesting
results.  Galaxies at 6 $< z <$ 8 appear to have bluer rest-frame
ultraviolet (UV)
colors than at lower redshift, likely due to a decrease in dust
attenuation, although the brightest/most massive galaxies do appear to
have comparable dust attenuation at $z =$ 4--7
\citep[e.g.,][]{stanway05,finkelstein10,bouwens10b,wilkins11,finkelstein12a,dunlop13,bouwens13}.
Lower mass galaxies have colors consistent with stellar populations harboring
significant metal content (though likely sub-Solar), and therefore the currently
detectable populations of galaxies are not dominated by the primordial
first generation of stars
\citep[e.g.,][]{finkelstein12a,dunlop12,dunlop13}.  The structures of
these galaxies are resolvable, though they show small sizes with
half-light radii $\leq$ 1 kpc,
consistent with the evolution previously detected at lower redshifts
\citep[e.g.,][]{ferguson04,oesch10b,ono13}.  
Finally, the  abundance of high-redshift star-forming galaxies may account
for the necessary photons to sustain an ionized intergalactic
medium (IGM) by $z \sim$ 6, and perhaps as high as $z =$ 7--8 if one
assumes that galaxies at least 5 magnitudes below
the detection limit of {\it HST} exist
\citep[e.g.,][]{finkelstein12b,robertson13}, though the unknown
ionizing photon escape fraction is a major systematic uncertainty.

One of the key measurements is the galaxy rest-frame
UV luminosity function (hereafter referred to as the
luminosity function), as it is one of the most useful tools to
study the evolution of a galaxy population.  This measure encapsulates the
relative abundances of galaxies over a wide dynamic range in
luminosity.  As the UV light probes recent star-formation
activity, the integral of the rest-UV luminosity function provides an
estimate of the cosmic star-formation rate density
\citep[e.g.,][]{madau96,bouwens12,madau14}, although this measurement is reliant
on dust corrections.
The luminosity function is typically parameterized with a
\citet{schechter76} function with a power-law slope at faint
luminosities, and an exponentially declining form at the bright end.
Comparing the shape of the luminosity function to the underlying
dark-matter halo mass function, previous studies have found that the
luminosity function at $z \leq$ 6, when normalized to the halo mass
function at the characteristic magnitude $M^{\ast}_\mathrm{UV}$, lies below the
halo mass function at both bright and faint luminosities.  This is
generally assumed to be due to feedback: dominated by accreting supermassive
black holes at the bright end (active galactic nuclei; AGN), and by
supernova or radiative-driven winds at the faint end
\citep[e.g.,][]{somerville08}.  Dust extinction can also play a role,
particularly if the level of attenuation is dependent on a galaxies
stellar mass or UV luminosity
\citep[e.g.,][]{finkelstein12a,bouwens13}.
Although luminous AGN are present at $z =$ 6 \citep[e.g.,][]{fan06},
they are exceedingly rare, and to date only a single quasar has been
observed at $z \geq$ 7 \citep{mortlock11}.  Therefore one may expect the
degree of the exponential decline at the bright end to become weaker with increasing
redshift.  In addition, robustly quantifying the bright end of the luminosity
function can allow us to gain physical insight into how these distant
galaxies turn their gas into stars, as the star-formation timescale is
a significant fraction of
the age of the Universe, therefore enough time has not yet elapsed for
feedback to bring these galaxies into equilibrium.  A change in the
star-formation timescale is therefore more readily apparent in the
shape of the bright end of the luminosity function \citep[e.g.,][]{somerville12}.

Thanks to the combination of observations from {\it GALEX} and the
{\it Hubble Space Telescope}
estimates of the UV luminosity function
exist now from $z <$ 1 \citep{arnouts05,cucciati12} out $z \geq$ 8
\citep[e.g,][]{bouwens07,mclure09,bouwens11,oesch12,oesch13,lorenzoni13}.
Earlier works have concluded that $M^{\ast}_\mathrm{UV}$ declines from around $-$21
at $z =$ 3 to fainter than $-$ 20 at $z =$ 8, with the faint-end slope $\alpha$
becoming steeper over this same redshift range
\citep[e.g.,][]{bouwens07,reddy09,bouwens11,mclure13,schenker13}.  However, in
order to adequately quantify the amplitude and form of the bright end,
large volumes need to be probed, as bright sources are relatively
rare.  This has been accomplished via a combination of ground and
space-based surveys at $z \leq$ 6, with a variety of studies showing
conclusively that a single power law does not fit the data, and that
some sort of cut-off is needed at the bright
end \citep[e.g.,][]{arnouts05,bouwens07,reddy09,mclure09}.  Although
previous luminosity functions have been published at $z \geq$ 6, the
space-based studies have been
based on small volumes \citep[e.g.,][]{bouwens11}, and
thus, while they can somewhat constrain the faint-end slope, they do
not have the capability to constrain the bright end.

Recent studies are starting to make progress at the bright end.
\citet{finkelstein13}, while selecting galaxies for spectroscopic
followup in the GOODS-N field, found an overabundance of bright
galaxies at $z =$ 7.  \citet{ono12} found a similar result, with their
discovery of the M$_\mathrm{UV} = -$21.8 galaxy GN-108036 at $z =$ 7.2 in
GOODS-N.  Likewise, \citet{hathi12} found two bright $z
>$ 6.5 candidate galaxies in a ground-based near-infrared survey of GOODS-N.
Thus, it appears that the abundance of galaxies at the bright
end of the luminosity function may not be decreasing towards higher
redshift as previously thought.  Although these studies were based in
a single field, 
further evidence comes from \citet{bowler14}, who
used new deep ground-based near-infrared imaging from the
UltraVISTA survey \citep{mccracken12} to discover 34 luminous $z \sim$ 7 galaxy candidates
over 1.65 deg$^2$.  They combined these galaxies with the results from
\citet{mclure13}, which included deep and wide {\it HST} imaging over
300 arcmin$^2$ in the GOODS-S, UDS and HUDF fields, to analyze the
rest-frame UV luminosity function at $z =$ 7.
They concluded that they did see evidence for a drop-off in the
luminosity function at the bright end, however, the drop-off was less
steep than that predicted by a Schechter function, leading them to
postulate that the $z =$ 7 luminosity function has the shape of a
double-power law, perhaps similar to that of the possible form of
far-infrared luminosity
functions \citep{sanders03,casey14}.

In this study, we measure the
rest-frame UV luminosity function at 4 $< z <$ 8 with solely
space-based data, using the largest {\it HST} project ever, the Cosmic Assembly Near-infrared
Deep Extragalactic Legacy Survey (CANDELS; PIs Faber \& Ferguson).
The large area observed by CANDELS allows us to probe large volumes of
the distant universe for the rare, bright galaxies.  With these data,
we investigate the form of the bright end of the luminosity function
and the implications on galaxy evolution.  In addition to the deep data in the HUDF, we use the CANDELS
data in the GOODS-S and GOODS-N fields, which have not only deeper
near-infrared imaging, but also imaging in more optical and
near-infrared filters than the other three CANDELS fields (UDS, EGS
and COSMOS).  We also include in our analysis the parallel fields from the first year
dataset of the Hubble Frontier Fields, near the Abell 2744 and MACS
J0416.1-2403 galaxy clusters.  The combination of these data allows us
to select a large sample of nearly 7500 galaxies, over a wide dynamic
range in UV luminosity at $z =$ 4--8 (Figure~\ref{fig:nums}).

\begin{deluxetable*}{ccccccccccc}
\tabletypesize{\small}
\tablecaption{Summary of Data -- Limiting Magnitudes}
\tablewidth{0pt}
\tablehead{
\colhead{Field} & \colhead{Area} & \colhead{$B_{435}$} & \colhead{$V_{606}$} & \colhead{$i_{775}$} & \colhead{$I_{814}$}  & \colhead{$z_{850}$} & \colhead{$Y_{098/105}$} & \colhead{$J_{125}$} & \colhead{$JH_{140}$} & \colhead{$H_{160}$}\\
\colhead{$ $} & \colhead{(arcmin$^{2}$)} & \colhead{(mag)} & \colhead{(mag)} & \colhead{(mag)} & \colhead{(mag)} & \colhead{(mag)} & \colhead{(mag)} & \colhead{(mag)} & \colhead{(mag)} & \colhead{(mag)}
}
\startdata
GOODS-S Deep&61.6&28.2&28.6&27.9&28.1&27.8&28.2&28.1&---&27.9\\
GOODS-S ERS&41.4&28.2&28.5&27.9&27.9&27.6&27.6&28.0&---&27.8\\
GOODS-S Wide&35.6&28.2&28.7&28.1&27.9&27.9&27.3&27.6&---&27.4\\
GOODS-N Deep&67.6&28.1&28.3&27.9&---&27.7&28.1&28.3&---&28.1\\
GOODS-N Wide&71.7&28.1&28.4&27.8&---&27.6&27.3&27.4&---&27.4\\
HUDF Main&5.1&29.5&30.0&29.7&--- &29.1&29.9&29.6&29.6&29.7\\
HUDF PAR1&4.7&---&29.0&28.8&---&28.5&28.9&29.0&---&28.8\\
HUDF PAR2&4.8&---&29.0&28.7&---&28.3&28.9&29.2&---&28.9\\
MACS0416 PAR&4.4&28.8&28.9&---&29.2&---&29.2&29.0&29.0&29.0\\
Abell 2744 PAR&4.3&29.0&29.1&---&29.2&---&29.1&28.8&28.8&28.9\\[1ex]
\hline\\[-1ex]
Zeropoints&---&25.68&26.51&25.67&25.95&24.87&26.27&26.23&26.45&25.95
\enddata
\tablecomments{The magnitudes quoted are 5$\sigma$ limits
  measured in 0.4\arcs-diameter apertures on non-PSF matched images.}
\end{deluxetable*}

This paper is organized as follows.  In \S~2 we discuss the
imaging data used and the catalog construction, and in \S 3 we present
our sample selection via photometric redshifts, and estimates of the
contamination.  In \S~4 we highlight our completeness
simulations, and in \S~5 we discuss the construction of the rest-UV
luminosity function at $z =$ 4, 5, 6, 7 and 8.  In \S~6 we discuss the
implications of our luminosity function results, while in \S~7 we
compare our results to semi-analytical models.  In \S~8 we present our
measurements of the cosmic star-formation rate density, and in \S~9 we
discuss the implications for galaxies at higher redshifts.  Our conclusions are
presented in \S~10.  Throughout this paper we
assume a WMAP7 cosmology \citep{komatsu11}, with H$_\mathrm{0} =$ 70.2
km s$^{-1}$ Mpc$^{-1}$, $\Omega_M =$ 0.275, $\Omega_{\Lambda} =$ 0.725
and $\sigma_8 =$ 0.816.
All magnitudes given are in the AB system
\citep{oke83}.  All error bars shown in the figures represent
1$\sigma$ uncertainties (or central 68\% confidence ranges), unless
otherwise stated.

\section{Observations and Photometry}

\subsection{Imaging Data}
Studying galaxies in the early universe requires extremely deep
imaging, necessitating space-based data.  Additionally, to probe a
large dynamic range in luminosities, we need to combine deep studies
over small areas with larger-area
surveys with shallower limiting magnitudes. Our study used imaging
data from a number of 
surveys covering both the Northern and Southern fields from the Great
Observatories Origins Deep Survey \citep{giavalisco04}, with both the
{\it Hubble Space Telescope} ({\it HST}) and the {\it Spitzer Space
  Telescope}.  

The deepest imaging comes from three surveys of the
Hubble Ultra Deep Field (HUDF): the original HUDF survey which
obtained optical imaging with the Advanced Camera for Surveys
\citep[ACS;][]{beckwith06}; and the more recent HUDF09 \citep[PI
Illingworth; e.g.,][]{bouwens10a, oesch10} and UDF12 surveys
\citep[PI Ellis;][]{ellis13,koekemoer13}, which obtained near-infrared imaging
with the Wide Field Camera 3 (WFC3).  The full {\it HST} dataset over the HUDF
comprises imaging in eight bands: F435W, F606W, F775W and F850LP with
ACS, and F105W, F125W, F140W and F160W with WFC3 (hereafter referred to as $B_{435},$ $V_{606}$,
$i_{775}$, $z_{850}$, $Y_{105}$, $J_{125}$, $JH_{140}$ and $H_{160}$,
respectively), which cover an area of $\sim$5 arcmin$^2$.  The HUDF09 survey also obtained deep WFC3 imaging over
two similarly-sized flanking fields, first observed with ACS in the UDF05 survey \citep[PI
Stiavelli;][]{oesch07}, referred to as the HUDF09-01 and HUDF09-02
fields \citep{bouwens11}.  These fields each have imaging in the $V_{606}$,
$i_{775}$, $z_{850}$, $Y_{105}$, $J_{125}$, and $H_{160}$ bands.

The majority of our candidate galaxy sample comes from the
Cosmic Assembly Near-infrared Deep Extragalactic Legacy Survey
\citep[CANDELS; PIs Faber and Ferguson;][]{grogin11,koekemoer11}.  CANDELS
is the largest {\it HST} project ever, comprising 902 orbits over five
extragalactic deep fields, including the two GOODS fields
\citep{giavalisco04}.  CANDELS, which finished
in August 2013, is composed of a deep and a wide survey.  The
deep survey covers the central $\sim$50\% of each of the two GOODS
fields, while the wide survey covers the remainder of the GOODS-N
field, and the southern $\sim$25\% of the GOODS-S field to depths
$\sim$ 1 mag shallower than the deep survey (the wide survey also
covers three additional fields not used in this study; see \S 6.4.1
and Figure~\ref{fig:colcol}).  We use
ACS imaging from the original GOODS survey in the $B_{435}$,
$V_{606}$, $i_{775}$ and $z_{850}$ bands.  We use the most recent
ACS mosaics in these fields,
which in GOODS-S includes all ACS imaging in that field prior to the ACS repair
on Servicing Mission 4 in 2009, and in the GOODS-N field
includes all ACS imaging from the GOODS survey (CANDELS internal
team release versions 3 and 2, respectively).  The
CANDELS imaging in both the deep and wide regions of both GOODS fields
includes the $Y_{105}$, $J_{125}$ and $H_{160}$ bands.  We add to our GOODS-S dataset imaging over the
northern $\sim$25\% of the GOODS-S field from the WFC3 Science
Oversight Committee's Early Release Science (ERS) program \citep[PI
O'Connell;][]{windhorst11}, which also includes $J_{125}$ and
$H_{160}$ imaging, as well as the F098M (hereafter referred to as
$Y_{098}$) band.  Unless otherwise distinguished, throughout the paper
we will refer to $Y_{098}$ and $Y_{105}$ together as the $Y$-band
(both filters probe observed 1$\mu$m light, but the $Y_{098}$ filter
is narrower and thus has a higher spectral resolution).

Finally, we complete our dataset with the recently obtained deep {\it
  HST} observations near the galaxy clusters Abell 2744 and MACS J0416.1-2403
(hereafter MACS0416) from the Hubble
Frontier Fields (HFF) program (PI Lotz).  
For this study, we use only the parallel (unlensed) fields.  Both
fields have been observed in the $B_{435}$, $V_{606}$, $I_{814}$, $Y_{105}$, $J_{125}$,
$JH_{140}$ and $H_{160}$ bands.  We use these data to complement our
candidate galaxy samples at $z =$ 5, 6, 7 and 8 (excluding $z =$ 4 due
to the reduced number of optical bands).

In parallel to the primary WFC3 observations, CANDELS obtained
extremely deep imaging in the F814W band (hereafter
$I_{814}$) in both of the GOODS fields.  As these data were obtained
recently, they suffers from poor
charge transfer efficiency.  Although algorithms have been devised to
correct for this \citep{anderson10}, as the CANDELS fields have
imaging in both the $i_{775}$ and $z_{850}$ bands, we do not include
the CANDELS $I_{814}$ photometry in the initial photometric redshift
fitting (though we do explore its inclusion in \S 3.6).  However, we
did use these very deep data during our visual
inspection step, which was highly useful at $z =$ 8, where true $z =$ 8 galaxies should be
completely undetected in the $I_{814}$-band.  In the HFF parallel
fields, where the $I_{814}$ band is the only imaging covering the red
end of the optical, we used these data in the full analysis.

The description of the CANDELS {\it HST} imaging reduction is available from
\citet{koekemoer11}.  These reduction steps were also followed for the
ERS, HUDF \citep{koekemoer13} and HFF data we use here.  
We use imaging mosaics with 0.06\arcs\
pixels, and make use of their associated weight and rms maps.
The combined imaging dataset covers an area of 301.2
arcmin$^2$, with 5$\sigma$ limiting magnitudes in the $H_{160}$ band
ranging from 27.4 to 29.7 mag (measured in 0.4\arcs\ diameter apertures).  These
datasets are summarized in Table 1.

\subsection{Point Spread Function Matching}

The {\it HST} imaging used here spans more than a factor of
three in wavelength, thus the differences in point-spread function
(PSF) full-width at half-maximum (FWHM) across that range are
significant.  For example, the PSF in the GOODS-S Deep field has a
FWHM $=$ 0.193\arcs\ in the $H_{160}$-band, but only 0.119\arcs\ in the
$B_{435}$-band.  A point-source will thus have more of its flux
contained within a 0.4\arcs\ aperture in the $B_{435}$-band compared
to the $H_{160}$-band.  As the selection of distant galaxies relies very heavily on accurate
colors, and we are using apertures of fixed sizes (determined
by the detection image, see \S 2.3) to measure photometry in all bands, this
changing PSF needs to be addressed.

We corrected for this by matching the PSF of the {\it
  HST} imaging to the $H_{160}$-band image (which has
the largest PSF FWHM) in each field.  We did this using the IDL \textit{deconv\_tool}
Lucy-Richardson deconvolution routine, in the same way as 
\citet{finkelstein10, finkelstein12a}.  This routine requires 
the PSF for a given band as well as a reference PSF (in this case,
the $H_{160}$-band), and it generates a kernel.  The PSFs were
generated by stacking stars in each field in each band, where the
stars were selected via identifying the stellar locus in a half-light
radius versus magnitude plane.  Each star was then visually inspected
to ensure that there were no bright near-neighbors, and then the stars were
stacked, subsampling by a factor of 10 to ensure an accurate
centroiding of each star (i.e., to avoid smearing the PSF during the stacking).
Using these PSFs, the deconvolution routine performed an iterative process, and relies on the user to determine the
number of iterations.  We did this by
making a guess as to the correct number of iterations, and then
changing this number until the stars in the PSF-matched images in a
given band had curves-of-growth which matched the $H_{160}$-band
curves-of-growth to within 1\% at a radius of 0.4\arcs.  The images
were then convolved with the final kernel to generate PSF-matched
images.  

\subsection{Photometry}
Photometry was measured on the PSF-matched dataset with a modified
version of the Source Extractor software
\citep[v2.8.6,][]{bertin96}.  Our modified version adds a buffer
between the source and the local background cell and removes spurious
sources associated with the distant wings of bright objects. 
Catalogs were generated independently in each of our ten sub-fields,
using Source Extractor
in two-image mode, where the same detection image was used to measure
photometry from all available {\it HST} filters.  For most of our
fields, we used a weighted sum of the F125W and F160W images as the
detection image, to increase our sensitivity to faint objects.  In
the HUDF main field and the MACS0416 and A2744 HFF parallel fields,
we supplemented this catalog with catalogs using 10 additional detection
images, derived by stacking all possible combinations of adjacent WFC3
filters.  In these three fields, a combined catalog was made up of all
unique sources in the catalogs, using a 0.2\arcs\ matching
radius.  This allowed very blue sources that may be too faint in the
$H_{160}$ image to be selected
in the original F125W+F160W-selected catalog to be included.  This
procedure was replicated in our completeness simulations (\S 4).
To derive accurate flux uncertainties, Source
Extractor relies on both an accurate rms map, and a realistic estimate
of the effective gain.  The provided rms map has been shown to produce
accurate uncertainties, and it has been corrected for pixel-to-pixel
correlations which occur as a result of the drizzling process
\citep[see][]{guo13} which are typically on the order of 10 - 15\% of the total rms.
The effective gains were computed for each band separately as the the
instrument gain (1 for ACS, 2.5 for WFC3/IR) $\times$ the
total exposure time for a given image.  We have previously
verified that the uncertainties measured in this manner on {\it HST}
imaging are accurate \citep{finkelstein12a}.  The zero-points to
convert the observed fluxes into AB magnitudes are given in Table 1,
and are appropriate for the dates when these data were
taken.

Following our previous work
\citep{finkelstein10,finkelstein12b,finkelstein12a,finkelstein13},
colors were measured in small Kron apertures with the Source Extractor
Kron aperture parameter PHOT\_AUTOPARAMS set to values of 1.2 and
1.7.  \citet{finkelstein12a} found that these apertures result in more
reliable colors for faint galaxies when compared to isophotal or
small circular apertures.  An aperture correction to the total flux
was derived in the $H$-band and was computed as the ratio between
this small Kron aperture flux, and the default Source Extractor
MAG\_AUTO flux, which is computed with PHOT\_AUTOPARAMS $=$ 2.5, 3.5.
These aperture corrections were then applied to the fluxes in all filters.
To see if our aperture corrections accurately recovered the total
flux, we examined our completeness simulations (discussed in \S 4), and
found that after applying this aperture correction recovered fluxes were
typically 5\% fainter in each band than their input fluxes (with the
exception of the HUDF main field, where the measured correction was 2\%).  We thus
increased the flux in each band by the appropriate factor to derive our best estimate of
the total flux.

\begin{figure*}[!t]
\epsscale{0.385}
\plotone{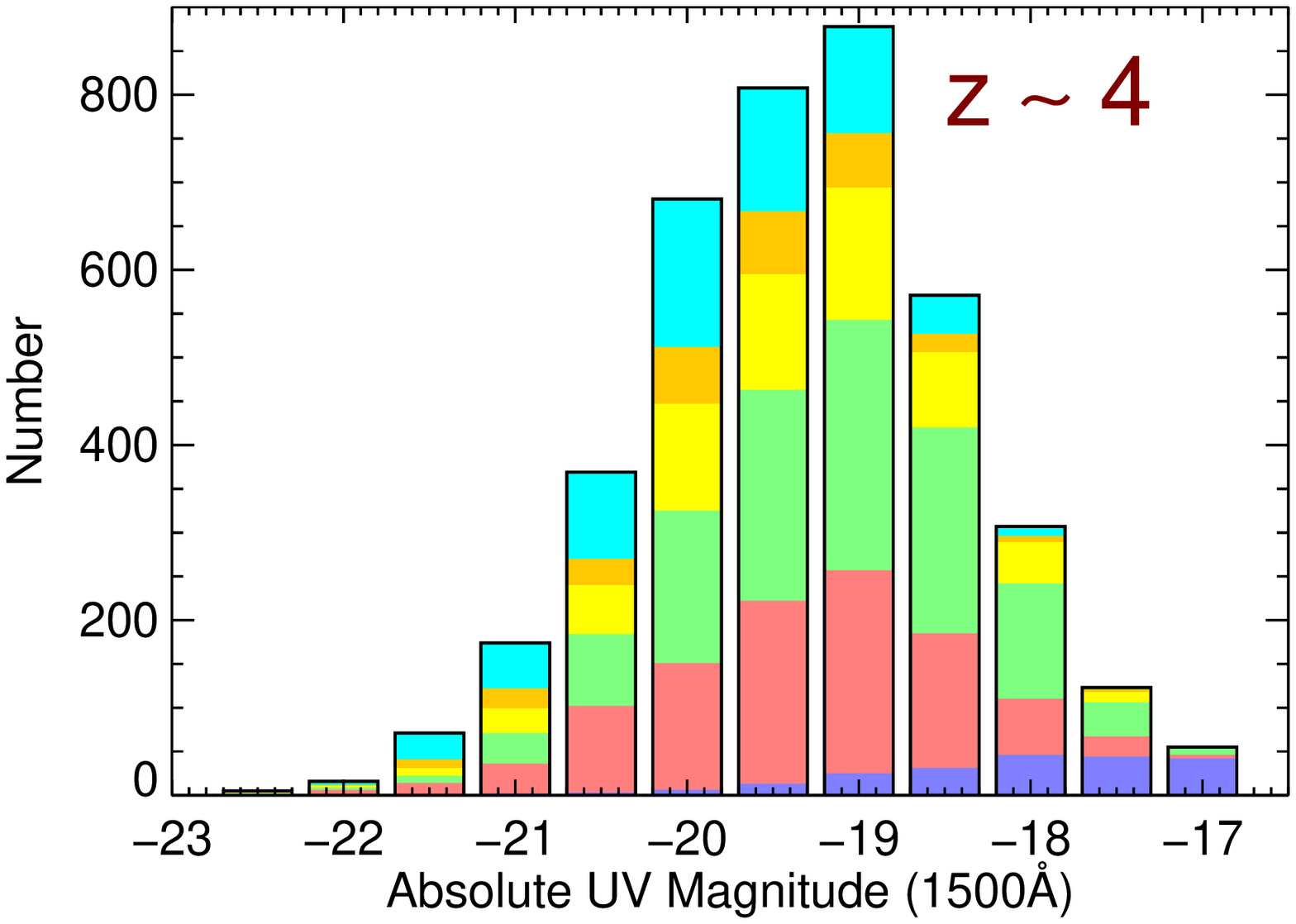}
\plotone{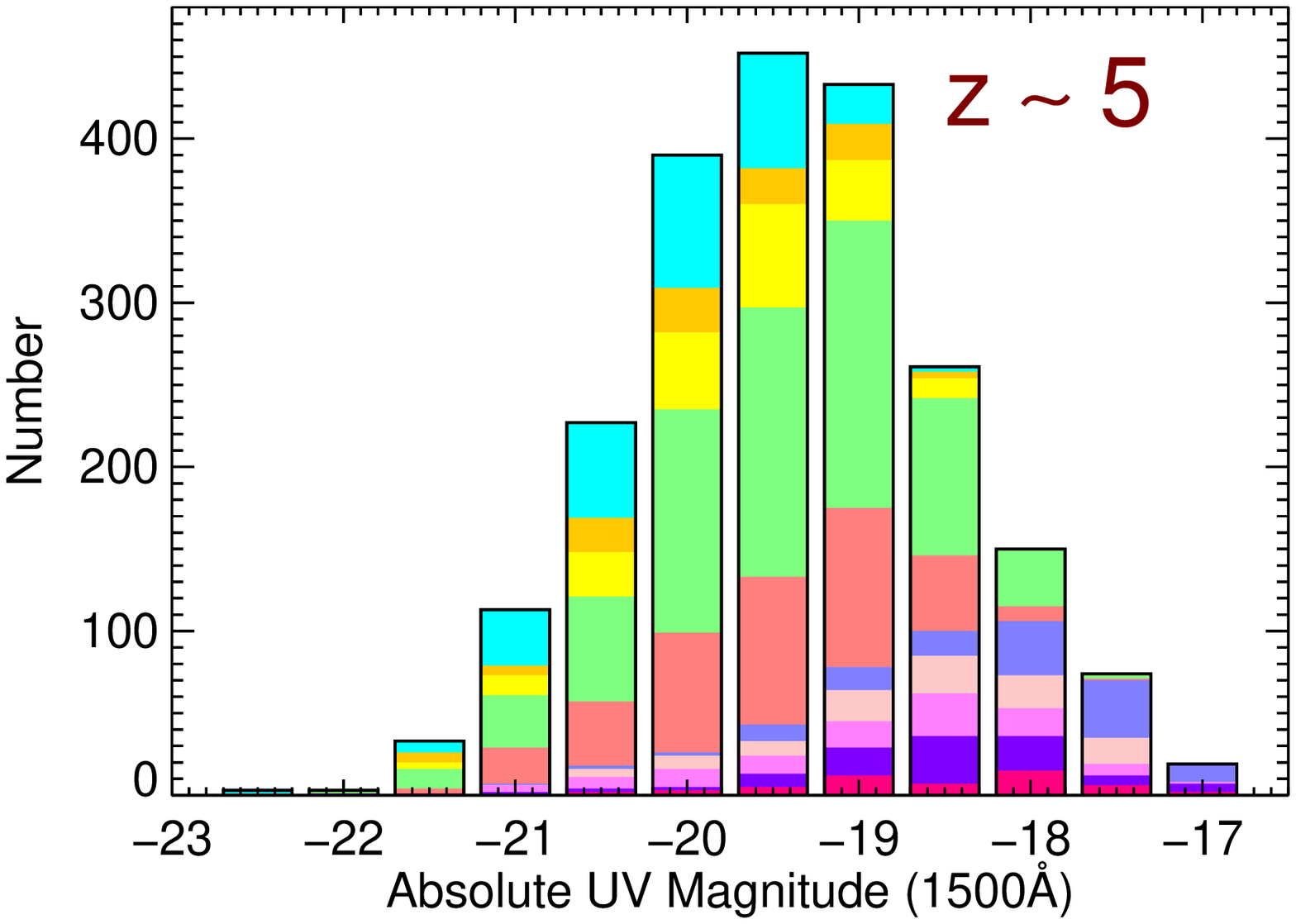}
\plotone{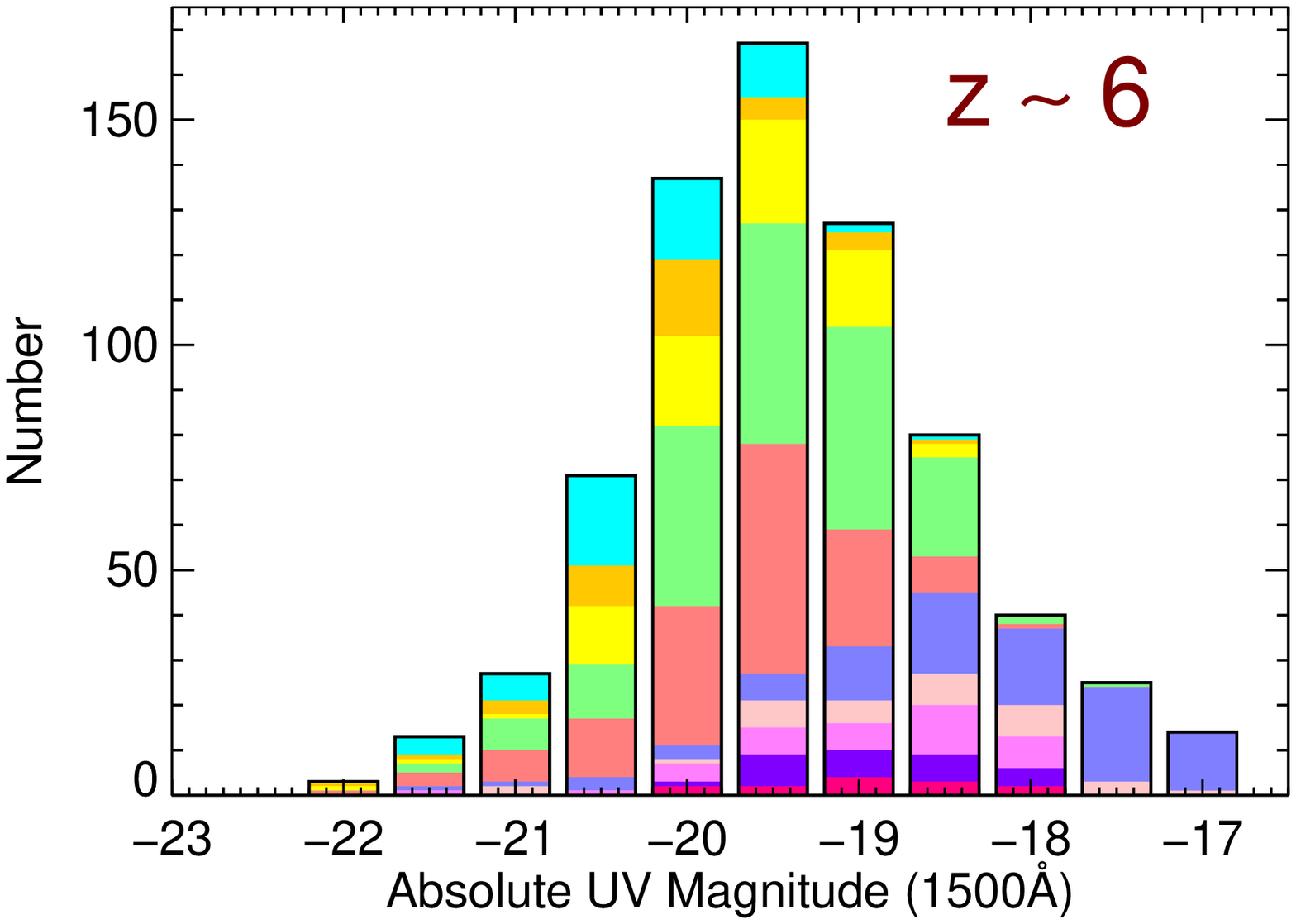}
\plotone{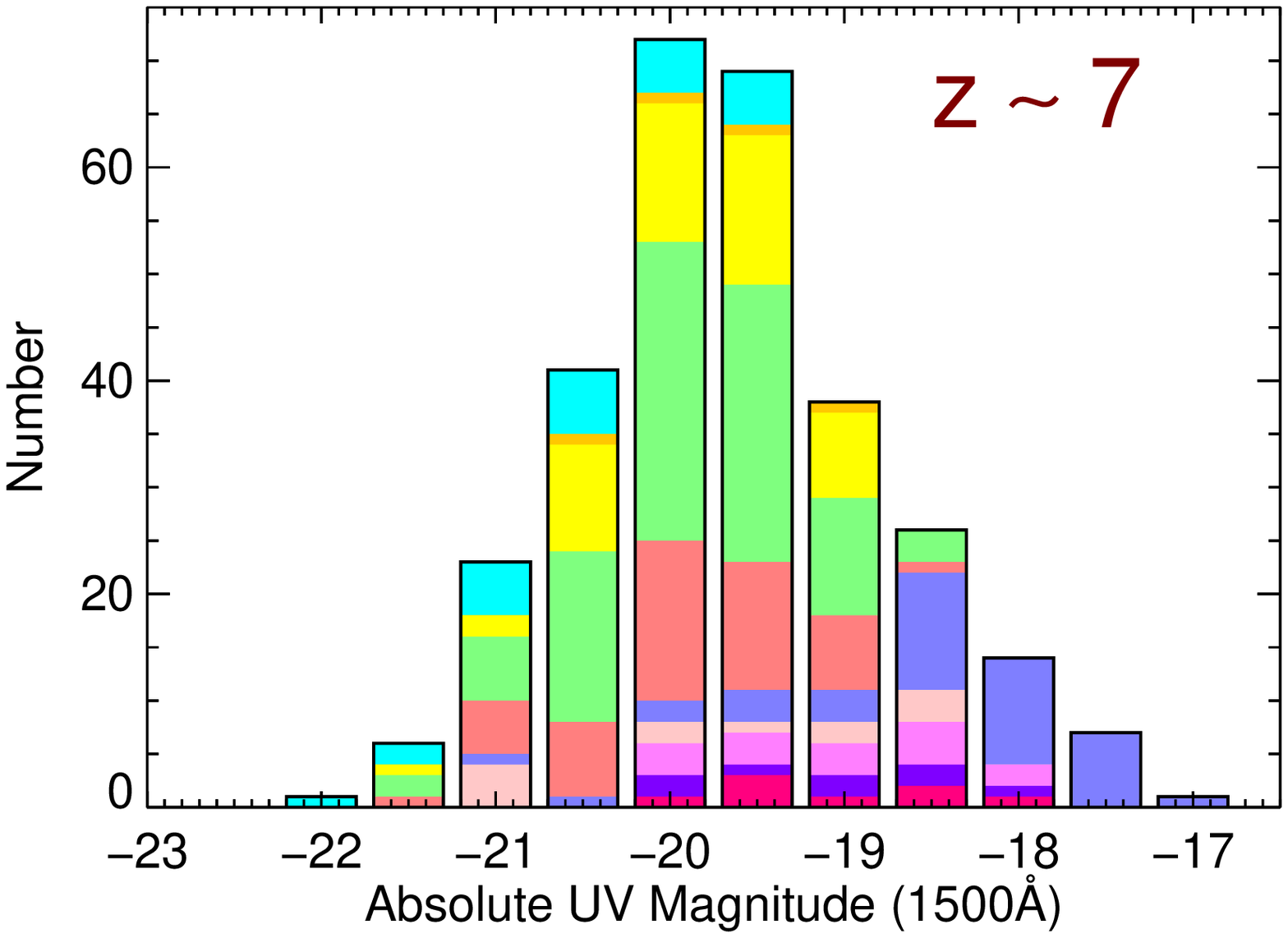}
\plotone{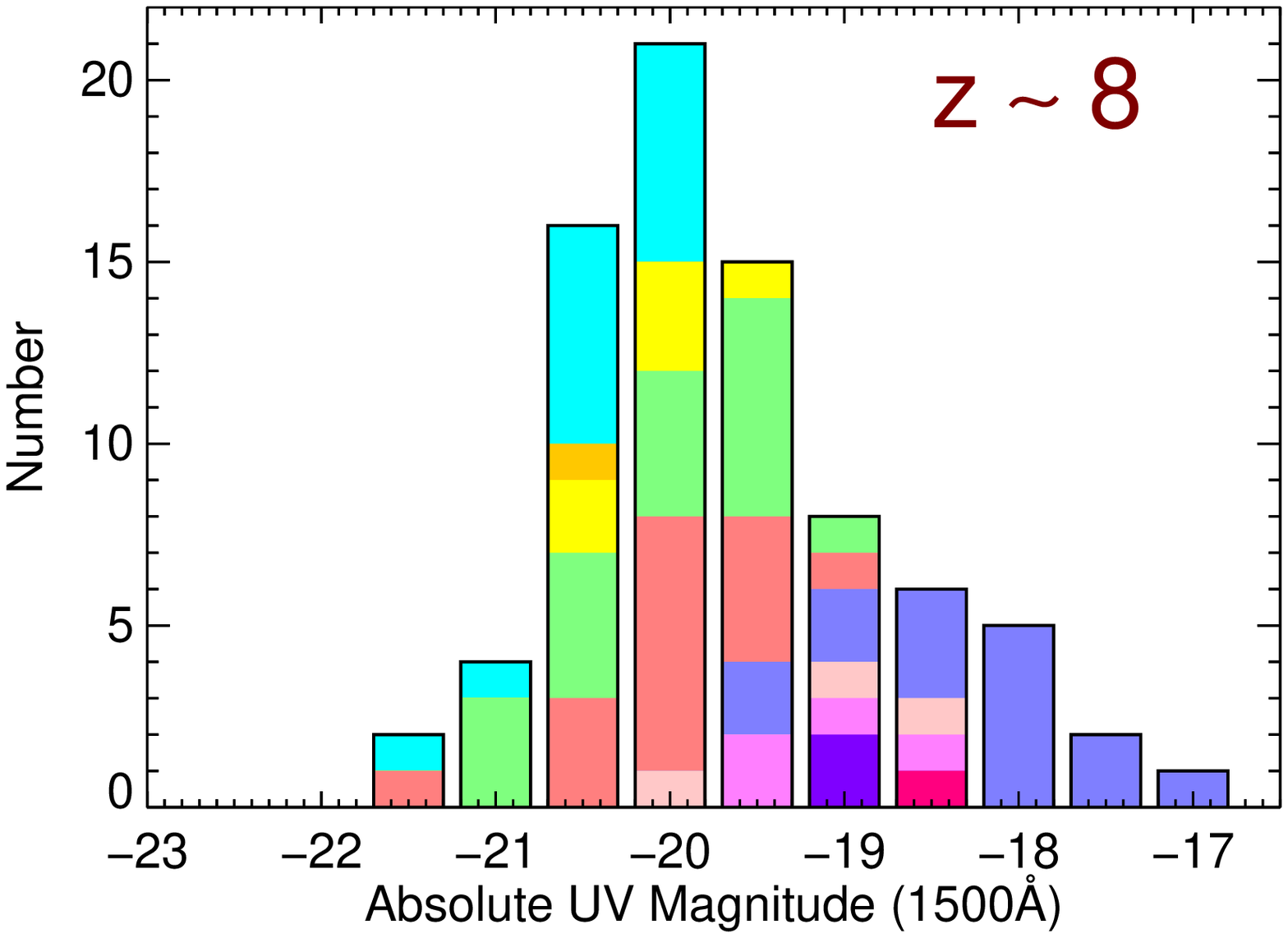}
\plotone{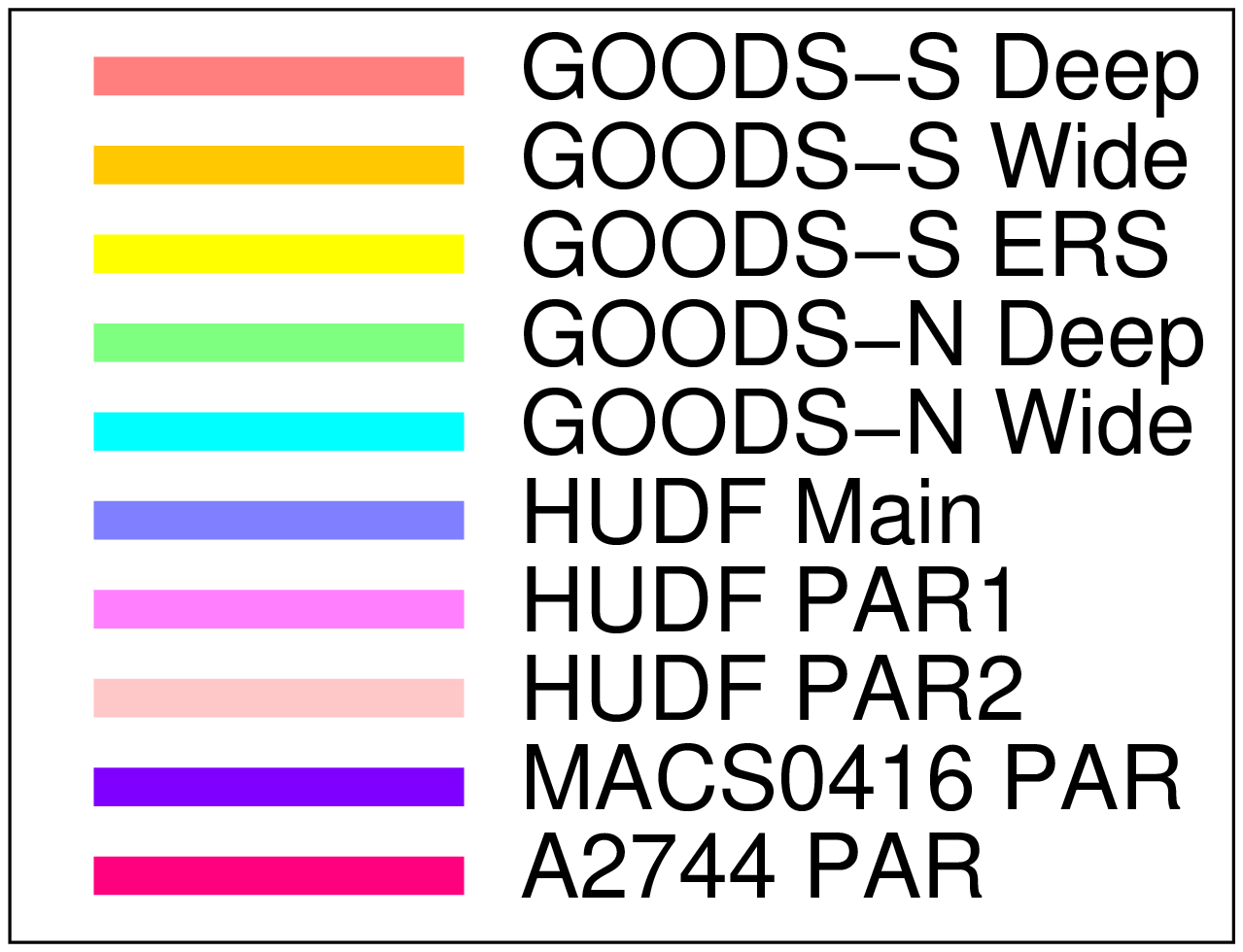}
\caption{The absolute magnitude distribution of all candidate galaxies in our redshift
  samples.  The shaded color denotes which of the sub-fields a given galaxy was detected in.
  This figure demonstrates that while the HUDF is useful for finding
  the faintest galaxies, the CANDELS imaging is necessary to
  discover much larger numbers, as well as to probe a large dynamic
  range in luminosity.}
\label{fig:nums}
\end{figure*}

The Source Extractor catalogs from each band were
combined into a master catalog for each field.  At this step, the
observed fluxes were corrected for Galactic extinction using the color
excess E(B-V) from \citet{schlafly11} appropriate for a given field\footnote[1]{http://ned.ipac.caltech.edu/forms/calculator.html}, and using the \citet{cardelli89} Milky Way reddening
curve to derive the corrections based on each filter's central
wavelength.  We used a mask image to remove objects in regions of bad
data, where the mask was generated using a threshold value from the
weight map.  This mask primarily trims off the noisier edges of the
imaging, but it also excludes the ``death star'' region on the WFC3
array where the number of dithers was low (i.e., in the CANDELS
Wide regions).  The areas quoted in Table 1 are those of the good
regions in these masks.  Objects were also removed from the catalog if
they had a negative aperture correction, which applied to a very
small number of sources, primarily restricted to areas near very
bright sources where the flux in the larger aperture was unreliable.  The remaining objects comprised our final catalog in
each field.

\section{Sample Selection}

\subsection{Photometric Redshifts}
We selected our candidate high-redshift galaxy sample via a
photometric redshift fitting technique.  This has the advantage in
that it uses all of the available photometry simultaneously, rather
than the multi-step Lyman break galaxy (LBG) method, which selects
galaxies using two colors, and then subsequently imposes a
set of optical non-detection criteria
\citep[e.g.,][]{bouwens07,bouwens11}.  Another advantage of
photometric redshifts is that one obtains a redshift probability distribution
function (PDF), which not only allows one to have a better estimate of the
redshift uncertainty ($\sigma_z$ typically $\sim$0.2--0.3 versus 0.5
for the LBG technique), but can also be used as a tool in the
construction of the sample itself.  A potential disadvantage of
photometric redshift techniques is that the results are based on a set
of assumed template spectra; if these templates do not encompass
properties similar to the galaxies being studying, systematic offsets
may occur (though we also note that similar templates are used to
construct LBG color selection criteria).  That being said, initial work
comparing the differences between galaxy samples selected via both LBG
and photometric redshift techniques found that the resulting sample
properties are fairly similar \citep[e.g.,][]{mclure13,schenker13}.

Photometric redshifts for all sources in the catalogs for each fields
were measured using the EAZY software \citep{brammer08}.  The input
catalog used all available {\it HST} photometry, with the exception of
the F814W imaging in the CANDELS fields, which was used solely for
visual inspection (see \S 2.1).
We used an updated set of templates provided with EAZY based on the
P\'{E}GASE stellar population synthesis models \citep{fioc97}, which
now include an increased contribution from emission lines, as recent evidence points
to strong rest-frame optical emission lines being ubiquitous amongst
star-forming galaxy populations at high-redshift
\citep[e.g.,][]{atek11,finkelstein11a,vanderwel11,smit14,stark13,finkelstein13}.  
EAZY assumes the intergalactic medium (IGM)
prescription of \citet{madau95}.  EAZY does have the option to include magnitude priors
when fitting photometric redshifts, which uses the
luminosity functions as a prior for whether a galaxy at a given
apparent magnitude resides at a given redshift.  As we show later,
there is still non-negligible uncertainty at the bright end of the
luminosity function, therefore we did not include these magnitude priors during our 
photometric-redshift fitting process.

\subsection{Selection Criteria}

We selected candidate galaxy samples in five redshift bins centered at $z
\sim$ 4, 5, 6, 7 and 8 with $\Delta~z =$ 1, using criteria similar to our previous work
\citep{finkelstein12a,finkelstein13}.  The cosmic time
elapsed between our last two bins at $z \approx$ 7 and $z \approx$ 8
is $\sim$125 Myr.  This time is much longer than the dynamical time of
the systems we study, and thus leaves significant time for evolution.
However, as studies of galaxy
evolution move towards higher redshift, this will not always be the
case (e.g., $\Delta~t_{z=13 \rightarrow 12} =$ 40 Myr) thus future
studies with the {\it James Webb Space Telescope}
will need to pay careful attention to the choice of sample redshifts
when studying galaxy evolution.

Rather than relying solely on the best-fit redshift
value, we used the full redshift probability distribution curves $P(z)$
calculated by EAZY (where $P(z) \propto$ exp($-\chi^2$), normalized to
unity).  Our selection criteria are:
\begin{itemize}

\item[] 1) A $\geq$ 3.5 significance detection in {\it both} the $J_{125}$ and
$H_{160}$ bands.  A requirement of a significant detection in two
bands removes nearly all spurious sources, as the chances of a noise
peak occurring in two images at the
same position are very small (\S 3.8.1).  This requirement also limits our
analysis to galaxies with $z <$ 8.5, as the Lyman break shifts into the
$J_{125}$ band at $z =$ 8.1.

\begin{figure*}
\epsscale{0.475}
\plotone{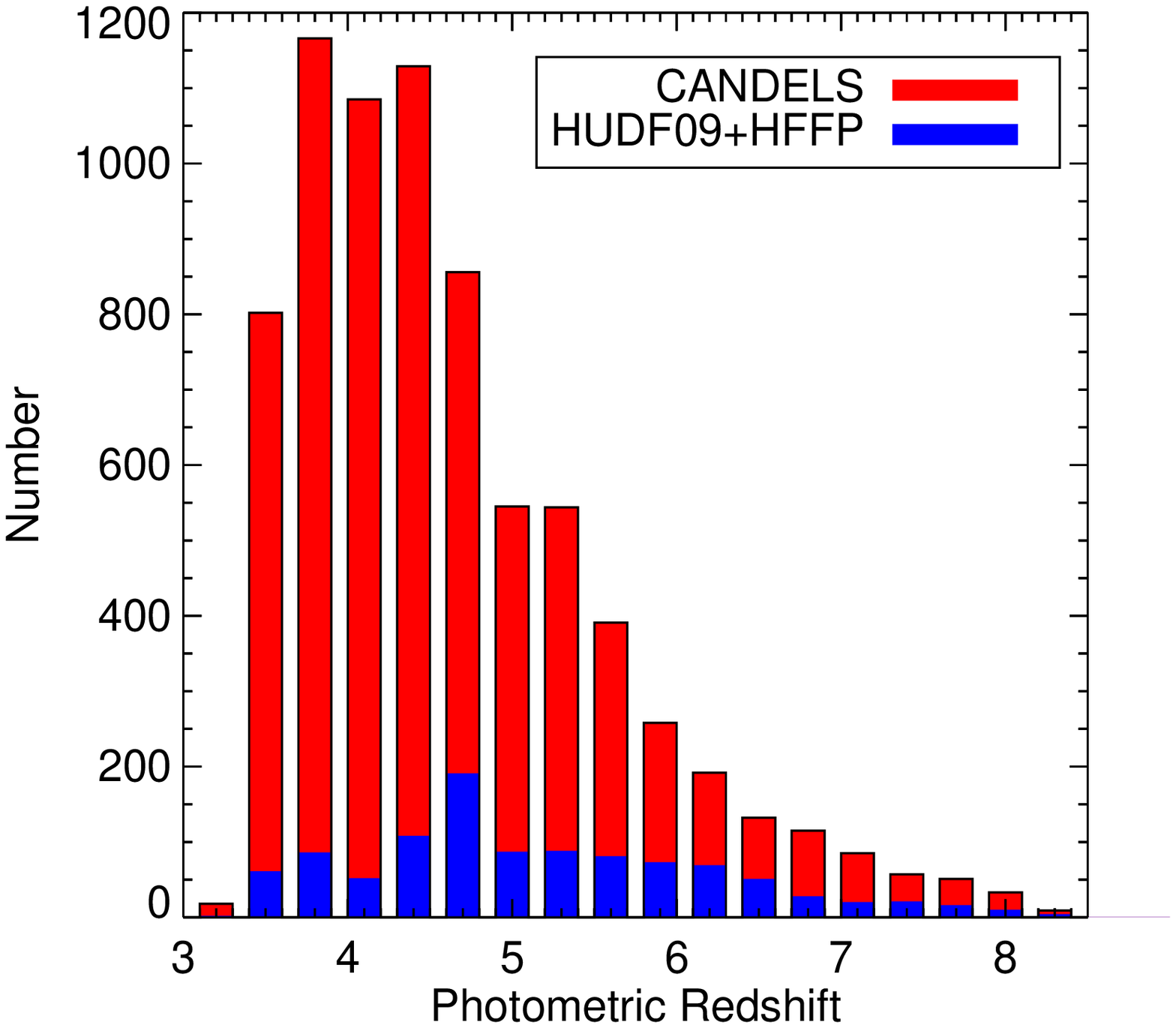}
\hspace{10mm}
\plotone{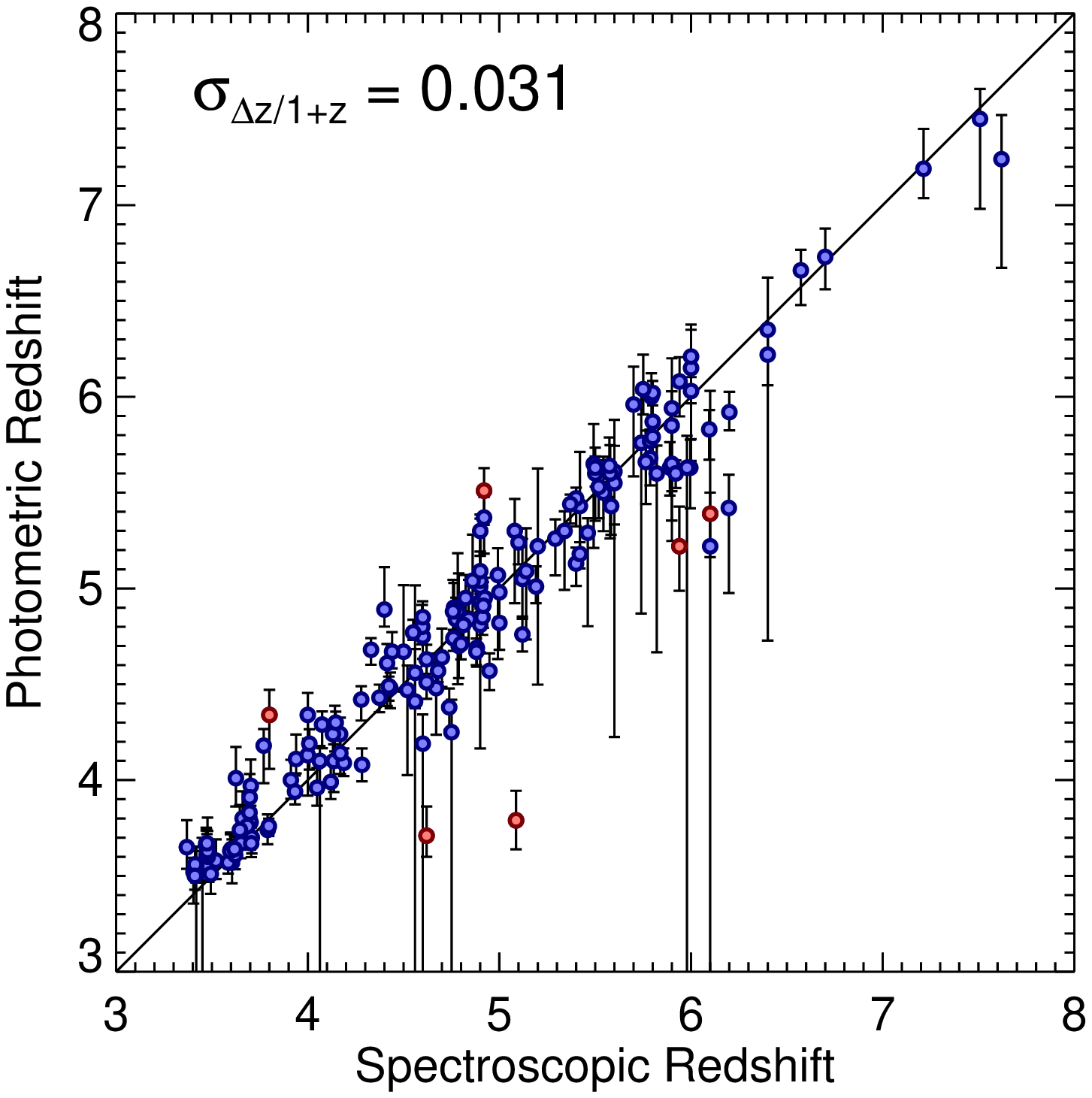}
\caption{Left) The distribution of photometric redshifts in our candidate galaxy
  sample.  The red shading denotes candidates discovered in the
  CANDELS GOODS fields (including the ERS), while the blue shading denotes those in the
  combined five deep fields from the HUDF09 and HFF parallel
  programs.  The shallower yet much wider CANDELS imaging dominate the
  numbers in every redshift bin, by a factor of $\sim$10 at $z =$
  4--5, and $\sim$2 at $z =$ 7--8, though the deep fields are necessary for
  constraints on the faint end of the luminosity function.
Right) A comparison between the spectroscopic redshift and our
  best-fit photometric redshifts for the 171 galaxies in our sample with
  spectroscopic redshifts in the literature.  The red circles denote
galaxies with $|z_{spec} - z_{phot}| >$ 1 at $\geq$3$\sigma$ significance.  There
are only six such galaxies, and all have spectroscopic redshifts at $z
\gtrsim$ 4.}
\label{fig:zcomp}
\end{figure*}

\item[] 2) The integral of the redshift probability distribution function under
the primary redshift peak must comprise at least 70\% of the total integral.  This
enforces that no more than 30\% of the integrated redshift PDF can be in a
secondary redshift solution.

\item[] 3) The integral under the redshift PDF in the redshift corresponding to a
given sample (i.e., 6.5 -- 7.5 for the $z =$ 7 sample) must be at
least 25\%, which ensures that the redshift PDF is not too broad.

\item[] 4) The area under the curve in the redshift range of interest must be
higher than the area in any other redshift range (i.e., for a galaxy
in the $z =$ 7 sample, the integral of $P(6.5 < z < 7.5)$ must be
higher than the integral in any other redshift bin).  This criterion
ensures that a given galaxy cannot be included in more than one redshift sample.

\item[] 5) At least 50\% of the redshift PDF must be above $z_{sample} - 1$
(i.e., $\int P(z > 6) >$ 0.5 for $z_{sample} =$ 7), and the best fit
redshift must be above $z_{sample} - 2$.

\item[] 6) The $\chi^2$ from the fit must be less than or equal to 60.  This
criterion ensures that EAZY provides a
reasonable fit, though in practice it does not reject many sources.

\item[] 7) Magnitude in the $H_{160}$ band must be $\geq$ 22.  This effectively
cleans many stars from our sample, but the limit is still more than
two magnitudes brighter than our brightest $z
\geq$ 6 galaxy candidate.  At $z =$ 4, we do have a few sources close
to this limit, but only two sources are brighter than $H =$ 22.4.
This fact, coupled with the observation that the very few sources at
$H <$ 22 that satisfy our $z =$ 4 selection criteria are either
obvious stars, or diffraction spikes, implies that
this criterion should not significantly affect our luminosity function results.
\end{itemize}
 
Of these criteria, items \#1 and \#2 are by far the most
constraining, as most galaxies which meet these criteria, with
$z_{best} >$ 3.5 make it into our sample.  Items \#3 and \#4 are
responsible for putting a candidate galaxy in a
given redshift sample.  While some of the cuts above are arbitrary,
these choices will be corrected for as we
apply these same criteria to our completeness simulations
discussed in \S 4.  In Figure~\ref{fig:zcomp} we compare the photometric
redshifts for 171 galaxies in our
sample to available spectroscopic redshifts in the literature\footnote[2]{The spectroscopic redshifts come from a compilation
made by N.\ Hathi (private communication) which include data from the following studies:
\citet{szokoly04,grazian06,vanzella08,vanzella09,hathi08,barger08,rhoads09,wuyts09,balestra10,ono12,kurk13,rhoads13,finkelstein13}.}.
The agreement is excellent, with
$\sigma_{\Delta z/(1+z)} =$ 0.031 (derived by taking an iterative
3$\sigma$-clipped standard deviation), though the number
of confirmed redshifts at $z >$ 6.5 is small (only five galaxies).  The number of outliers
is also small, with only six out of 171 galaxies (3.5\%) having a photometric redshift
differing from the spectroscopic redshift by $\Delta z >$ 1 at
$\geq$3$\sigma$ significance.  All of these six galaxies have
$z_{spec} \gtrsim$ 4, thus no galaxies in our sample have a
catastrophically lower spectroscopic redshift.  In comparison, defining outliers in
the same way, we find that the published CANDELS team
photometric-redshift catalog has 13 outliers out of 174 total
spectroscopic redshifts, for a somewhat higher outlier fraction of
7.5\% \citep{dahlen13}.
Although the fraction of galaxies with confirmed redshifts is relatively small, the
available spectroscopy confirms that our selection methods yield an
accurate high-redshift sample.  In the remainder of this paper, we
will therefore refer to our candidate galaxies solely as galaxies,
with the caveat that spectroscopic followup of a much larger sample,
particularly at $z >$ 6, is warranted.

\subsection{Visual Inspection}
As the candidate selection process is automated, for a truly robust
galaxy sample, we required a visual inspection of each of our $\sim$7500
candidate high-redshift galaxies.
During the visual inspection, we examined the following features:
\vspace{1mm}

\begin{figure*}[!t]
\epsscale{1.15}
\plotone{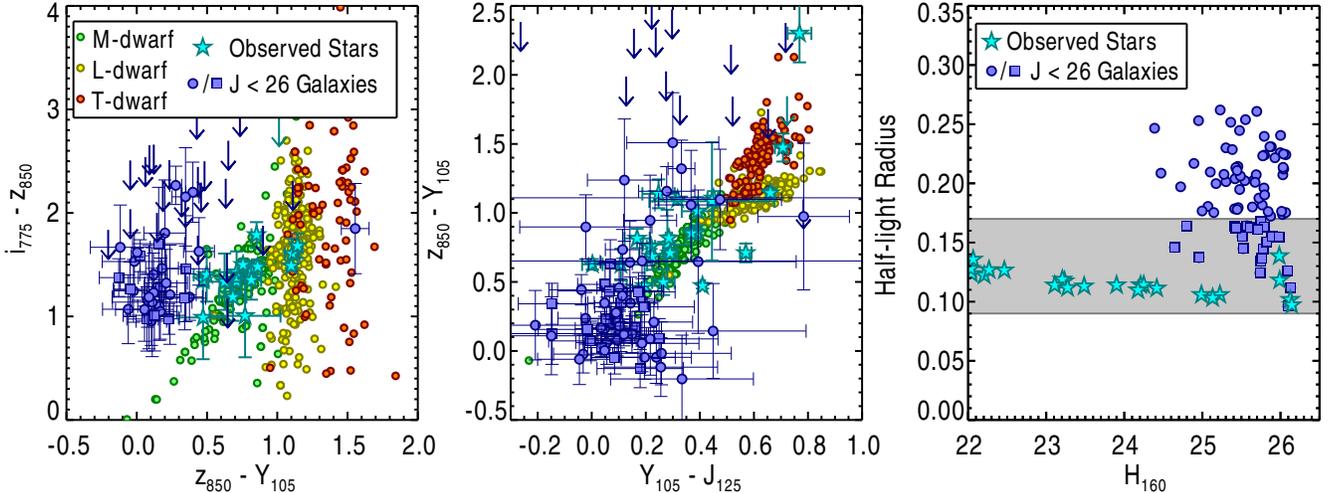}
\caption{Left and Center: Color-color plots.  Blue circles and squares
  denote objects accepted
 as $J<26$, $z \gtrsim 6$ galaxies, with squares indicating the ones
 with half-light radii $<$0.17\arcs.  Arrows represent 1$\sigma$
 limits.  Cyan stars denote candidates
 originally selected as galaxies but reclassified as stars based on
 their sizes
 and colors.  Small circles denote known stars from the the SpeX Prism 
Spectral Libraries from the 3m
NASA Infrared Telescope Facility, with spectral types as indicated in the legend.
 Right: Half-light radius versus magnitude for $J<26$ candidates.
 Symbols are the same as in the other panels.  No compact galaxies
 have colors similar to known stars in both color-color plots.  Similar plots were
used to exclude stellar contaminants at $z =$ 4 and 5.}
\label{fig:stars}
\end{figure*}  

$\bullet$ Is the source a real galaxy?  Objects were inspected to
ensure that they were not an artifact, examples of which are a part of a
diffraction spike (which frequently appear in different places in the
ACS and WFC3 imaging due to different roll angles during the
respective observations), oversplit regions of bright galaxies, or
noise near the edge of the images.
\vspace{1mm}

$\bullet$ Is the aperture drawn correctly?  While the small Kron
apertures yield the most reliable colors, they are also susceptible to
``stretching'' (i.e., becoming highly elongated) in regions of high
noise or near very bright objects.  For each source, we compared the
ratio of the flux between the Kron aperture and a
0.4\arcs-diameter circular aperture to that same quantity for objects
of a similar magnitude from the full photometry catalog.  If an object had a
value $\gtrsim$30\% higher than similarly bright sources in the full photometry
catalog {\it and} the aperture looks to have been affected by
noise/bright sources, we adjusted the photometry of the object in
question accordingly, using the 0.4\arcs-to-total correction of
similarly-bright objects in the catalog.  In practice, these issues
affected $<$10\% of galaxies in our high-redshift sample.
\vspace{1mm}

$\bullet$ Is there significant optical flux that did not get measured
correctly?  Primarily due to the issues with inaccurate apertures
discussed in the above bullet, a very small number of sources appeared
to have optical flux when visually inspected that was not
measured to be significant in our catalog (i.e., in the case of a too-large aperture,
the flux is concentrated in a small number of pixels, while the flux
error comes from the full aperture, so the signal-to-noise is low).
In these cases, objects were removed from our sample.  
This step is somewhat qualitative, as there are cases of objects where
the aperture appears correct, yet there is still a $\sim$1--2$\sigma$
detection in a single optical band.  In the majority of these cases, as we are
confident in our photometric redshift analysis, we left these objects
in the sample.  During this step, we also examined $I_{814}$
photometry for each source in the CANDELS fields, which
primarily benefits the selection of $z =$ 8 galaxies, which should not
be visible at this wavelength.  Three $z =$ 8 candidates with observable
$I_{814}$ flux were removed from our sample.
\vspace{1mm}

\subsection{Stellar Contamination}

The most crucial step in our visual inspection is the classification
and removal of stellar sources, 
as stellar contamination would dominate the bright end of the
luminosity function if these contaminants were not considered.  In
particular, M-type stars as well as L and T brown dwarf stars can
have similar colors (including optical non-detections) as our
high-redshift galaxies of interest, particularly at $z \geq$ 6.
While some studies use dwarf star colors during their
selection \citep[e.g.,][]{bowler12,mclure13,bowler14}, many use primarily
the Source Extractor ``stellarity'' parameter to diagnose
whether a compact object is a star or a galaxy
\citep[e.g.,][]{bouwens11,bouwens15}.  However, the stellarity
parameter loses its ability to discern between a point source and a
resolved source for faint objects.  To test this further, we examined
the stellarity of sources in the CANDELS GOODS catalogs.  At very bright magnitudes
($J_{125} <$ 24), there is a clear separation between stars and
galaxies, with objects either having a stellarity near unity (i.e.,
stars), or having stellarity near zero (i.e., galaxies).  However,
this separation becomes less clear at 
$J_{125} >$ 25, where the stellar and galaxy sequences begin to blend
together.  Therefore, stellarity can be an unreliable star-galaxy separator
at $J_{125} >$ 25, which is similar to the brightness of our brightest $z
\geq$ 7 galaxies.

While the GOODS fields cover relatively small regions on the sky, the
potential number of brown dwarf contaminants, even at $J_{125} >$ 25, is significant.
The Galactic structure model of Ryan \& Reid (in prep) predicts the
surface density of brown dwarfs
in our covered fields.  In the GOODS-S region, using the area covered
by the CANDELS, ERS and HUDF09 observations, we would expect $\sim$6
stars of spectral type M6--T9 with $J_{125}$-band magnitudes between
25 and 27.  The surface density of M6-T9 stars in GOODS-N is similar,
with an expected number of stars in the field of $\sim$5.  Thus, the
expected number of 25 $< J_{125} <$ 27 stars of spectral type M6-T9 in
our whole surveyed region is
$\sim$11.  While this number is small, the numbers of brown dwarfs
are expected to fall off toward fainter magnitudes, thus the majority of these are
likely have  $J_{125}$ close to 25.  This magnitude is similar to
those of the brightest
galaxies in our sample, which dominate the shape of the bright end of
the galaxy luminosity function.  As stellarity is an unreliable method
of identifying these sources, we must find an alternative method.

Although brown dwarfs can have similar colors to $z >$ 6 galaxies, and
can be included in the initial sample, they fall on well-defined color
sequences, and can thus be distinguished from true galaxies.  
Figure~\ref{fig:stars} shows two color-color plots\footnote[3]{This
  research has benefitted from the SpeX Prism Spectral Libraries,
  maintained by Adam Burgasser at
  http://pono.ucsd.edu/$\sim$adam/browndwarfs/spexprism}
which we used 
in tandem with the size information, examining not only stellarity,
but also the FWHM and half-light radius as measured by Source
Extractor, to identify stars lurking our sample (similar plots were
used at $z =$ 4 and 5).  If a galaxy appeared un-resolved (defined as having a
stellarity $>$ 0.8, or a half-light radius and/or FWHM similar to that
of stars in the field) then we examined that object in the color-color
plots as shown in Figure~\ref{fig:stars}.  If the object also had colors similar to
a dwarf star, then we removed it from our sample.  Over all of our
fields, we had a total of 23 objects flagged as stars (many with $J <$
25) in our $z \geq$
6 samples; 18 from our initial
$z \sim$ 6 galaxy sample, and 5 from our initial $z \sim$ 7 galaxy
sample.  These objects were removed from our sample.
One of these stars removed from our $z \sim$ 7 sample was
previously flagged as a probable T-dwarf by \citet{castellano10}.
We examined the subset of eight of these stars which were detected in
the FourStar Galaxy Evolution (zFourGE) medium band imaging survey of
a portion of GOODS-S, and found that all eight have
$z -J1$ and $J1-J3$ colors consistent with brown dwarf stars [][where
$J1$ and $J3$ refer to two of the three medium bands which comprise
the $J$ band]\citep{tilvi13}.
Of these, six stars have $J_{125} >$ 25, meaning that our
high-redshift galaxy selection criteria also originally selected
$\sim$ half of the expected number
of faint brown dwarfs in this field.  Four of these six
stars have Source Extractor stellarity measurements $<$ 0.8, thus a
stellarity-only rejection method would have failed to remove them.
We conclude that our visual inspection step efficiently removed
stellar contaminants from our sample, but we emphasize that the color
examination portion was crucial to exclude the faintest stars from our
sample.

\subsection{Active Galactic Nuclei}
We screened for the presence of bright active galactic nuclei (AGN) in our
sample by searching for counterparts in {\it Chandra X-ray
  Observatory} point source catalogs.  In the GOODS-S field, we used
the 4 Msec Chandra Deep Field -- South (CDF-S) catalog of
\citet{xue11}, and in GOODS-N, we used the 2Msec Chandra Deep Field -- North
catalog of \citet{alexander03}.  These catalogs have average positional
accuracies of 0.42\arcs\ and 0.3\arcs, respectively.  To be
conservative, we searched for matches in each catalog out to a radius of
1\arcs.  We then visually inspected each of the 34 galaxies in our
sample with a match.  Seven objects, all with {\it Chandra} catalog separations
$>$0.6\arcs, had nearby {\it HST} counterparts with positions consistent with
the {\it Chandra} catalog, and thus these interlopers are likely providing the
X-ray emission; none of these sources had spectroscopic redshifts in
the CDF-S catalog.  These seven sources thus remained in our sample.  
The remaining 27 sources, all with separations $\leq$0.6 \arcs, had {\it Chandra}
positions consistent with the X-ray emission coming from the galaxies
in our sample.  Secure spectroscopic redshifts were available for four of
these 27 galaxies in the CDF-S catalog, of $z =$ 3.06, 3.66, 3.70 and
4.76.  These 27 galaxies (25 from our $z =$ 4 sample, and
two from our $z =$ 5 sample) were removed from our galaxy sample.
This removal is conservative, as although the X-ray detections imply
the presence of an AGN, it does not prove that the AGN dominates the
UV luminosity.

\subsection{Photometric Redshifts with {\it Spitzer}/IRAC Photometry}

As we will discuss below, one of the main results of this work is an
apparent constant value of $M^{\ast}_\mathrm{UV}$ at $z >$ 5, brighter than many
previous works.  It is thus imperative that we have high confidence
that our bright galaxies are all in fact at high-redshift, and not
lower-redshift contaminants.  To provide a further check on our bright
sources, we re-examined the photometric redshifts of our bright
galaxies with the addition of {\it Spitzer Space Telescope} Infrared
Array Camera \citep[IRAC;][]{fazio04} imaging over our fields.  
This imaging probes the rest-frame optical at these wavelengths, and
thus provides significant constraining power because the most likely
contaminants are red, lower-redshift galaxies, which
would have very different fluxes in the mid-infrared than true
high-redshift galaxies. 
We examined sources with $M_{1500} < -$21, which is approximately the value of
$M^{\ast}_\mathrm{UV}$ at these redshifts, and provides samples of 164, 85, 29, 18 and
3 bright galaxies at $z =$ 4, 5, 6, 7 and 8, respectively. 

During the cryogenic mission, the
GOODS fields were observed by the GOODS team (Dickinson et al., in prep) at
3.6, 4.5, 5.8, and 8.0\,$\mu$m.  Later, during Cycle 6 of the warm
mission, broader regions encompassing the GOODS footprints were
covered by the {\sl Spitzer} Extended Deep Survey \citep[SEDS][]{ashby13} to 3$\sigma$ depths of 26\,AB mag at both 3.6 and 4.5\,$\mu$m.
A somewhat narrower subset of both fields was subsequently covered by
{\sl Spitzer}-CANDELS \citep[S-CANDELS][]{ashby15}, to
even fainter levels; reaching $\sim0.5$\,mag deeper than SEDS in both
of the warm IRAC bandpasses.  The HUDF09 fields were observed by {\sl Spitzer} program 70145 \citep[the
IRAC Ultra-Deep Field][]{labbe13}, reaching 120, 50 and 100 hr in the
HUDF Main, PAR1 and PAR2 fields, respectively.  Finally, program
70204 (PI Fazio) observed a region in the ERS field to 100 hr depth.
The present work is based on mosaics constructed by coadding all the
above data following the procedures described by Ashby et al. (2013).
The combined data have a depth of $\gtrsim$50 hr over both CANDELS GOODS
fields and $>$100 hr over the HUDF main field.

As the IRAC PSF is much broader than that of {\it HST}, our galaxies
may be blended with other\
nearby sources.  We measure \textit{Spitzer}/IRAC 3.6 and 4.5
$\mu$m photometry by performing PSF-matched photometry on the
combined IRAC data, which reach at least  26.5 mag (3$\sigma$) at 3.6
$\mu$m and 4.5 $\mu$m \citep{ashby15}.  We utilized the {\tt {\tt
    TPHOT}} software (Merlin et al. in prep.), an updated version of {\tt TFIT} (Laidler et
al. 2007), to model low-resolution images (IRAC images) by convolving
{\it HST} imaging with empirically derived IRAC PSFs and simultaneously
fitting all IRAC sources.  Specifically, we used the light profiles and isophotes in the
detection ($J+H$) image
obtained by Source Extractor, and convolved them with a transfer
kernel to generate model images
for the low-resolution data.  These models were then fit to the real
low-resolution images, dilating the segmentation maps of the model
images to account for missing flux on the edges of
galaxies \citep[][]{galametz13}. The fluxes of sources are determined by the
model which best represents the real data.
As the PSF FWHM of the high-resolution image ($H$-band) is negligible
($\sim$0.19\arcs) when compared to those of the low-resolution IRAC images
($\sim$1.7\arcs), we use the IRAC PSFs as transfer kernels. We derive empirical
PSFs by stacking isolated and moderately bright
stars in each field. As our own WFC3 catalog was used as the input for
{\tt TPHOT}, all of our galaxies have IRAC measurements in the {\tt TPHOT} catalogs.
We visually inspected the positions of each of our high-redshift
galaxy candidates in the IRAC images
to ensure no significant contamination from the residuals of nearby
bright galaxies.  If an object was on or near a strong residual, we ignored
the IRAC photometry in the
subsequent analysis.   This was the case for 18/164 galaxies at $z =$
4, 23/85 at $z =$ 5, 3/29 at $z =$ 6, 6/18 at $z =$ 7 and 1/3 at $z =$
8.  With these contaminated fluxes removed, we found that all
remaining $M_{1500} < -$21 galaxies at $z =$ 4--8 had 3.6 $\mu$m
detections of at least 3$\sigma$ significance, with a magnitude range
at $z \geq$ 6 of 22.7 $\leq$ m$_{3.6} \leq$ 25.8.
The full description of our {\tt TPHOT} IRAC
photometry catalog will be presented by M.\ Song et al.\ (in prep).

We reran EAZY for this subsample of bright galaxies, including the {\it
  Spitzer}/IRAC fluxes, as well as photometry from the ACS F814W
filter, which was not included in the original photometric redshift
calculation (see \S2.1).  We examined these updated photometric
redshift results, searching for galaxies in our $z =$ 4 and 5 samples
with $z_{new} <$ 2.5, and in our $z =$ 6, 7 and 8 samples with
$z_{new} <$ 4.  We found 14 out of 164 galaxies in our $z =$ 4 sample
and 14 out of 85 galaxies in our
$z =$ 5 sample with $z_{new} <$ 2.5.  We found one galaxy out of 29 at $z =$ 6
that appears to be better fit with a low-redshift solution, of
$z_{new} =$ 0.9, while zero galaxies in our $z =$ 7 or 8 samples had
preferred low-redshift solutions with the inclusion of IRAC photometry.

\begin{figure}[!t]
\epsscale{1.1}
\plotone{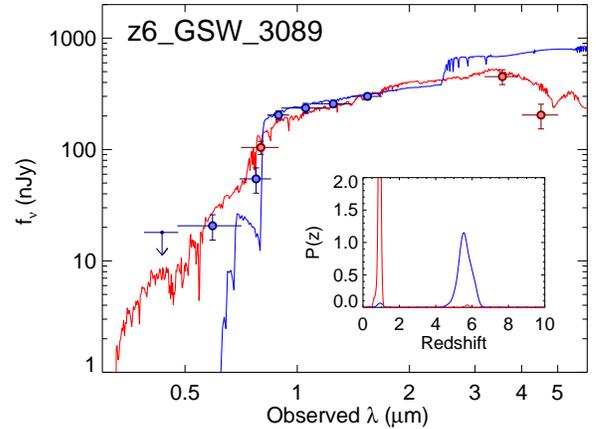}
\caption{The SED of the only galaxy in our 50-object sample of bright
  ($M_{1500} \leq -$21) $z \gtrsim$
  6 galaxies which had a photometric redshift which
  preferred a low-redshift solution after the inclusion of IRAC and
  F814W photometry.  The blue curve shows the original high-redshift
  best-fitting stellar population model and photometric redshift
  probability distribution function, while the red curve shows the
  results including IRAC and F814W.  This galaxy was removed from our
  sample, as the IRAC photometry is consistent with the
  stellar emission peak at $z \sim$ 1.  The inferred contamination
  rate of 2\% (one out of 50 galaxies) is even lower than our
  estimates for $z \gtrsim$ 6 in \S 3.8.}
\label{fig:plotlowz}
\end{figure}  

Examining these results, out of the 28 $z =$ 4 or 5 galaxies with
preferred low-redshift solutions, 23
had photometry consistent with a true low-redshift galaxy.
Four galaxies, however, had photometry which appeared to be
consistent with a high redshift galaxy
with a strong emission line (H$\alpha$ or [O\,{\sc iii}]) in
one IRAC band.  
Systems with lines such as these (i.e., EW$_\mathrm{[OIII]}$ $>$500 \AA) are
rare locally, but appear to be more common at
high-redshift \citep[e.g.,][]{vanderwel11,finkelstein13,smit14}.
Although typical emission lines strengths are now included in the EAZY
templates, these do not account for extreme emission lines, thus it is
not surprising that EAZY does
not return a high-redshift solution.  We elect to keep these four
galaxies in our sample, noting that
the lack of strong rest-frame optical lines in the EAZY templates does
not affect our initial sample selection, which does not make use of
the IRAC photometry.
A fifth galaxy (z5\_GNW\_13415) has a high-quality published
spectroscopic redshift of $z =$ 5.45, thus we also keep it in our sample.  Perhaps unsurprisingly, the
low-redshift fit for this galaxy had a very poor quality-of-fit, with $\chi^2 =$ 127,
implying that the EAZY templates are a poor match for this galaxy.
The remaining 23 galaxies at $z =$ 4 and 5 were removed from our sample.
The sole $z \geq$ 6 galaxy with a preferred low-redshift solution with
the inclusion of IRAC photometry, z6\_GSW\_3089, is shown in
Figure~\ref{fig:plotlowz}.  The red {\it HST} colors imply a much brighter IRAC flux
than is seen.  A solution at $z =$ 0.93 yields a better fit, as the
peak of stellar emission at that
redshift better matches the observed IRAC fluxes.  We have thus
removed this galaxy from our sample.   This galaxy has a spectroscopic redshift of $z =$ 5.59
from the observations of
\citet{vanzella09}.  However, this object has a spectroscopic quality
flag of ``C'', which indicates that the spectroscopic redshift is
unreliable.  This combined with the
$\sim$ 4$\sigma$ detection in the $V_{606}$ band leaves us confident
that a low redshift solution is more likely.

\begin{deluxetable}{ccccc}
\tabletypesize{\small}
\tablecaption{Summary of Final High-Redshift Galaxy Samples}
\tablewidth{0pt}
\tablehead{
\colhead{Redshift} & \colhead{N$_{all}$} & \colhead{N$_{M < -21}$} &
\colhead{V$_{eff}$ (M$_{1500} = -$22)} & \colhead{V$_{eff}$ (M$_{1500} = -$19)}\\
\colhead{$ $} & \colhead{$ $}& \colhead{$ $} & \colhead{(10$^5$ Mpc$^3$)} & \colhead{(10$^5$ Mpc$^3$)}
}
\startdata
4 (3.5 -- 4.5)&4156&150&12.2&4.11\\
5 (4.5 -- 5.5)&2204&\phantom{1}77&8.98&3.36\\
6 (5.5 -- 6.5)&\phantom{1}706&\phantom{1}28&7.93&2.50\\
7 (6.5 -- 7.5)&\phantom{1}300&\phantom{1}18&6.99&0.30\\
8 (7.5 -- 8.5)&\phantom{11}80&\phantom{11}3&5.88&0.16
\enddata
\tablecomments{The total number of sources in our final galaxy sample,
after all contaminants were removed.  The final two columns give the
total effective volume at each redshift for two different values of
the UV absolute magnitude.}
\end{deluxetable}

With the removal of these likely contaminants, we retain a total
sample of 150, 77, 28, 18 and 3 $M_{1500} < -$21 galaxies at $z =$ 4, 5, 6, 7
and 8, respectively.  The fraction of contaminants at $z
\geq$ 6 (one out of 50 $z =$ 6, 7, and 8 galaxies, or 2\%) is
consistent with (albeit somewhat less than) the expected low value
calculated in \S 3.8 below.

Our final galaxy sample is summarized in Table 2, and a catalog of all
galaxies in our sample is provided in Table 3.  In Figure~\ref{fig:nums} we show
the absolute UV magnitude distribution of our final samples,
highlighting that we cover a dynamic range of five magnitudes.  In
particular, the CANDELS data are crucial, as galaxies
from these data dominate the total number of galaxies in our
sample, and approximately double the luminosity
dynamic range which we can probe.  This is highlighted in the left
panel of Figure~\ref{fig:zcomp}, which shows that galaxies discovered
in the CANDELS GOODS fields dominate the total number at all redshifts
in our sample.

\begin{deluxetable*}{lllccc}
\tabletypesize{\small}
\tablecaption{Catalog of Candidate Galaxies at 3.5 $\lesssim z
  \lesssim$ 8.5}
\tablewidth{0pt}
\tablehead{
\colhead{Catalog ID} & \colhead{IAU Designation} & \colhead{RA} & \colhead{Dec} &
\colhead{$z_{phot}$} & \colhead{M$_{1500}$}\\
\colhead{$ $} & \colhead{$ $} & \colhead{(J2000)}& \colhead{(J2000)} & \colhead{$ $} & \colhead{(AB mag)}
}
\startdata
z4\_GSD\_27037&HRG14 J033240.8$-$275003.1&\phantom{1}53.169922&$-$27.834183&3.54 (3.45 to 3.63)&$-$20.45 ($-$20.57 to $-$20.42)\\
z4\_ERS\_3675&HRG14 J033235.0$-$274117.5&\phantom{1}53.145882&$-$27.688189&4.08 (3.79 to 4.28)&$-$20.39 ($-$20.47 to $-$20.15)\\
z4\_GND\_29830&HRG14 J123718.1+621309.7&189.325211&\phantom{$-$}62.219368&4.01 (3.85 to 4.17)&$-$19.68 ($-$19.81 to $-$19.54)\\
z4\_GND\_30689&HRG14 J123721.4+621259.2&189.339355&\phantom{$-$}62.216450&3.66 (3.59 to 3.75)&$-$21.09 ($-$21.16 to $-$21.03)\\
z5\_GSD\_8969&HRG14 J033216.2$-$274641.6&\phantom{1}53.067379&$-$27.778219&5.00 (4.87 to 5.14)&$-$20.62 ($-$20.68 to $-$20.51)\\
z5\_GND\_31173&HRG14 J123731.0+621254.2&189.379272&\phantom{$-$}62.215046&4.85 (4.37 to 5.09)&$-$19.59 ($-$19.77 to $-$19.43)\\
z5\_MAIN\_3271&HRG14 J033243.5$-$274711.4&\phantom{1}53.181351&$-$27.786510&5.50 (4.58 to 5.67)&$-$16.95 ($-$17.05 to $-$16.78)\\
z5\_PAR2\_3762&HRG14 J033304.8$-$275234.7&\phantom{1}53.270004&$-$27.876295&4.47 (3.71 to 4.70)&$-$18.83 ($-$18.97 to $-$18.57)\\
z6\_GND\_16819&HRG14 J123718.8+621522.7&189.328232&\phantom{$-$}62.256317&5.55 (5.41 to 5.65)&$-$21.56 ($-$21.62 to $-$21.47)\\
z6\_GNW\_16070&HRG14 J123549.0+621224.8&188.954025&\phantom{$-$}62.206898&5.88 (5.64 to 6.02)&$-$20.83 ($-$20.88 to $-$20.64)\\
z6\_MAIN\_2916&HRG14 J033244.8$-$274656.8&\phantom{1}53.186806&$-$27.782433&6.42 (5.79 to 6.76)&$-$18.39 ($-$18.55 to $-$18.19)\\
z6\_MACS0416PAR\_145&HRG14 J041632.2$-$240533.3&\phantom{1}64.134117&$-$24.092587&5.91 (5.08 to 6.23)&$-$18.49 ($-$18.64 to $-$18.16)\\
z7\_GSD\_12285&HRG14 J033206.7$-$274715.8&\phantom{1}53.028114&$-$27.787714&7.30 (6.44 to 7.89)&$-$19.57 ($-$19.79 to $-$19.31)\\
z7\_ERS\_6730&HRG14 J033216.0$-$274159.2&\phantom{1}53.066677&$-$27.699766&6.74 (5.64 to 6.87)&$-$20.31 ($-$20.31 to $-$19.96)\\
z7\_GND\_16759&HRG14 J123619.2+621523.2&189.079834&\phantom{$-$}62.256454&6.69 (6.33 to 6.89)&$-$20.89 ($-$20.98 to $-$20.76)\\
z7\_A2744PAR\_4276&HRG14 J001357.5$-$302358.3&\phantom{10}3.489512&$-$30.399530&6.51 (6.26 to 6.78)&$-$19.37 ($-$19.47 to $-$19.22)\\
z8\_GSD\_16150&HRG14 J033213.9$-$274757.7&\phantom{1}53.057983&$-$27.799349&7.91 (6.21 to 8.54)&$-$20.14 ($-$20.38 to $-$19.91)\\
z8\_MAIN\_5173&HRG14 J033241.5$-$274751.0&\phantom{1}53.172874&$-$27.797487&8.11 (6.29 to 8.64)&$-$17.67 ($-$17.98 to $-$17.41)\\
z8\_GND\_32082&HRG14 J123727.4+621244.4&189.364258&\phantom{$-$}62.212334&7.64 (7.02 to 8.16)&$-$20.27 ($-$20.33 to $-$20.02)\\
z8\_GND\_8052&HRG14 J123704.8+621718.8&189.270020&\phantom{$-$}62.288559&8.10 (7.04 to 8.41)&$-$20.68 ($-$20.84 to $-$20.44)
\enddata
\tablecomments{A catalog of our 7446 $z =$ 4--8 galaxy candidates, with
their derived properties. We include the IAU designation for
continuity with previous and future works, with a designation
prefix HRG14 denoting ``High Redshift Galaxy 2014''.  The values in parentheses represent the
68\% confidence range on the derived parameters. Here, we show 20
representative galaxies, four from each redshift
bin.  This table is available in its entirety in a machine-readable
form in the online journal. A portion is shown here for guidance
regarding its form and content.}
\end{deluxetable*}

\subsection{Stellar Population Modeling}
To derive the rest-frame absolute magnitude at 1500 \AA\ ($M_{1500}$), as well as
the UV spectral slope $\beta$ \citep[f$_{\lambda} \propto
\lambda^{\beta}$][]{calzetti94}, we fit spectral energy distributions
(SEDs) from synthetic stellar population models to the observed
{\it HST} photometry of our high-redshift candidate galaxies.  The technique
used here is similar to our previous works
\citep{finkelstein10,finkelstein12a,finkelstein12b,finkelstein13}.  We
used the updated (2007) stellar population synthesis
models of \citet{bruzual03} to generate a grid of spectra, varying the
stellar population metallicity, age, and star-formation
history\footnote[4]{These models may overestimate the
  contribution of thermally pulsating asymptotic giant branch (TP-AGB)
stars.  However, these stars typically begin to dominate the emission at
population ages $\gtrsim$1 Gyr.  
Additionally, though the TP-AGB contribution may impact the SED in
post-starburst galaxies at wavelengths as low as 0.5 $\mu$m \citep{kriek10},
our longest wavelength filter (1.6 $\mu$m) at our
lowest redshift ($z =$ 4) probes only 0.3 $\mu$m, and all other
filter/redshift combinations probe bluer rest-frame wavelengths.
Thus, our choice to use the updated models should have no effect on our results.}.
Metallicities spanned 0.02 -- 1 $\times$ Solar, and ages spanned 1 Myr to
the age of the Universe at a given redshift.  We allowed several
different types of star-formation histories (SFHs), including a single
burst, continuous, as well as both exponentially decaying and rising
(so-called ``tau'' and ``inverted-tau'' models).
To these spectra, we added dust attenuation using the starburst attenuation
curve of \citet{calzetti00}, with a range of 0 $\leq$ E(B-V) $\leq$
0.8 (0 $\leq$ A$_\mathrm{V}$ $\leq$ 3.2 mag).
We also included nebular emission lines using the
prescription of \citet{salmon15}, which uses the line ratios
from \citet{inoue11}, based on the number of ionizing photons from a
given model, and assuming the ionizing photon escape fraction is
$\approx$ zero. We then redshifted these models to 0 $< z <$ 11
and added intergalactic medium (IGM) attenuation \citep{madau95}.
These model spectra were integrated through our {\it HST} filter
bandpasses to derive synthetic photometry for comparison with our
observations.  For each model, we computed the value of $M_{1500}$ by
fitting a 100 \AA-wide synthetic top-hat filter to the spectrum
centered at rest-frame 1500 \AA.  Likewise, for each model we measured the value
of $\beta$ by fitting a power law to each model spectrum using the
wavelength windows specified by \citet{calzetti94}, similar to \citet{finkelstein12a}.

The best-fit model was found via $\chi^2$ minimization, including an
extra systematic error term of 5\% of the object flux for each band to
account for such items as residual uncertainties in the zeropoint
correction and PSF-matching process.  The stellar mass was computed as
the normalization between the best-fit model (which was normalized to
1 M\sol) and the observed fluxes,
weighted by the signal-to-noise in each band.  These best-fit values of
$M_{1500}$ are used in our luminosity function analysis
below, while $\beta$ is used to correct for incompleteness in a
color-dependent fashion. The uncertainties in the
best-fit parameters were derived via Monte Carlo simulations,
perturbing the observed flux of each object by a number drawn from a Gaussian 
distribution with a standard deviation equal to the flux uncertainty in a
given filter.  For each galaxy, 10$^3$ Monte Carlo simulations were
run, providing a distribution of 10$^3$ values for
each physical parameter.  The 68\% confidence range for each parameter
was calculated as the range of the central 68\% of results from these
simulations.  In these simulations, the best-fit redshift
was allowed to vary following the redshift probability distribution
function, thus folding in the uncertainty in redshift into the
uncertainty in the physical parameters \citep[most notably, the stellar mass
and $M_{1500}$;][]{finkelstein12b}.  During this process,
we only allowed the redshift to vary within $\Delta z = \pm$ 1 of the best-fit
photometric redshift.  This excludes any low-redshift solution from
biasing the uncertainties on a given parameter.  The amount of the
integrated $P(z)$ at $z >$ 3 excluded via this step was typically
$\leq$ 10\% (at $z =$ 6).

\begin{figure*}[!t]
\epsscale{1.0}
\plotone{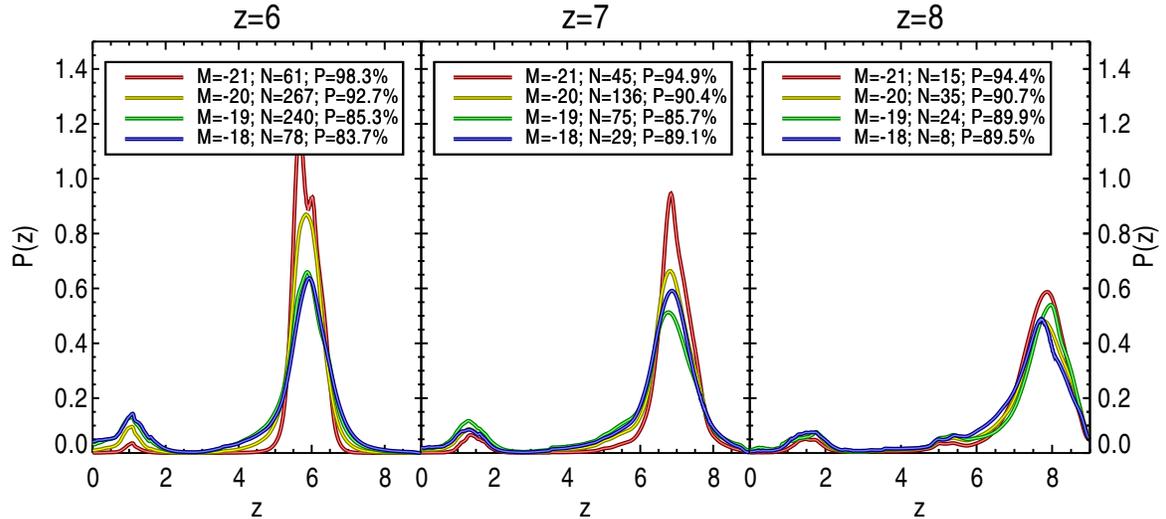}
\caption{Redshift PDFs for galaxies in each of our three
  redshift samples, stacked in bins $\Delta$M$_{UV}$ $=$ 1.  The legends give
the number of galaxies in each stack, as well as the fraction
of the redshift PDF at $z >$ 4 (denoted as P).  Even in the worst case (which is for
faint galaxies at $z =$ 6) $\lesssim$16.3\% of the sample could possibly be at
lower redshift.}
\label{fig:stackpz}
\end{figure*}  

\subsection{Contamination}
A key issue in any study of high redshift galaxies is the risk of sample
contamination, either by spurious sources or by lower-redshift
interlopers.  The gold standard for eliminating contamination is to obtain
spectroscopic redshifts.  This is clearly unfeasible for all galaxies
in our sample (until the next generation of space and ground-based telescopes),
but there is significant archival spectroscopic data.
As discussed in \S 3.2, we find
excellent agreement between available spectroscopic redshifts and our
photometric redshifts, with $\sigma_{z/(1+z)} =$ 0.031.  In
particular, the four bright\footnote[5]{Object z7\_MAIN\_2771 has a
tentative spectroscopic redshift of $z_{spec} =$ 7.62, based on a
4$\sigma$ possible Ly$\alpha$ emission line from
\citet{schenker14}.  This object is quite faint, with
$J_{125} =28.8$, resulting in a somewhat large 95\% photometric redshift
confidence range of 6.02--7.74, though consistent with the tentative
spectroscopic redshift.} galaxies (24.9 $< J_{125} <$ 25.7) with
confirmed $z_{spec} >$ 6.5 have an excellent agreement with
spectroscopic redshifts: 
z7\_GNW\_24443 with $z_{phot} =$ 6.66 and $z_{spec} =$ 6.573 \citep{rhoads13},
z7\_GSD\_21172 with  $z_{phot} =$ 6.73 and $z_{spec} =$ 6.70 \citep{hathi08},
z7\_GNW\_4703 with  $z_{phot} =$ 7.19 and $z_{spec} =$ 7.213 \citep{ono12},
and z7\_GND\_42912 \footnote[6]{This source was originally called
  z8\_GND\_5296 in our previous catalog (Finkelstein et al.\ 2013).  Our new catalog uses an
  updated version of the CANDELS GOODS-N data, thus the catalog
  numbering is different.  Additionally, the slightly updated
  photometry pushes the photometric redshift of this galaxy slightly
  below $z =$ 7.5, placing it in the $z =$ 7 sample.} with $z_{phot} =$ 7.45 and $z_{spec} =$ 7.51
\citep{finkelstein13}.  The brightest
source in our $z =$ 6 sample,
z6\_GSW\_12831 with $M_{1500} = -$22.1 and z$_{phot} =$ 5.77, is
confirmed with $z_{spec} =$ 5.79 \citep{bunker03}.  This galaxy has a
3$\sigma$ detection in the $V_{606}$-band, which could have resulted
in its exclusion from a typical LBG color-color selection sample
\citep[e.g.,][]{bouwens07}, though the observed flux can be explained
by non-ionizing UV photons transmitted through the Ly$\alpha$ forest.

In general, spectroscopic followup of sources selected on the basis of
their Lyman breaks (either color-color selection, or photometric
redshift selection) finds a very small contamination by low-redshift
sources \citep[e.g.,][]{pentericci11}.  However, given the apparent difficulty in detecting
Ly$\alpha$ emission at $z >$ 6.5
\citep[e.g.,][]{pentericci11,ono12,finkelstein13}, the true
effect of contamination at these
higher redshifts is not empirically known.  In this subsection, we
will attempt to estimate our contamination fraction by other means.

\subsubsection{Properties of the Image Noise}
Two key components of our selection
processes should eliminate contamination by spurious sources in our
sample.  First, we restricted our sample to galaxies 
detected at $\geq$3.5$\sigma$ in \emph{two} imaging bands: $J_{125}$
and $H_{160}$.  Formally,
requiring a 3.5$\sigma$ detection in a single band should yield only a
0.05\% contamination by noise.  However, the wings of the noise
distribution are highly non-Gaussian.   We examined this by measuring
the fluxes at 2$\times$10$^5$ random positions in the $J_{125}$ and
$H_{160}$ images in the GOODS-S Deep field (see \citet{schmidt14} for
a similar analysis).  
To avoid biasing from real objects in the image, we only considered
negative fluctuations \citep[e.g.,][]{dickinson04}, where the contamination
percentage was computed as the ratio
of the number of apertures with a flux $<$ $-$1$\times$3.5$\sigma$
to the total number of apertures with negative fluxes.
We found that in each of
these bands individually, the fraction of positions measuring at $>$3.5$\sigma$ was
$\sim$1.4\%, much higher than predicted based on an assumption of
Gaussian noise.  If we instead
look at the number of 3.5$\sigma$ fluctuations in both the $J_{125}$
and $H_{160}$ images at the
same location, we find that this contamination drops to nearly zero,
at 0.05\%.  Thus, we conclude that as we require significant
detections in two bands, the contamination in our sample by noise is negligible.
Spurious sources other than noise spikes are
eliminated by our detailed visual inspection of each source, described
in \S 3.3.

\subsubsection{Estimates from Stacked Redshift PDFs}
Contamination by low-redshift interlopers is a more complicated
issue.  While extreme emission line galaxies at lower-redshift could
theoretically be an issue \citep{atek11}, our requirement of detections in two bands
(as well as the frequent detections in more than two bands for all but the highest
redshift objects in the $z =$ 8 sample) makes a significant
contamination by these sources unlikely.  The most likely possible
contaminants are faint red galaxies at $z \leq$ 2 \citep[e.g.,][]{dickinson00}.  These galaxies can
be too faint to be detected in our optical imaging, but their red SEDs
yield detections in the WFC3/IR bands.  Although faint sources that
are very red will have a disfavored high-redshift solution with our
current photometric selection, we have information on their
likelihoods encoded in our redshift PDFs.  Figure~\ref{fig:stackpz}
shows the redshift PDFs of galaxies in each of our three highest redshift samples,
stacked in magnitude bins of $\Delta$M $=$ 1 mag.  At all
redshifts and all magnitudes, $\gtrsim$85\% of the redshift PDF is at $z >$
4, implying that there is not significant contamination by
lower-redshift galaxies.  

The position of the secondary redshift peak is consistent with the
detection of a 4000 \AA\ break rather than
the Lyman break (at $z =$ 6, 7 and 8, this gives $z_{secondary} =$ 1.1, 1,4 and 1.7). 
At $z =$ 8, the possible
contamination from galaxies at $z <$ 4 is $<$10.5\%, primarily due to
the fact that at $z =$ 8, a
galaxy will have to be undetected in most of the filters we consider
here (there is an additional $\sim$8\% chance the true redshift is in
the range $4 < z < 6$).  The worst case is for faint galaxies at $z =$ 6, as $z =$ 6
galaxies are typically detected in all but two filters, though even here,
the indicated contamination by $z <$ 4 galaxies is $\lesssim$15\%.  

\begin{figure*}[!t]
\epsscale{0.8}
\plotone{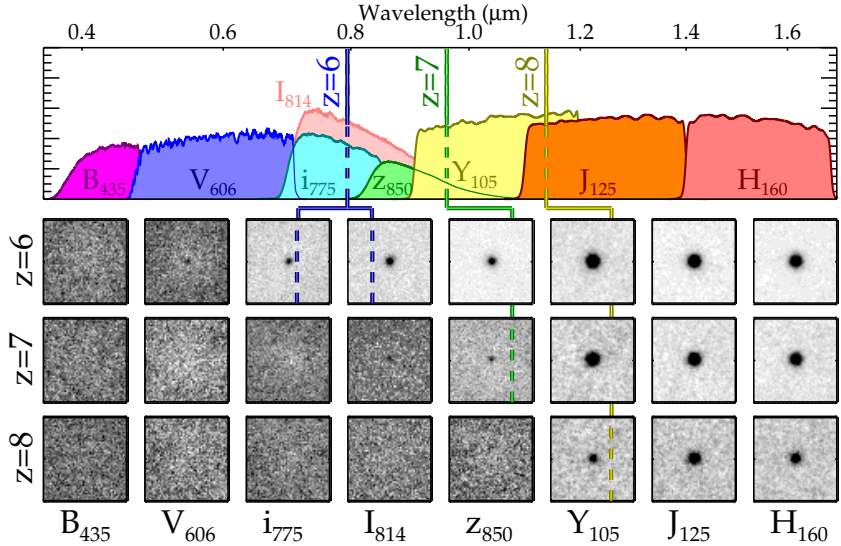}
\caption{Top: Filter transmission curves for the filter set used in
  this study ($Y_{098}$, which was used in the GOODS-S ERS field only, is not
  shown).  The vertical lines denote the relative position of the Ly$\alpha$
break (rest 1216 \AA) in a given filter for galaxies at the center of
our three highest redshift bins.  Bottom)
Negative mage stacks of galaxies in our three highest redshift galaxy samples.  
If our sample had a significant fraction of lower-redshift
interlopers, significant flux would be seen blueward of the break (e.g.,
$B_{435}$ at $z =$ 6, $i_{775}$ and blueward at $z =$ 7, and $I_{814}$
and blueward at $z =$ 8).  This is not observed at any redshift, thus
we conclude that our sample does not contain a dominant population of
low-redshift interlopers..}
\label{fig:stack}
\end{figure*}  

\subsubsection{Stacking Imaging}
The limits from the previous subsection are likely upper limits
on the contamination fraction.  When fitting photometric
redshifts, to rule out all low-redshift solutions, the Lyman
break needs to be detected at high significance, which is the case for
only the brightest galaxies (e.g., at $z =$ 6, the brightest bin has a
contamination of $<$2\%).  Additionally, these results are dependent
on the templates used, which by
definition do not account for unknown galaxy populations.  We therefore
consider two empirical tests of contamination.  The first is to stack
all galaxies in a sample, in order to search for detections below the Lyman
break.  The results from this test for $z =$ 6, 7 and 8 are shown in
Figure~\ref{fig:stack}.  As expected for galaxies at the expected
redshifts, there is no visible signal in the $B_{435}$-band at $z =$
6, $i_{775}$-band and blueward at $z =$ 7, and $I_{814}$-band
and blueward at $z =$ 8.  This confirms that the majority of the flux
in our sample does not arise from lower-redshift sources.

\subsubsection{Estimates from Dimmed Real Sources}
As a final test, we estimated the contamination by artificially
dimming real lower redshift sources in our catalog, to see if the increased
photometric scatter allows them to be selected as high redshift candidates.
This empirical test is useful as it does not rely on known spectral
templates to derive the contamination, though it does assume that the
fainter objects which could potentially contaminate our sample have
similar SEDs to the bright objects which we dim.  We performed this
exercise twice, once using the combined catalog of the  GOODS-S and
GOODS-N Deep fields, and once in the HUDF main field, to probe fainter
magnitudes.  In the GOODS Deep fields, we selected all real sources with 21 $< H_{160} <$
24 and $z_{phot} < 3$, and reduced their observed fluxes by a factor of 20.  The
same was done for sources drawn from the HUDF Main field, here extending the
magnitude range to be 21 $< H_{160} <$ 26.  The limits on these
magnitudes were chosen to exclude any real high-redshift sources. 
We replaced the true flux uncertainties of these objects with flux uncertainties from a
randomly drawn real source from the full catalog from a given field
with a similar magnitude
as the dimmed source.  We then added scatter to the dimmed fluxes,
perturbing them by a random amount drawn from a Gaussian distribution
with a standard deviation equal to the flux uncertainties of the object.  We included
two realizations of the HUDF field to increase the number of dimmed objects.

The total number of sources in our artificially dimmed catalog was
4066 in the Deep fields, and 1254 in the HUDF field.
We measured photometric redshifts of these sources with
EAZY in an identical manner as on our real catalogs, and then we applied our
sample selection to this dimmed catalog.  In the Deep fields, we found
a total number of 149, 134, 54, 23 and 8 dimmed objects satisfied our
$z =$ 4, 5, 6, 7 and 8 selection criteria.  Investigating the original (not
dimmed or perturbed) colors of these sources, we found that they are
unsurprisingly red, with the bulk of sources having $V - H >$ 2 mag.
It is therefore this parent population of red sources which are responsible
for the majority of the possible contamination.  The contamination
fraction in our high-redshift sample is then defined as
\begin{equation}
\mathcal{F} = \frac{\frac{N_{dimmed,select}}{N_{dimmed,red}}*N_{total,red}}{N_{z}}
\end{equation}
where $N_{dimmed,select}$ was the number of dimmed sources satisfying
our high-redshift sample selection, $N_{dimmed,red}$ was the total
number of sources in the dimmed catalog with original colors of $V-H
<$ 2, $N_{total,red}$ was the number of sources in the full object
catalog with 25 $< H <$ 27, $z_{phot} <$ 3, and $V-H
<$ 2, and $N_{z}$ was the number of true galaxy candidates in a given
redshift bin.  For example, at $z =$ 6, where we found 54 dimmed
galaxies satisfied our selection criteria ($N_{dimmed,select}$ =54),
$N_{dimmed,red} =$ 1023, $N_{total,red} =$ 695, and $N_{z} =$ 322,
giving an estimated contamination fraction $\mathcal{F} =$ 11.4\%.
Thus for sources with 25 $< H <$ 27, we found an estimated
contamination fraction of $\mathcal{F} =$ 4.5\%, 8.1\%, 11.4\%, 11.1\%
and 16.0\% at $z =$ 4, 5, 6, 7 and 8, respectively.  We performed
the same exercise in the HUDF, here for fainter sources with 26 $< H <$ 29,
finding 30, 21, 8, 8 and 0 sources satisfied our $z =$ 4, 5, 6, 7 and 8
selection criteria, giving a contamination fraction of 9.1\%, 11.6\%,
6.2\%, 14.7\% and $<$4.9\% at $z =$ 4, 5, 6, 7 and 8.

\begin{figure*}[!t]
\epsscale{0.385}
\plotone{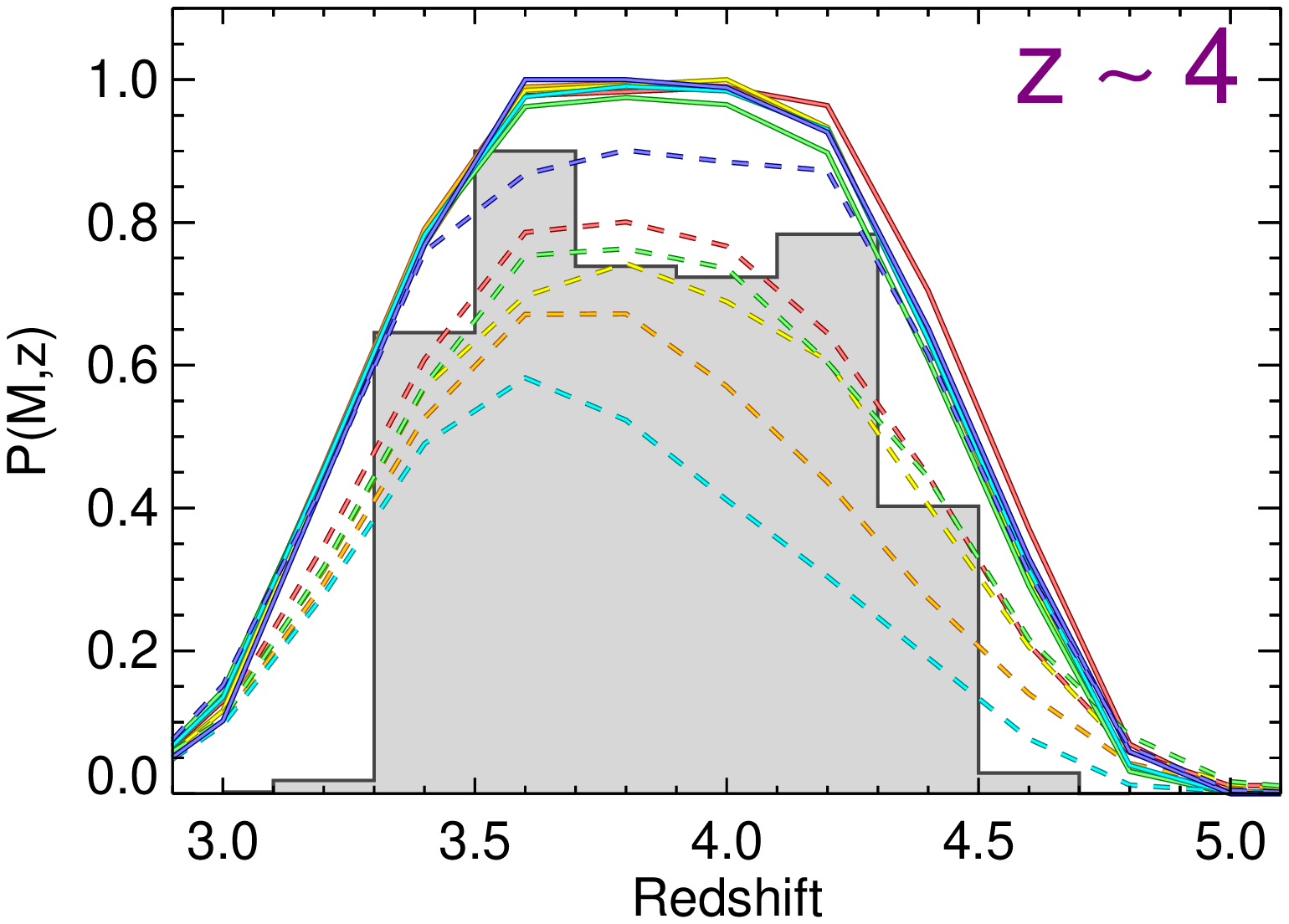}
\plotone{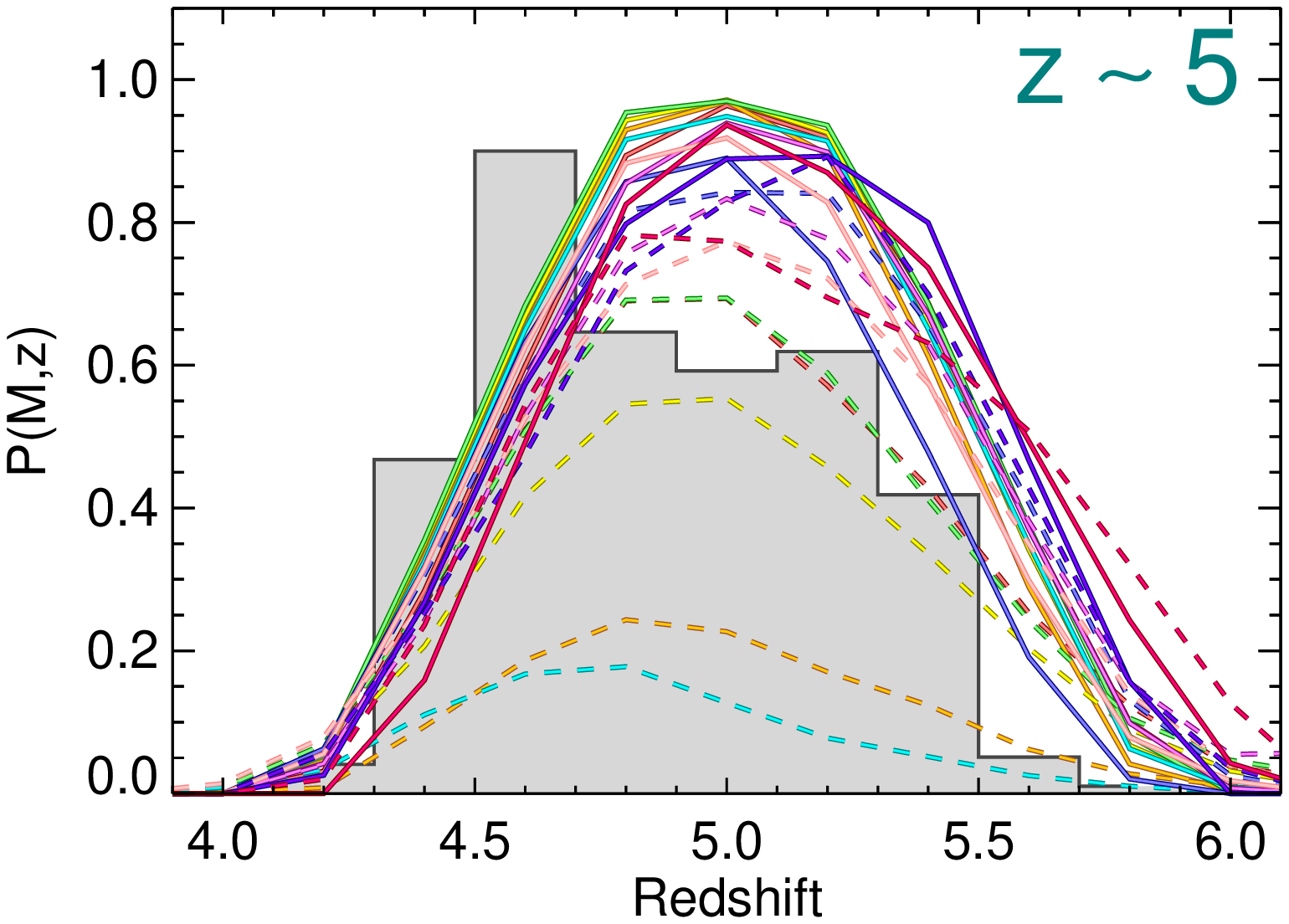}
\plotone{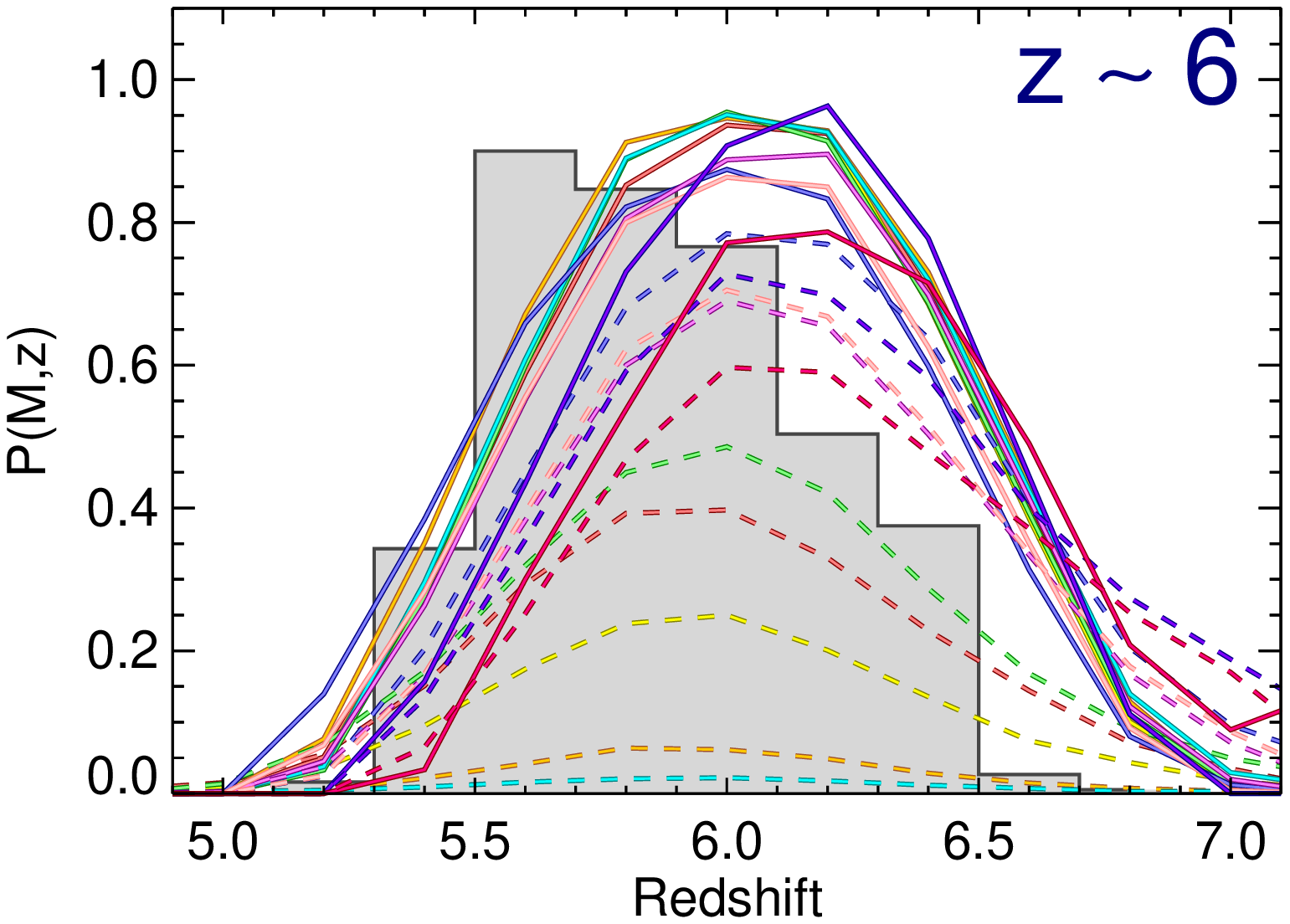}
\plotone{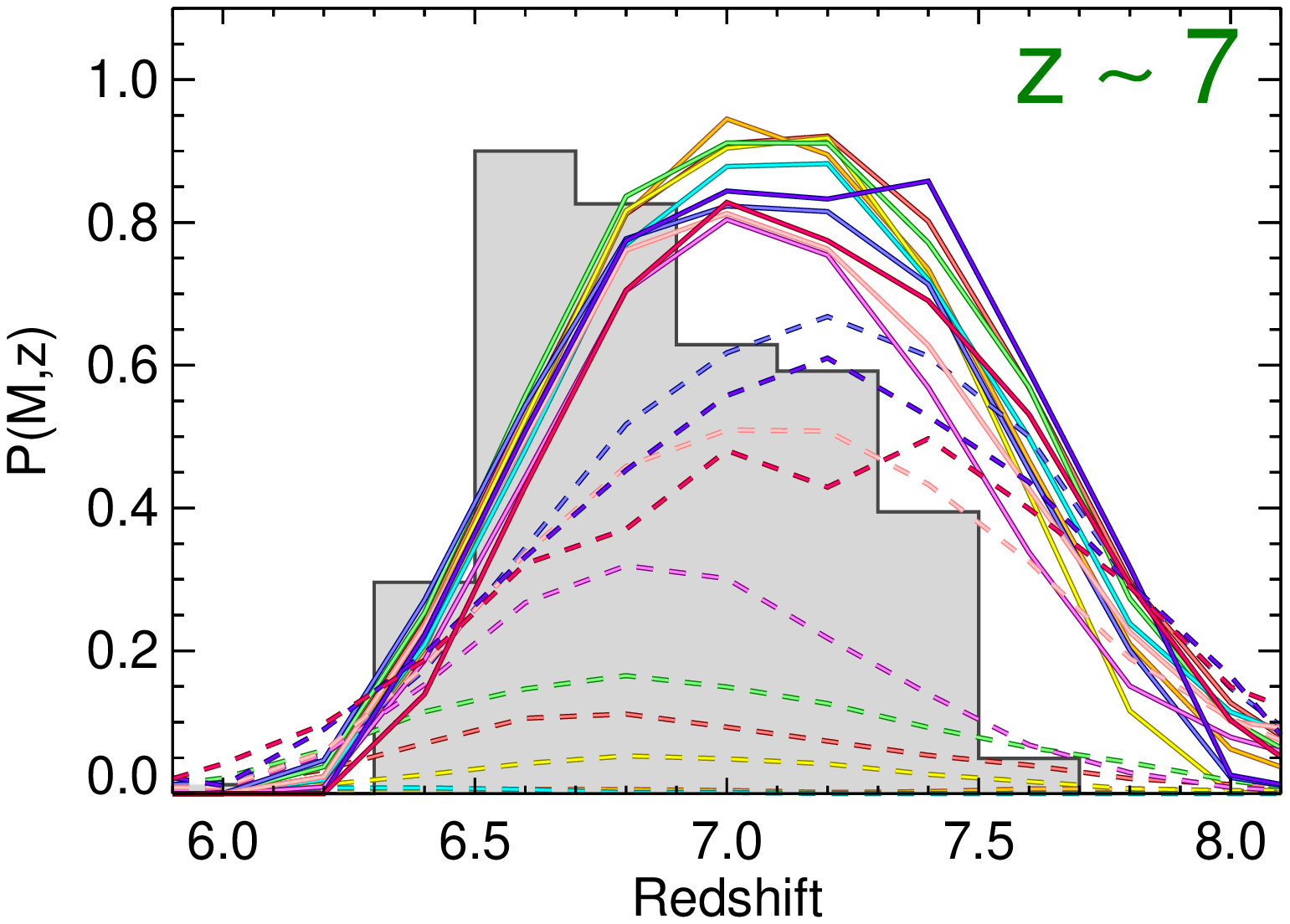}
\plotone{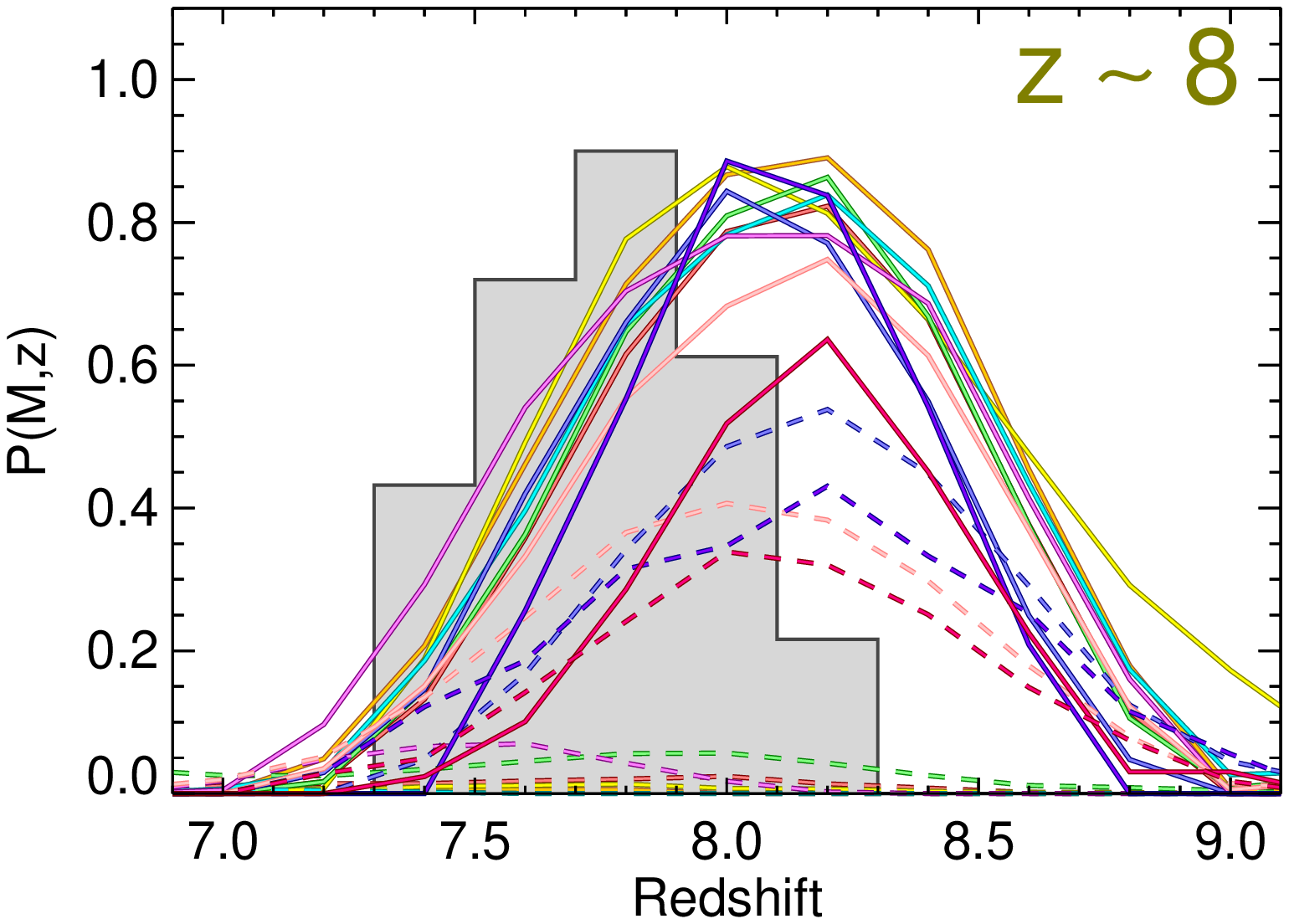}
\plotone{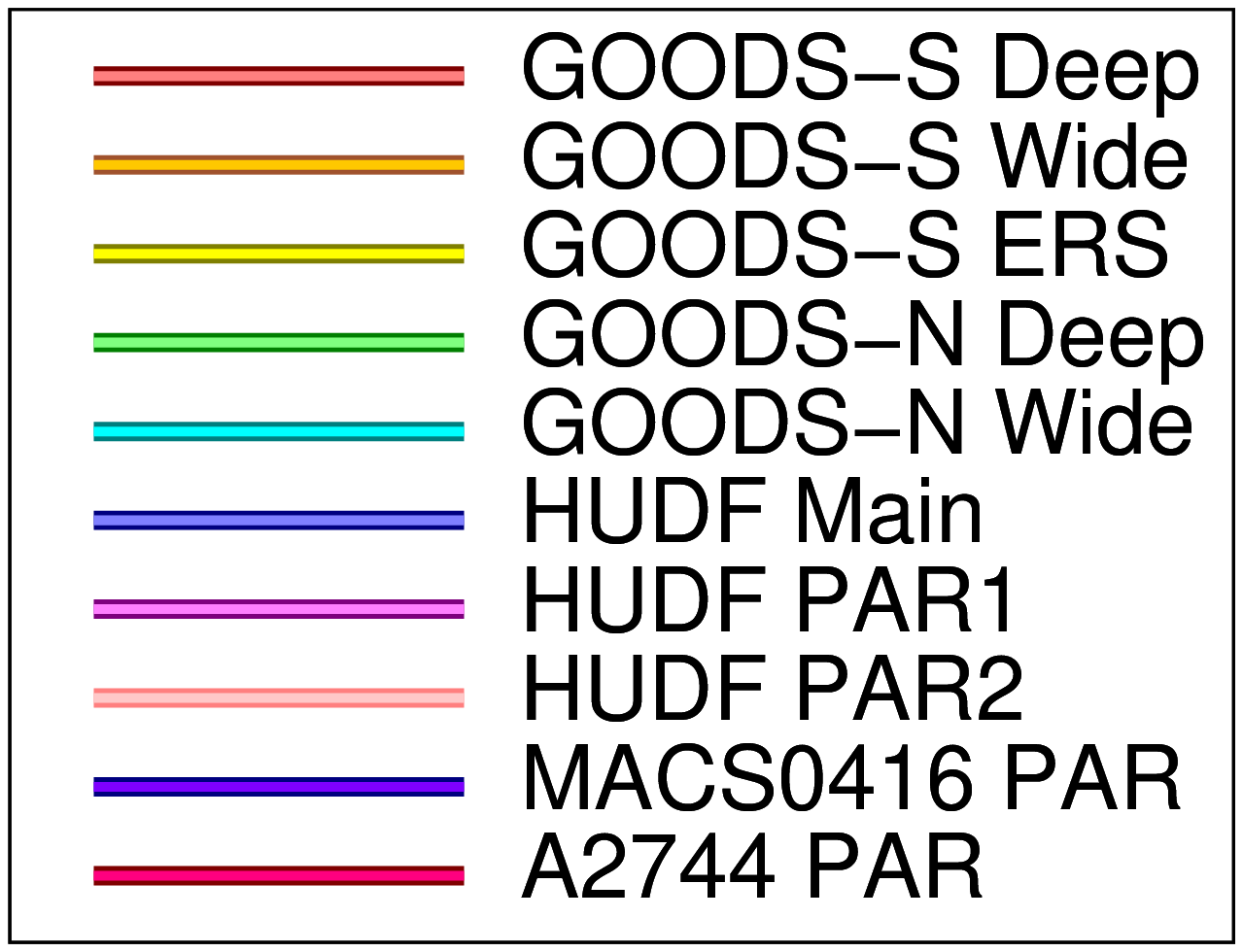}
\caption{The results of our completeness simulations, showing the
  probability that a given simulated source was recovered as a
  function of its input redshift.  The solid lines denote sources with
$M_{1500} = -$22, while the dashed lines denote $M_{1500} = -$19.
These lines assume a half-light radius of r$_h =$ 0.18\arcs\ and $\beta = -$2.0.
The background histogram shows the (normalized) distribution of best-fit
photometric redshifts for the real galaxies in each redshift subsample.
Although our selection criteria combined with the wavelengths probed
by our filter-set results in a
completeness that peaks at close to $z =$ 4, 5, 6, 7 or
8, the evolving luminosity function as well as our sensitivity to
bright galaxies results in our samples having mean redshifts slightly
lower than the bin center, particularly in the higher redshift samples.}
\label{fig:probs}
\end{figure*}    

Broadly speaking, we estimate relatively small contamination fractions
of $\sim$5-15\%, in line with the estimates above from the stacked
$P(z)$ curves.  As the bulk of contaminants appear to be red
galaxies, it is interesting to compare to the space density of these
potentially contaminating sources.  This was recently estimated by 
\citet{casey14b}, who find that dusty star-forming galaxies at $z
 <$ 5 will contaminate $z >$ 5 galaxies samples at a rate of $<$1\%.
This is much less than our contamination estimates, thus we may have
overestimated the contamination rate, though it may not be
inconsistent once photometric scatter is applied to faint, red
galaxies, making it easier for them to scatter into our sample.
In any case, the expected contamination rate is quite small, therefore
we do not reduce our observed number densities for the expected
minimal contamination.

\section{Completeness Simulations}

We performed an extensive set of simulations to estimate the effective volume for
each source in our sample, accounting for both
image incompleteness and selection effects.  We inserted mock
galaxies into the imaging data, repeating the same analysis for source
detection, photometry, photometric redshift measurement, and sample selection as was
done on the real data.  We then compared the fraction
of recovered and selected mock sources to the
total number of input sources in a given bin of absolute magnitude and
redshift to determine our completeness in that bin.

While the effective volumes are typically computed as a function of
magnitude and redshift, other key factors in these simulations are the
choices of galaxy size and
color.  At a constant magnitude, a very extended galaxy may not make
it into our sample, as it may fall below our surface brightness
sensitivity.  Additionally, very red galaxies may not satisfy our
selection criteria, as red colors typically enhance the amplitude of
the low-redshift solutions, particularly at low signal-to-noise levels.
Thus the effective volume depends not only on magnitude and redshift, but
also on the size and rest-frame UV color.  To see what effect this has, we have 
computed our completeness as a function of four properties:
redshift, absolute magnitude, half-light radius, rest-UV color, where
we have parametrized the latter via the UV spectral
slope $\beta$ \citep{calzetti94}.  A large number
of simulated objects are needed to fill out this four-dimensional space; our
completed simulations recovered $\sim$5.4 million out of 7 million objects input across
all of our fields (where the recovered objects were detected in our
photometry catalogs; this number does not account for the photometric redshift
selection, which we discuss below).

Our simulations were run separately on each of our 10 sub-fields
defined in Table 1.  To ensure that the mock galaxies did not affect the
background estimation, only a small number of galaxies were added during
each simulation.  To optimize the simulation runtime, the mock
galaxies were added to cutouts from the full images.  In the GOODS
sub-fields (i.e., CANDELS Deep and Wide,
and the ERS), 200 mock galaxies were added to a 2000$\times$2000 pixel
(2$^{\prime} \times 2^{\prime}$) region of the images, while for the single-pointing HUDF and HFF
fields, 100 galaxies were added to a 1000$\times$1000 pixel
region.  As the depth across our imaging data can vary, during each
simulation the position of the cutout varied, such that when we
combine all of our simulations, we
average over any differences in the depth across a given field.
 
To determine the colors of the mock galaxies, we created distributions
in redshift, dust attenuation (parameterized by E[B-V]), stellar
population age and stellar metallicity.  The redshift distribution was defined to
be flat across 3 $< z <$ 9, such that we simulate objects well
above and below the redshift ranges of interest.  The dust attenuation
E(B-V) was defined to have a Gaussian distribution with a mean of 0.1
and a $\sigma =$ 0.15 (with a minimum of zero).  The age was defined as a log-normal
distribution, with a peak near 10 Myr, and a tail extending out to
the age of the Universe at a given redshift.  The metallicity
distribution was also log-normal, with a
peak of $Z = 0.2Z$\sol, and a tail towards higher values.  The exact values of these
distributions are not crucial given our methodology (as
opposed to a multivariate analysis, where the distributions are very important), as they combine to create a
distribution of rest-frame UV slope $\beta$.  We crafted these
distributions to provide a distribution of $\beta$ encompassing the
expected values for our real objects \citep[e.g.,][]{finkelstein12a,bouwens13}.  We then used the updated (2007) stellar
population models of \citet{bruzual03} to calculate the colors of a
stellar population given the distributions above.  To convert these
colors into magnitudes, we assumed a distribution of $H$-band magnitudes
designed to have many faint ($H >$ 26) galaxies (which is where we
expect to become incomplete), and relatively few at bright
magnitudes.  To ensure enough bright galaxies to calculate a robust
incompleteness, every 10th simulation used a flat distribution of
$H$-band magnitudes of 22 $< H <$ 25.  These $H$-band magnitudes were
combined with the mock galaxy colors to generate magnitudes in each
filter for a given field.

To generate the galaxy images themselves, we used the GALFIT
software \citep{peng02}.  We assumed a log-normal distribution of
half-light radii with a peak at 1-pixel, and a high tail towards larger
radii, giving an interquartile range of half-light radii of 1.4--4.9
pixels.  This corresponds to $\sim$0.4--1.6 kpc, spanning the range of
the majority of resolved galaxies at $z >$ 4
\citep[e.g.,][]{oesch10b,grazian12,ono13,curtislake14}.  GALFIT also
requires a Sersic index ($n$), axis ratio and position
angle; the Sersic index was assumed to be a log-normal distribution at 1 $< n
<$ 4, with the majority of the mock galaxies having disk-like morphologies ($n
<$ 2); the axial ratio was also log-normal, with a peak at 0.8, and a
tail toward lower values, and the position angle was a uniformly distributed
random value between 0 and 360 degrees.  GALFIT was then used to
generate a 101$\times$101 pixel (6$^{\prime\prime} \times
6^{\prime\prime}$) stamp for a given mock galaxy, which
was then added to the image at a random location.  Because our data
are PSF-matched to the $H$-band, we had GALFIT
convolve the mock galaxy images with the $H$-band
PSF prior to adding them to the data for all filters.

Once the set of mock galaxies for a given simulation were added to the
data, photometric catalogs were generated using Source Extractor in
the exact same manner as was done on the data (i.e., using a weighted
$J+H$ image as the detection image).  These catalogs were read in and
combined, again in the same methodology as with the data, including aperture
corrections (the exception here is that a correction for Galactic
extinction was {\it not} applied in the simulation, as the simulated
objects did not have Galactic extinction included).  The photometric
catalog was then compared to the input catalog to generate the list of
recovered objects (i.e., mock galaxies which were recovered by Source
Extractor); an object was regarded as being recovered if it had a positional match
within 0.2\arcs\ of one of the input mock galaxies.  The recovered
object catalogs were processed through EAZY to generate photometric redshifts,
and then run through our SED-fitting routine to measure absolute
UV magnitudes ($M_{1500}$), stellar masses and UV spectral slopes.
These simulations were then repeated until a large sample of recovered
galaxies was available, which were then compiled in a single database
per field.  The completeness was defined as the number of galaxies
recovered versus the number of input galaxies, as a function of input
absolute magnitude, redshift, half-light radius and UV spectral slope
$\beta$.  Figure~\ref{fig:probs} shows the results from our
simulations.  

\begin{figure*}[!t]
\epsscale{0.385}
\plotone{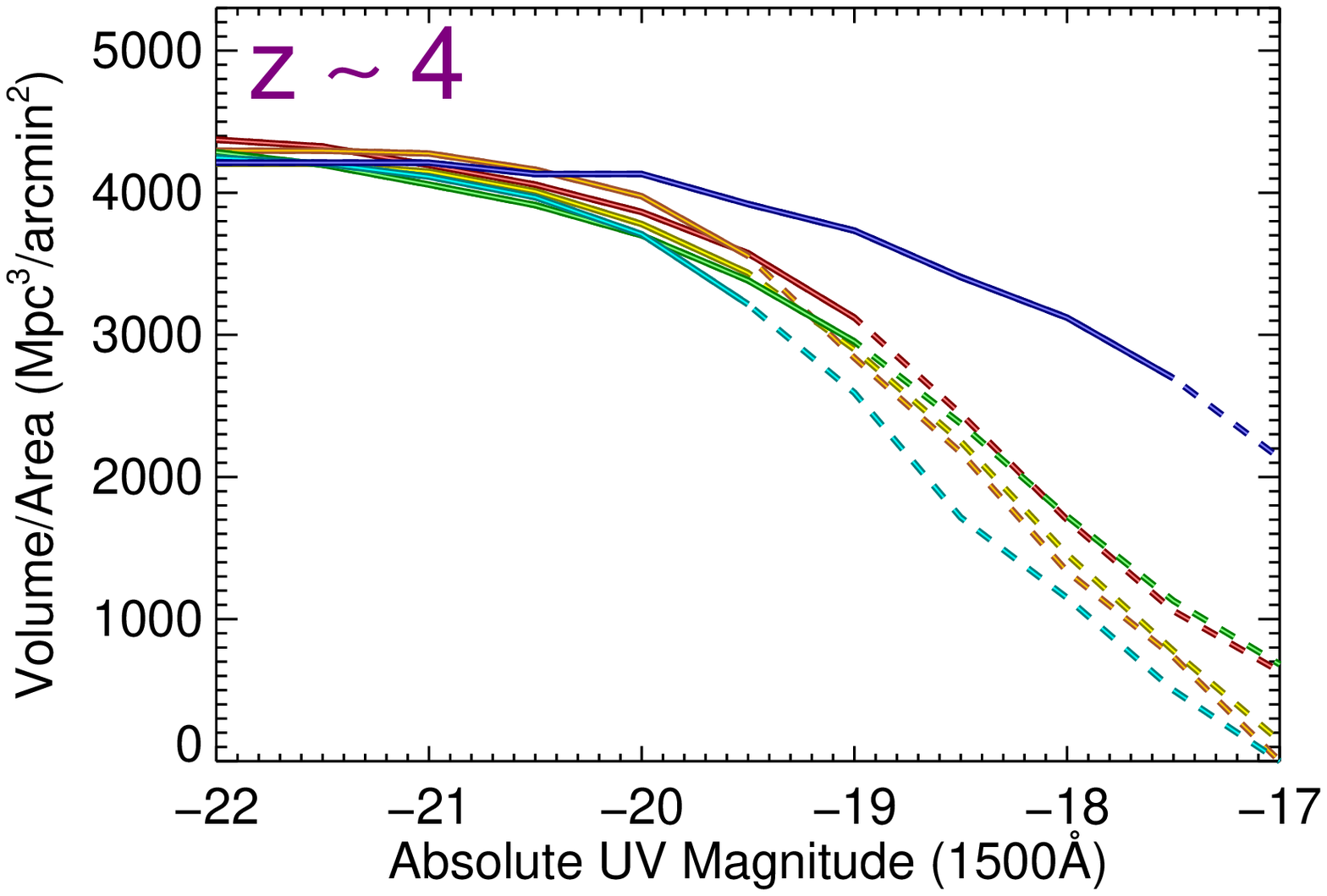}
\plotone{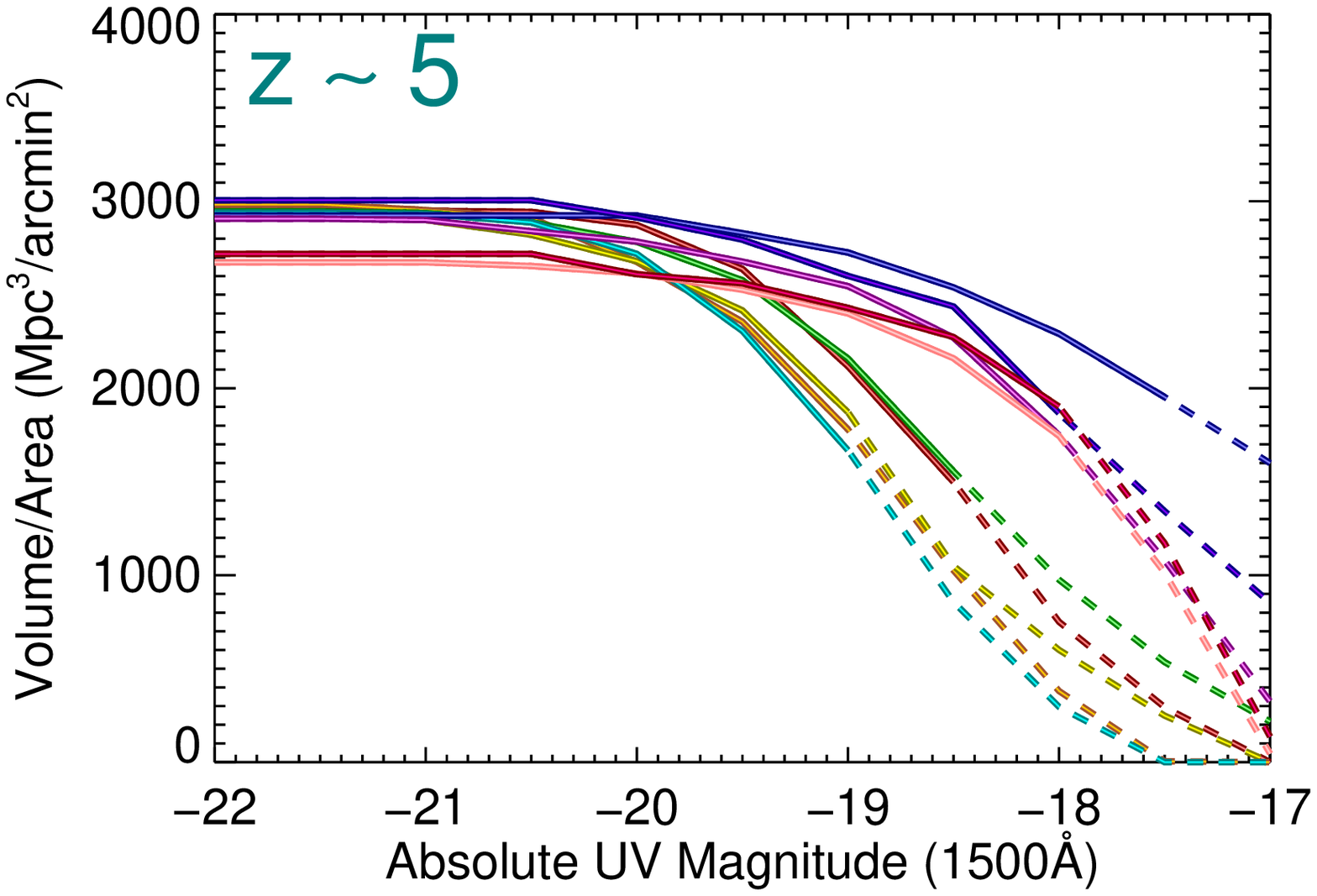}
\plotone{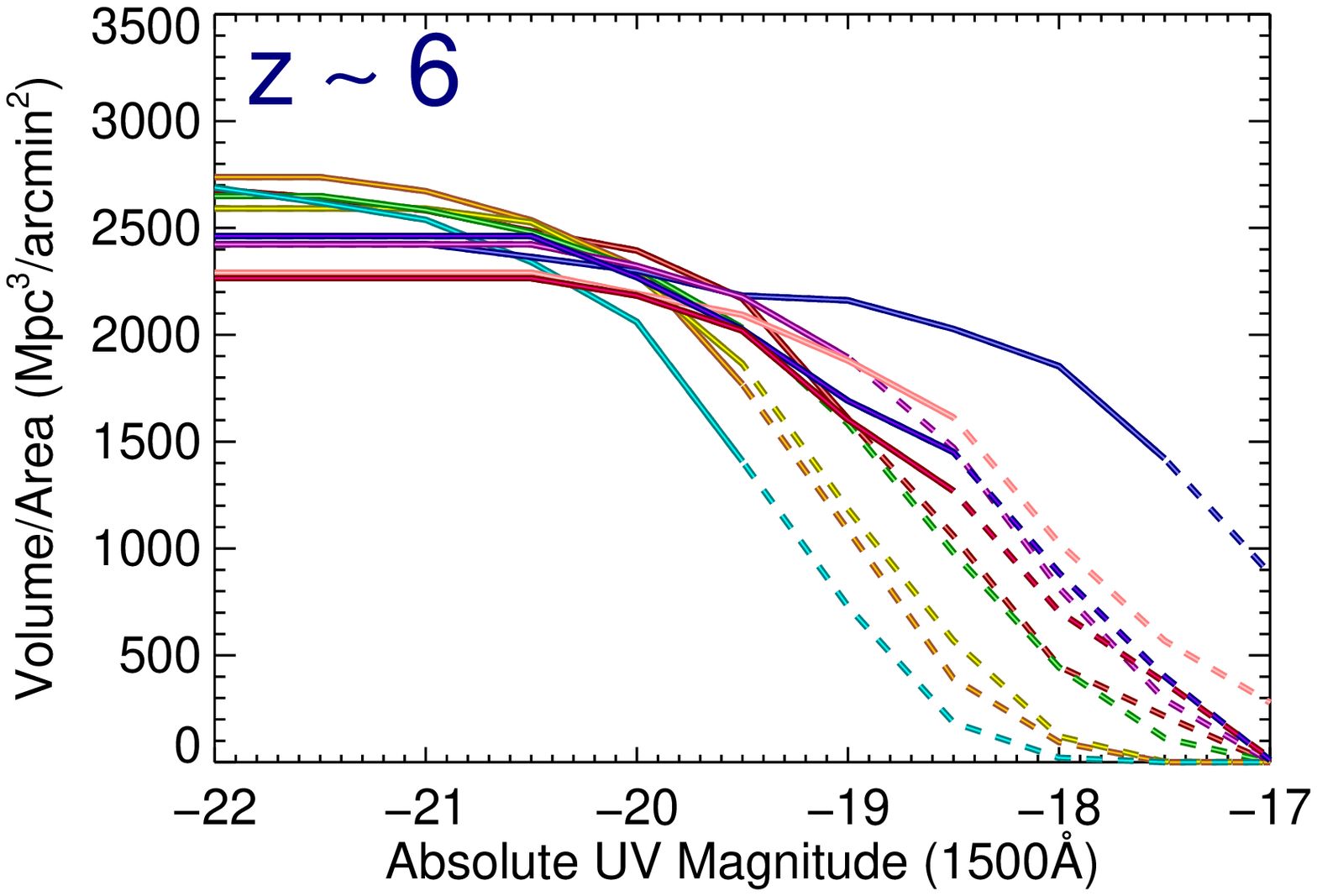}
\plotone{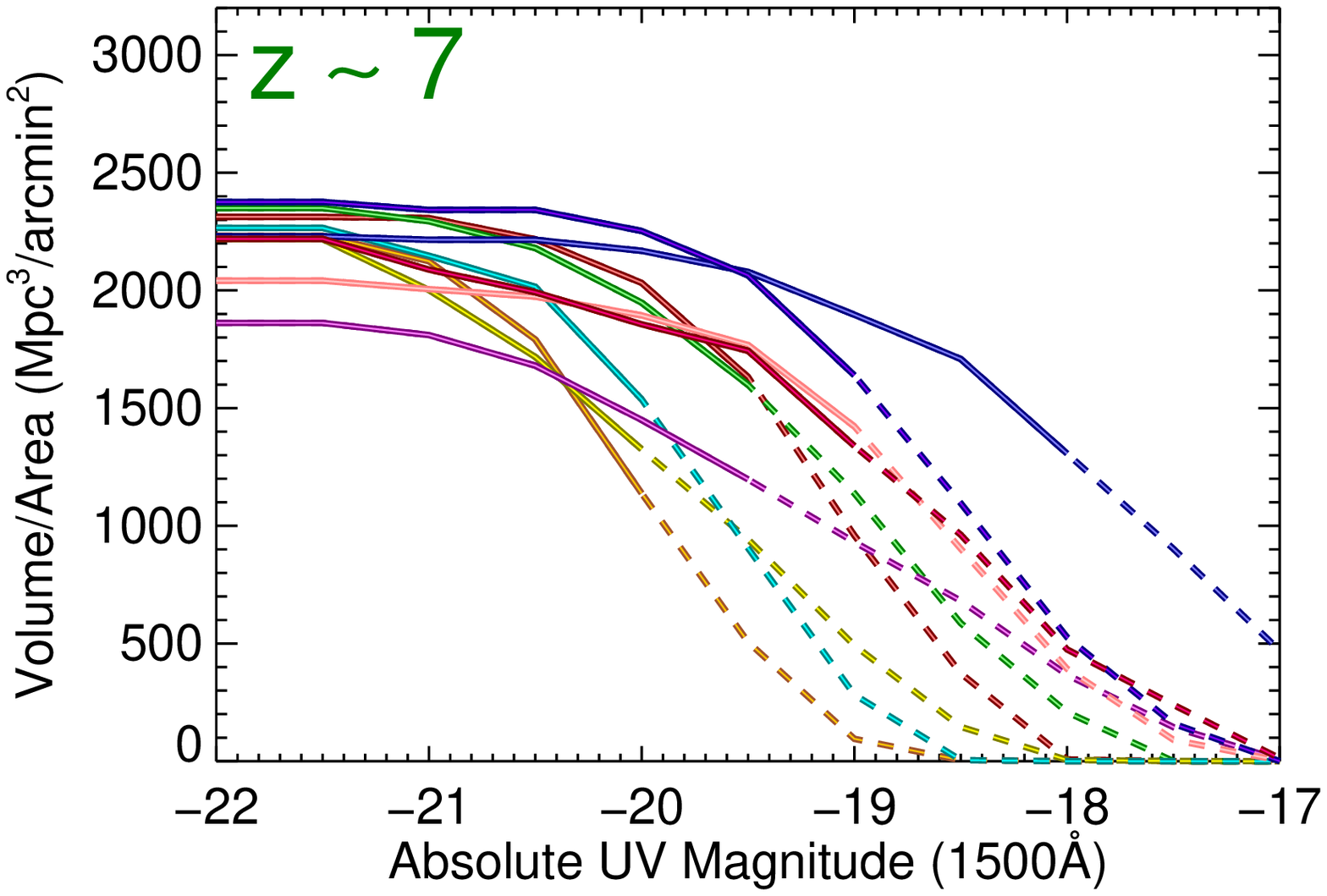}
\plotone{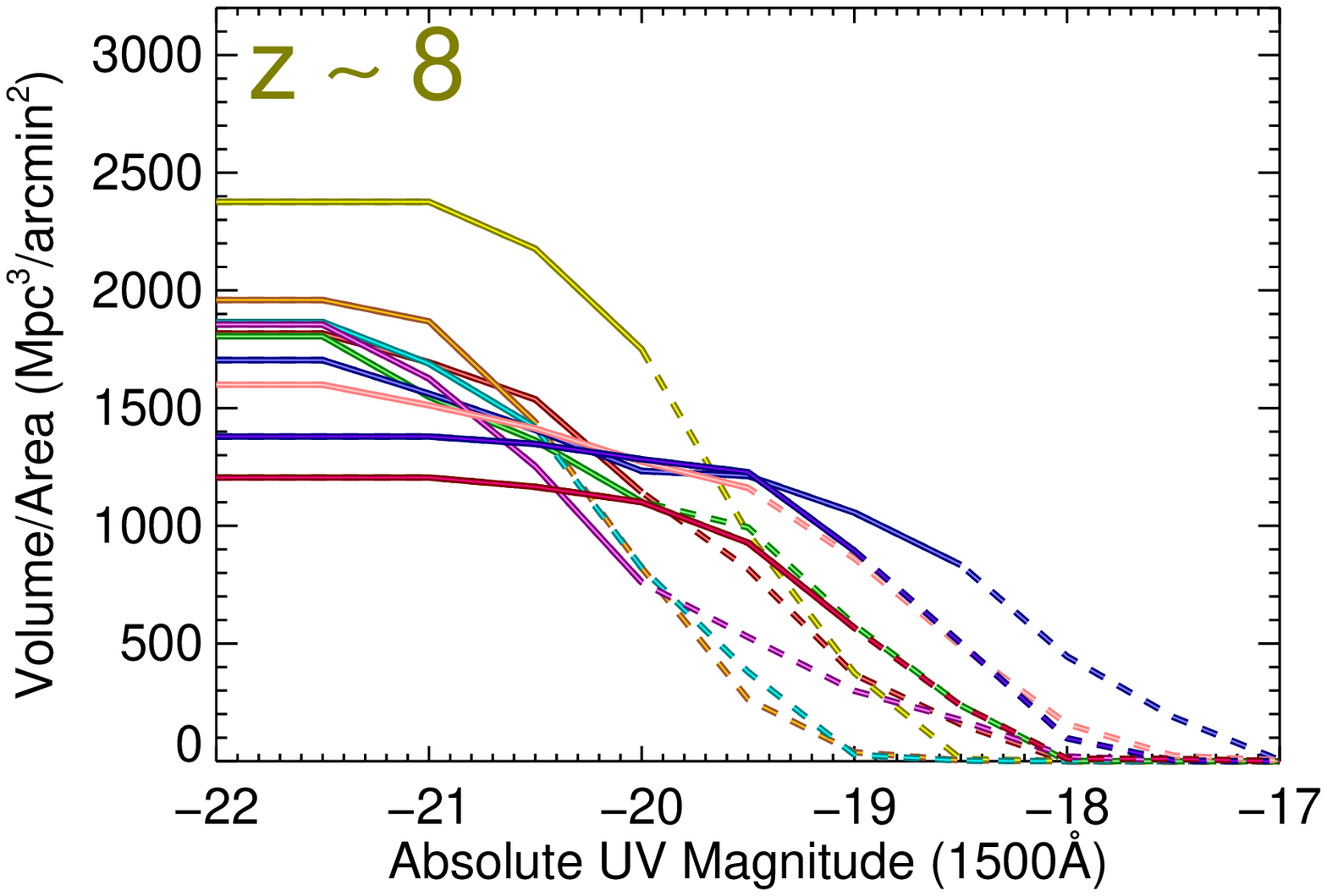}
\plotone{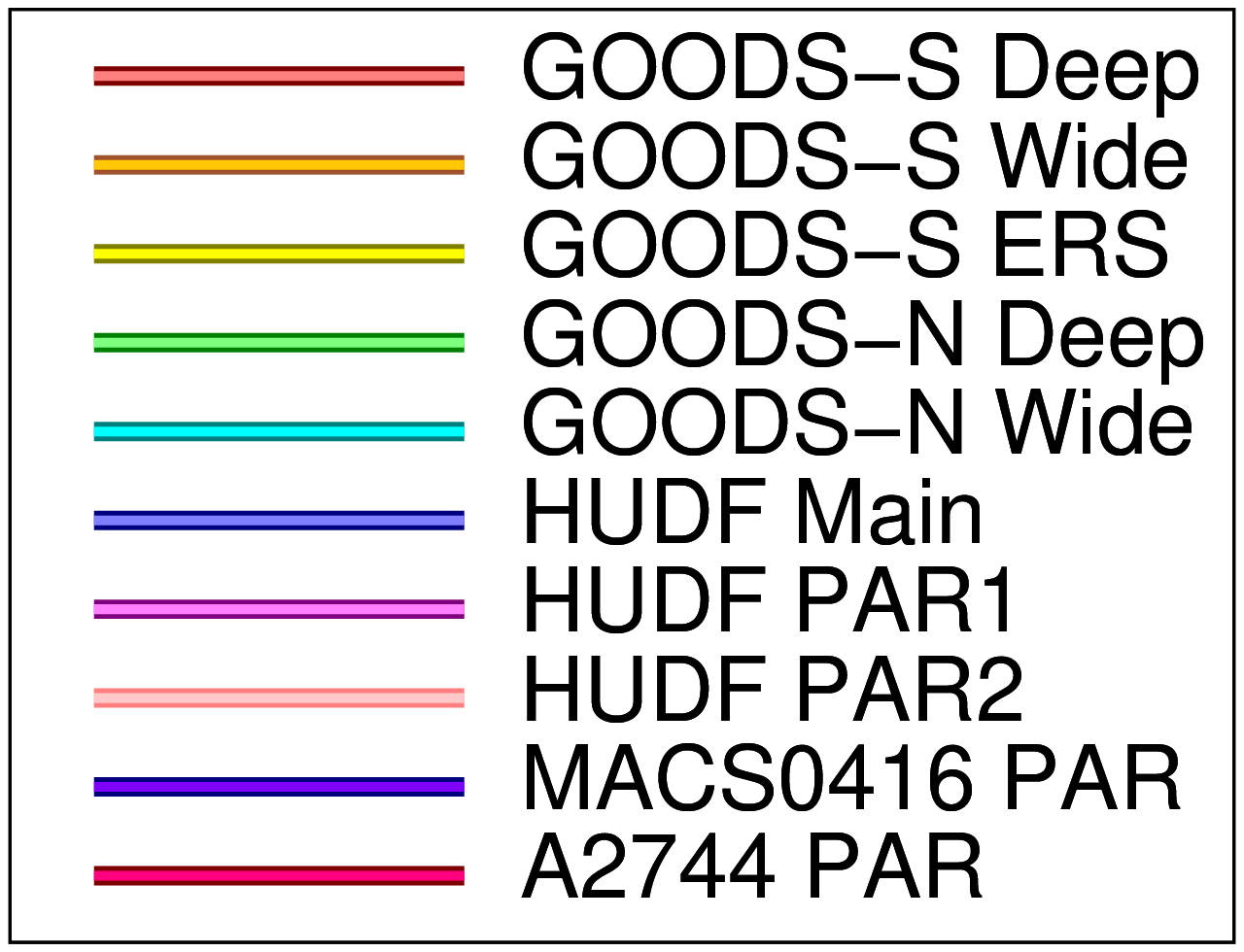}
\caption{The effective volume per unit area of our survey for high-redshift galaxies
  in each of our redshift bins.  Here we divide out the area of a given
 sub-field such that one can easily compare the completeness as a
 function of magnitude of the various fields.  The solid lines give
 way to dashed lines when the volume per unit area falls below 50\% of
the maximum value.  For the luminosity function, we only consider magnitude bins in each field
brighter than these 50\% completeness points, to avoid having data
dominated by incompleteness corrections.  The ERS has a
different $Y$-band filter ($Y_{098}$), which gives a better spectral
resolution around 1$\mu$m, likely responsible for the increased
selection efficiency at $z =$ 8 in that field.}
\label{fig:volumes}
\end{figure*}

\begin{figure}[!t]
\epsscale{1.15}
\plotone{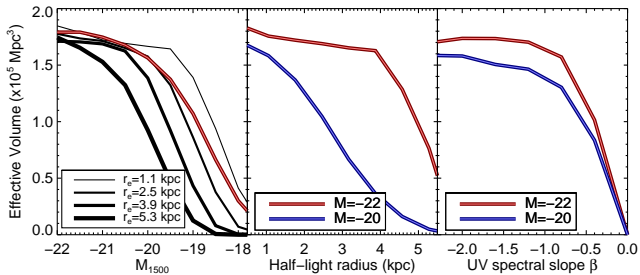}
\caption{Left) The effective volume as a function of UV absolute
  magnitude for galaxies in our $z =$ 6 sample in the GOODS-N Deep
  field.  The red line shows the mean effective volume for
  this field, weighted by the number of galaxies at a given radius and
UV color.  The black lines show how the effective volume changes as a
function of effective radius (r$_{e}$) for a fixed UV color ($\beta =
-$2).  Our weighted mean volume is similar to the effective volume
assuming r$_{e}$ = 2.5 kpc for bright galaxies, and r$_{e} =$ 1.1 kpc
for faint galaxies.  Middle) The dependence of the effective volume on
effective radius in two magnitude bins.  At fainter magnitudes, the
effective volume drops steeply with increasing size, as the surface
brightness drops below detectable levels.  Right) Same as middle,
except here showing the dependence on UV color.  The dependence on
color is much weaker than that on size, as we remain sensitive to
galaxies until $\beta$ becomes redder than $-$1, which is much redder
than the colors of observed high-redshift galaxies \citep[e.g.,][]{finkelstein12a,bouwens13}}
\label{fig:veff}
\end{figure} 

In our original simulations
the recovered redshift was typically $\sim$0.2 lower than the input
redshift, independent of magnitude.  This is likely not a fault in our photometric redshift
estimates, as Figure~\ref{fig:zcomp} shows that these 
agree well with existing spectroscopic redshifts for real galaxies.
Rather, it is likely a mismatch between our simulated SEDs and those
of the templates used in EAZY.
Upon further investigation, we found that the cause of this offset was
Ly$\alpha$ emission in the mock galaxies.  While Ly$\alpha$ photons
were attenuated by dust in the same manner as adjacent UV photons, we
did not include any additional Ly$\alpha$ attenuation for, e.g.,
geometric or kinematic effects.  This led to very high Ly$\alpha$
escape fractions, which were not matched in the templates.  This high
Ly$\alpha$ emission reduced the amplitude of the
photometrically-measured Lyman break,
resulting in a (slightly) lower photometric redshift.  After reducing the amount of 
Ly$\alpha$ flux to 25\% of the intrinsic value, our photometric
redshifts matched the input redshifts (which matches expectations for
the global Ly$\alpha$ escape fraction;
\citet[e.g.,][]{hayes11,blanc11}.  Rather than rerun all of our
completeness simulations, we elected to simply reduce the input
redshift by 0.2 when interpreting our simulations, which corrects for
this effect (this changes the distance modulus by $<$0.1 mag).  The
exception was the simulations for the HFF parallel fields, which were
run after this effect was noticed.  In those fields, the input models
had their Ly$\alpha$ flux reduced to 25\% of the intrinsic value, and
no change to the model redshift was needed.

It is important to examine whether the choice of computing
the completeness as a function of input properties affects our
result.  As mentioned in \S 2.3, we used the results from these
simulations to correct for offsets in the recovered versus input
magnitudes (i.e., to be sure the fluxes we use represent the total flux).  Additionally, we
examined whether there exist biases in the half-light radius or
$\beta$ measurements from the simulations.  
Recovered objects were typically measured to have a half-light radius
$\sim$0.03\arcs\ (0.5 pixels) smaller than the input value, and were
measured to be slightly redder ($\Delta\beta \lesssim$ 0.1).  However, these
corrections make effectively no change to the effective volumes
derived from the simulations, and so were not applied.

In each redshift bin, the effective volume for galaxies in a given field was then calculated
via
\begin{equation}
V_{eff}(M_{1500},r_h,\beta) = \int \frac{dV}{dz}~P(M_{1500},z,r_h,\beta)~dz
\end{equation}
where $dV/dz$ is the comoving volume element, and
$P(M_{1500},z,r_h,\beta)$ is the result
from our completeness simulations.  The integral was done over
$\Delta z =$ 1, centered on the center of each redshift bin.  In each
field, we used a weighted mean of this three-dimensional effective volume $V_{eff}(M_{1500},r_h,\beta)$  to calculate
$V_{eff}(M_{1500})$, where the weighting is based on the number of real
objects in a magnitude bin with a given value of r$_{h}$ and $\beta$, as
\begin{equation}
V_{eff}(M_{1500}) = \frac{\sum\limits_{r_h}\sum\limits_{\beta}
  V_{eff}(M_{1500},r_h,\beta)~N(M_{1500},r_h,\beta)}{\sum\limits_{r_h}\sum\limits_{\beta} N(M_{1500},r_h,\beta)}
\end{equation}
This assumes that the completeness corrections estimated using our
observed size and color distributions are similar to what we obtained
if we could measure the true sizes and colors, motivated by our
measurement of minimal size and color biases when comparing the input
to recovered values.

This weighted volume is the most representative of the
  true volume we are sensitive to, as we explicitly account for the
  incompleteness as a function of size and color.
  Figure~\ref{fig:veff} highlights the dependence of the effective
  volume on these quantities for galaxies in our $z =$ 6 sample in the
  GOODS-N Deep field.  The
  effective volume has a strong dependence on the surface brightness
  of galaxies, as the volume drops steeply both for larger sizes and
  fainter magnitudes.  The central and right panels highlight that while the
  effective volume (and thus sample completeness) is sensitive to both
  size and color, the color has a relatively minor role.  We remain sensitive to fairly
  red galaxies ($\beta = -$1), similar to previous results from
  \citet{bouwens12}.  Although the effective volume has
a strong dependence on size, the relatively small sizes of galaxies in
our sample yields a volume similar to that obtained when assuming a
constant small size.  Thus, although our volumes are the most
accurate, had we assumed a fixed effective radius of, e.g., r$_{e} =$
1 kpc, our results would not change significantly. This is consistent
with the conclusions of \citet{grazian12} who found, 
accounting for the size-luminosity relation when deriving the $z =$ 7
luminosity function, similar results as previous studies that neglected
the size-luminosity relation.
Our final effective volumes are shown in Figure~\ref{fig:volumes}.

\section{The Luminosity Function}

\subsection{Parametric Approach}

Possessing our final galaxy sample with measured values of
$M_{1500}$, as well as the effective volumes from our completeness simulations
in the previous section, we can now proceed to measure the rest-frame
UV luminosity function at $z =$ 4, 5, 6, 7 and 8.
We calculate the luminosity function in two ways: a
parametric version assuming that the luminosity function takes the
form of a \citet{schechter76} function, and a non-parametric
step-wise maximum likelihood (SWML) calculation.

The fitting of a Schechter function is well motivated, as it
successfully matches the observed rest-UV luminosity functions at lower
redshifts \citep[e.g.,][]{reddy09,bouwens06}.  This function is
characterized by a power-law at the faint end with slope $\alpha$, and
an exponential cut-off at the bright end, transitioning between the
two regimes at the characteristic magnitude M$^{\ast}$.  The
parameter $\phi^{\ast}$ sets the normalization of this function.
The number density at a given magnitude is then given by
\begin{equation}
\phi(M) = 0.4~\mathrm{ln}~(10) \!~\phi^{\ast}\!~10^{-0.4(M-M^{\ast})(\alpha + 1)}\!~e^{-10^{-0.4(M-M^{\ast})}}
\end{equation}

For the measurement of the luminosity function assuming a Schechter
functional form, we calculated the likelihood that the number
of observed galaxies in a given magnitude bin is equal to that
for an assumed value of the Schechter parameters $M^{\ast}$ and
$\alpha$.  Rather than performing a grid-based search, we performed a Markov Chain
Monte Carlo (MCMC) search algorithm, to better span the parameter
space, as well as to better characterize the uncertainties on the
Schechter parameters.  We performed this calculation in bins of
absolute magnitude with $\Delta$M $=$ 0.5 mag, ranging from $-$24
$\leq$ $M_{1500} \leq -$17.  At the bright end we are in the limit
of small numbers, therefore we model the probability distribution as a
Poissonian distribution \citep[e.g.,][]{cash79,ryan11}, with:
\begin{equation}
\mathcal{C}^2(\phi) = -2~\mathrm{ln}~\mathcal{L}(\phi)
\end{equation}
\begin{equation}
 C^2(\phi) = -2\sum\limits_{i}\! \sum\limits_{j}\!~N_\mathrm{j,obs}\!~\mathrm{ln}(N_\mathrm{j,model}) \!~-\!~ N_\mathrm{j,model}\!~-\!~ \mathrm{ln}(N_\mathrm{j,obs}!) 
\end{equation}
where $\mathcal{L}(\phi)$ is the likelihood that the expected number
of galaxies ($N_\mathrm{model}$) matches that observed
($N_\mathrm{obs}$) for a given value of $M^{\ast}$ and $\alpha$, and
$C^2$ is the goodness-of-fit statistic.  The subscripts $i$ and $j$
represent the sub-fields and magnitude bins, respectively.  The final
goodness-of-fit is the sum over all fields and magnitudes in a given
redshift bin.
We use the effective volume results for a given redshift, magnitude bin,
and field to convert from the model number density to the expected
number, calculating $\phi^{\ast}$ as the normalization such that the
total expected number of galaxies over all
magnitude bins matches the total number of observed galaxies.

\begin{figure*}[!t]
\epsscale{0.58}
\plotone{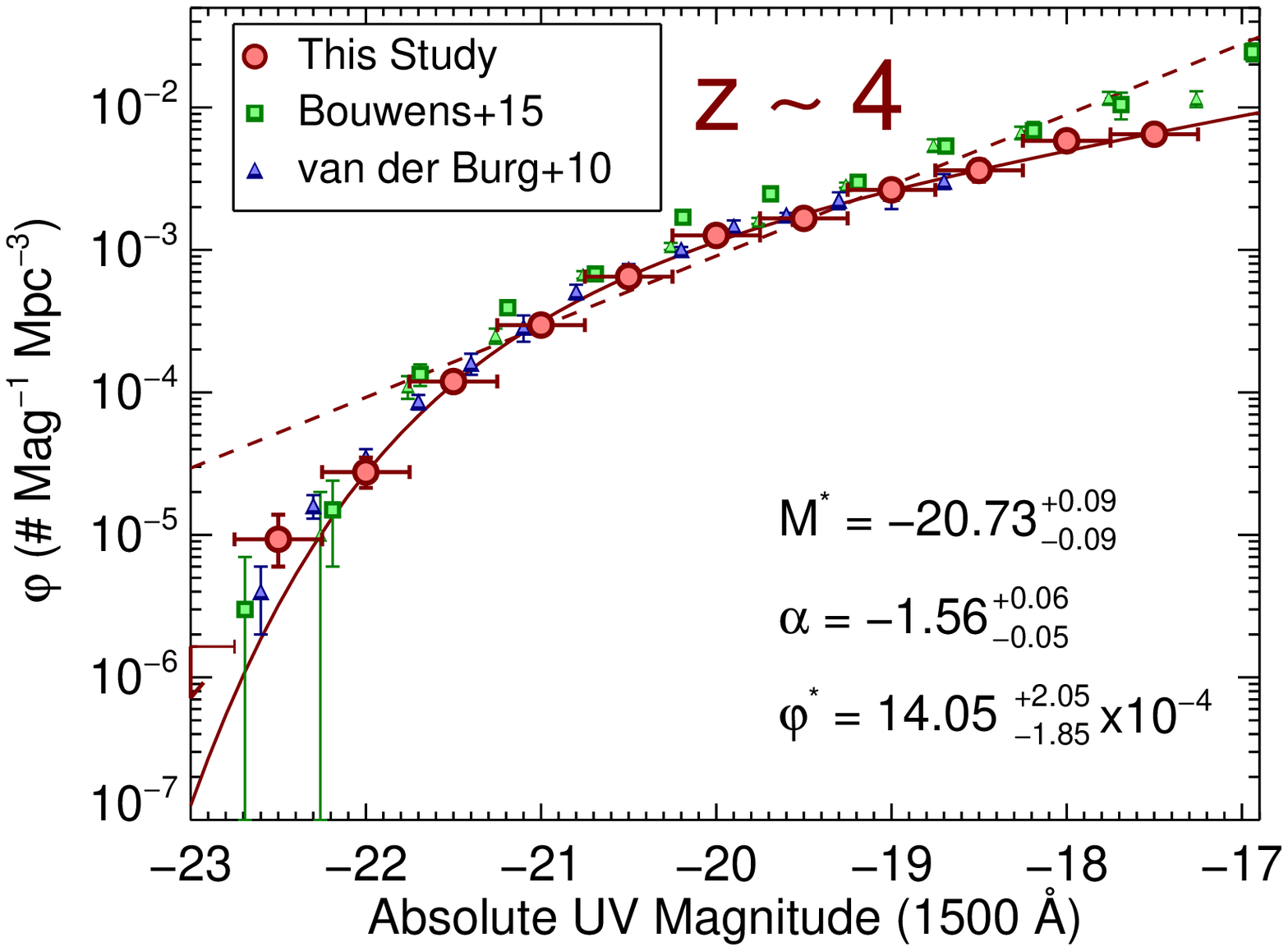}
\plotone{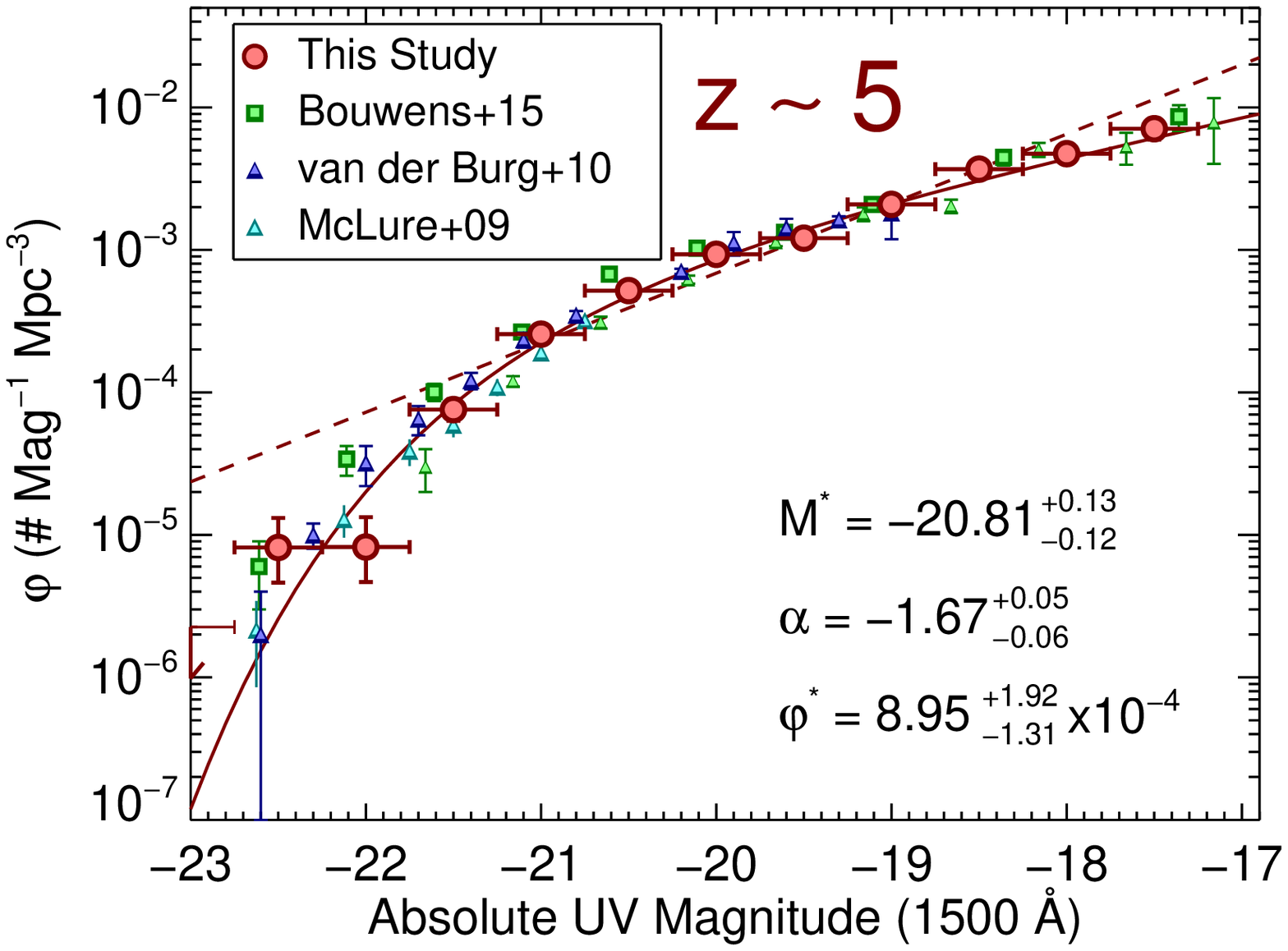}
\plotone{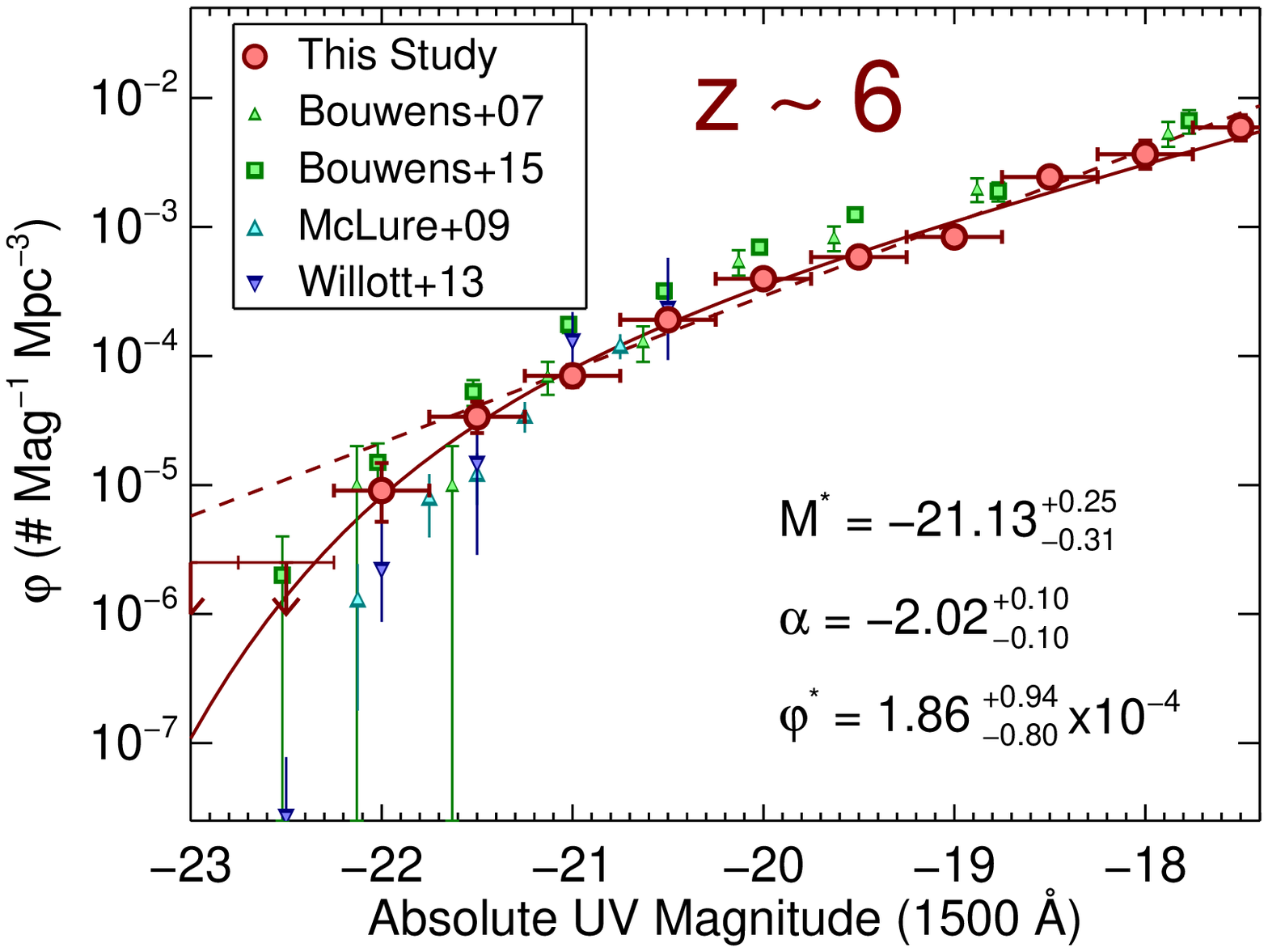}
\plotone{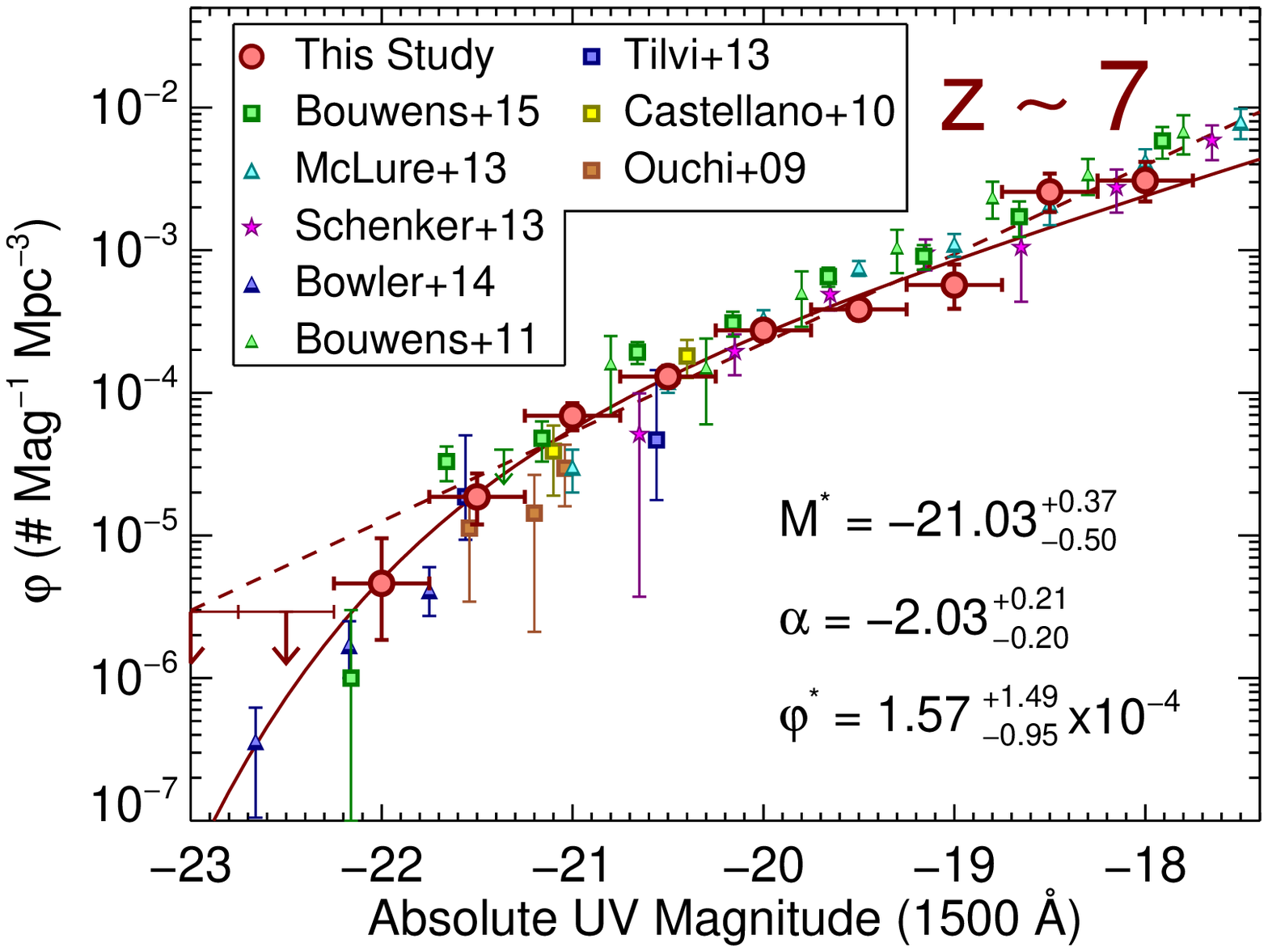}
\plotone{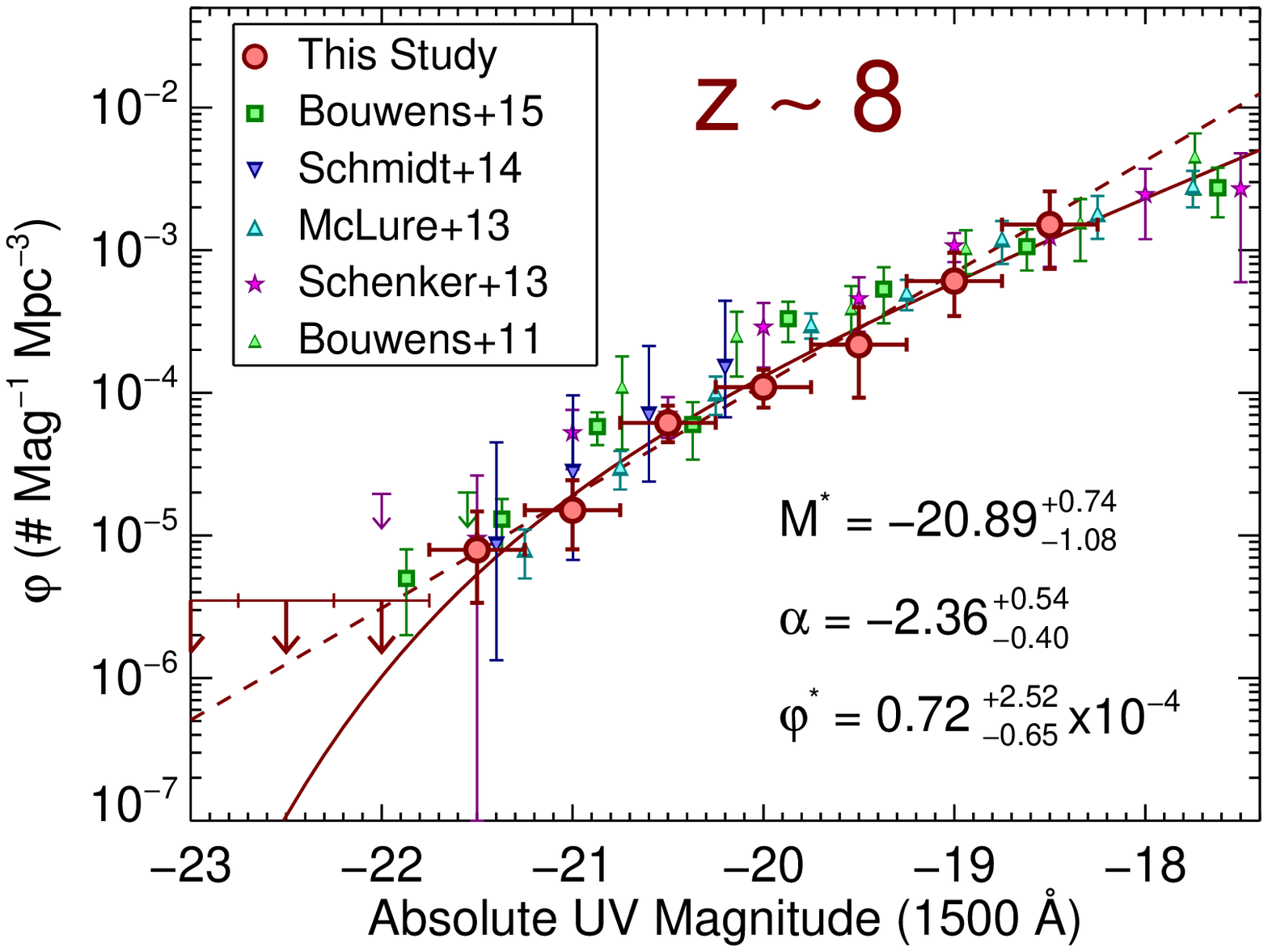}
\caption{The rest-frame UV luminosity functions for our $z =$ 4--8
  galaxy samples.  The large red circles denote our step-wise maximum
  likelihood luminosity function, while the solid red line denotes
  our best-fitting Schechter function, with the best-fit values
  given by the inset text.  We do not make use of our data below the
 determined 50\% completeness level in each field.  As the HUDF is our
 deepest field, the magnitude of our
 last data point denotes the 50\% completeness limit in the HUDF.  The dashed lines show the
  best-fit single power law at each redshift.  We also show several luminosity
  functions from the literature as indicated in the legends.}
\label{fig:lf}
\end{figure*}  

For each magnitude bin, we performed 10 independent MCMC chains utilizing a
Metropolis-Hastings algorithm, each
of 10$^5$ steps, building a distribution of $M^{\ast}$,
$\alpha$ and $\phi^{\ast}$ values for each
field.  During each step of
the chain, the likelihood of a
given model was computed for each of our observed fields, and then
added together to compute the likelihood for the sample as a whole
(we also recorded the individual field values, see \S 6.5).  
Prior to each recorded chain, we performed a burn-in run with a number of
steps equal to 10\% of the number of steps in each chain.  The starting point
for the burn is a brute-force $\chi^2$ fit of a grid of $\alpha$ and
$M^{\ast}$ values to our data.  At the end of the burn, the final values of the
parameters from the last step were then the starting
points for each chain.  The burn-in results were not otherwise recorded.
During each step, new values of $M^{\ast}$ and
$\alpha$ were chosen from a random Gaussian distribution, with the
Gaussian width tuned to generate an approximate acceptance rate of
23\%.  During each step $\phi^{\ast}$ was calculated as the
normalization.  If the difference between the likelihood of the model for
the current step exceeds that from the previous step by more than a
randomly drawn value ($\equiv$ 2~ln~($n$); where $n$ is a uniform random
number between zero and unity), then the current values of the
Schechter function parameters were recorded.  If not, the chain
reverted to the value from the previous step.  

By running 10 independent chains, we mitigate against being trapped by local minima
in the parameter space.  Our final result concatenates these 10
chains together, giving a distribution of 10$^{6}$ values of the
Schechter function parameters at each redshift.  The results were
visually inspected to confirm that the chains reached convergence.  
For each Schechter function parameter, the best-fit values were taken to be the
median of the distribution, with the uncertainty being the central
68\% of the distribution.  These results are given in Table 4.
For the $z =$ 8 Schechter function
fit, we imposed a top-hat prior forcing $M^{\ast}_\mathrm{UV}$ to be fainter than
$-$23.  Without this prior, the fit preferred a much brighter value of
$M^{\ast}_\mathrm{UV}$, such that the observed data points all lay on the
faint-end slope (i.e., a single power law).  We discuss the implications of this in \S 6.6.

Although we computed the volumes down to very faint
magnitudes, should we include these faint galaxies and calculate the luminosity
function down to $M_{1500} = -$17 or fainter, we would be highly incomplete
(Figure~\ref{fig:volumes}).  In practice, it is our deepest field (the HUDF) which
determines how faint we can constrain the luminosity function.  The
HUDF drops below 50\%
completeness at magnitudes fainter than
$M_{1500} \sim -$17.5 at $z =$ 4, 5 and 6, $-$18 at $z =$ 7, and
$-$18.5 at $z =$ 8.  Thus, in our calculation of the luminosity function, we
only include a given field's contribution at a given magnitude if it
is above the 50\% completeness limit for that magnitude and redshift.
Extending the analysis fainter will give
results dominated by the incompleteness correction.  

As shown in Figure~\ref{fig:volumes}, while the volume
per unit area for the different fields is very tight at $z =$ 4, there
is a progressively larger scatter apparent when moving towards higher
redshift, representing a systematic uncertainty in the effective volume
calculation.  One likely culprit is the fact that the volumes depend on
the distribution of the sizes and colors of objects in a given field.
For fields with few sources (i.e., the smallest fields at the highest
redshifts), there may be only a single object in a given magnitude
bin.  To mitigate significant variances in the effective volume at the
bright end, where numbers of sources are small, we set the effective
volume in a given redshift bin and field in bright bins with less than three
objects equal to the value in the brightest bin with more than three
objects (i.e., if there are no magnitude bins with more than three
objects at M $< -$ 21, the effective volumes for all brighter bins are
set equal to the value at M $= -$ 21).  This change has no discernable
effect on our luminosity function results as this is well above the 90\%
completeness limit for any of our fields, and is thus only done to
keep small numbers of galaxies from significantly affecting the
volumes. 

\begin{deluxetable}{cccc}
\tabletypesize{\small}
\tablecaption{Schechter Function Fits to the Luminosity Function}
\tablewidth{0pt}
\tablehead{
\colhead{Redshift} & \colhead{M$^{\ast}$} & \colhead{$\alpha$} & \colhead{$\phi^{\ast}$}\\
\colhead{$ $} & \colhead{$ $} & \colhead{$ $} & \colhead{(Mpc$^{-3}$)}
}
\startdata
4&$-$20.73$^{+0.09}_{-0.09}$&$-$1.56$^{+0.06}_{-0.05}$&(14.1$^{+2.05}_{-1.85}$) $\times$10$^{-4}$\\[1ex]
5&$-$20.81$^{+0.13}_{-0.12}$&$-$1.67$^{+0.05}_{-0.06}$&(8.95$^{+1.92}_{-1.31}$) $\times$10$^{-4}$\\[1ex]
6&$-$21.13$^{+0.25}_{-0.31}$&$-$2.02$^{+0.10}_{-0.10}$&(1.86$^{+0.94}_{-0.80}$) $\times$10$^{-4}$\\[1ex]
7&$-$21.03$^{+0.37}_{-0.50}$&$-$2.03$^{+0.21}_{-0.20}$&(1.57$^{+1.49}_{-0.95}$) $\times$10$^{-4}$\\[1ex]
8&$-$20.89$^{+0.74}_{-1.08}$&$-$2.36$^{+0.54}_{-0.40}$&(0.72$^{+2.52}_{-0.65}$) $\times$10$^{-4}$
\enddata
\tablecomments{The final values for each parameter are the median of the
  parameter distribution from the MCMC analysis.  The
  quoted errors represent the 68\% confidence range on each parameter.}
\end{deluxetable}

\begin{deluxetable*}{cccccc}
\tabletypesize{\small}
\tablecaption{Rest-Frame Ultraviolet Luminosity Function: Stepwise Maximum Likelihood Method}
\tablewidth{0pt}
\tablehead{
\colhead{M$_{1500}$} & \colhead{$\phi~(z \approx$ 4)}& \colhead{$\phi~(z \approx$ 5)}& \colhead{$\phi~(z \approx$ 6)} & \colhead{$\phi~(z \approx$ 7)} & \colhead{$\phi~(z \approx$ 8)}\\
\colhead{$ $} & \colhead{(10$^{-3}$ Mpc$^{-3}$ mag$^{-1}$)} &
\colhead{(10$^{-3}$ Mpc$^{-3}$ mag$^{-1}$)} & \colhead{(10$^{-3}$ Mpc$^{-3}$ mag$^{-1}$)} & \colhead{(10$^{-3}$ Mpc$^{-3}$ mag$^{-1}$)} & \colhead{(10$^{-3}$ Mpc$^{-3}$ mag$^{-1}$)}
}
\startdata
-23.0&$<$0.0016&$<$0.0023&$<$0.0025&$<$0.0029&$<$0.0035\\[1ex]
-22.5&0.0093$_{-0.0033}^{+0.0045}$&0.0082$_{-0.0035}^{+0.0050}$&$<$0.0025&$<$0.0029&$<$0.0035\\[1ex]
-22.0&0.0276$_{-0.0062}^{+0.0074}$&0.0082$_{-0.0036}^{+0.0051}$&0.0091$_{-0.0039}^{+0.0057}$&0.0046$_{-0.0028}^{+0.0049}$&$<$0.0035\\[1ex]
-21.5&0.1192$_{-0.0132}^{+0.0145}$&0.0758$_{-0.0125}^{+0.0137}$&0.0338$_{-0.0085}^{+0.0105}$&0.0187$_{-0.0067}^{+0.0085}$&0.0079$_{-0.0046}^{+0.0068}$\\[1ex]
-21.0&0.2968$_{-0.0219}^{+0.0230}$&0.2564$_{-0.0240}^{+0.0255}$&0.0703$_{-0.0128}^{+0.0148}$&0.0690$_{-0.0144}^{+0.0156}$&0.0150$_{-0.0070}^{+0.0094}$\\[1ex]
-20.5&0.6491$_{-0.0347}^{+0.0361}$&0.5181$_{-0.0338}^{+0.0365}$&0.1910$_{-0.0229}^{+0.0249}$&0.1301$_{-0.0200}^{+0.0239}$&0.0615$_{-0.0165}^{+0.0197}$\\[1ex]
-20.0&1.2637$_{-0.0474}^{+0.0494}$&0.9315$_{-0.0482}^{+0.0477}$&0.3970$_{-0.0357}^{+0.0394}$&0.2742$_{-0.0329}^{+0.0379}$&0.1097$_{-0.0309}^{+0.0356}$\\[1ex]
-19.5&1.6645$_{-0.0618}^{+0.0630}$&1.2086$_{-0.0666}^{+0.0488}$&0.5858$_{-0.0437}^{+0.0527}$&0.3848$_{-0.0586}^{+0.0633}$&0.2174$_{-0.1250}^{+0.1805}$\\[1ex]
-19.0&2.6392$_{-0.1165}^{+0.1192}$&2.0874$_{-0.1147}^{+0.1212}$&0.8375$_{-0.0824}^{+0.0916}$&0.5699$_{-0.1817}^{+0.2229}$&0.6073$_{-0.2616}^{+0.3501}$\\[1ex]
-18.5&3.6169$_{-0.6091}^{+0.6799}$&3.6886$_{-0.3725}^{+0.3864}$&2.4450$_{-0.3515}^{+0.3887}$&2.5650$_{-0.7161}^{+0.8735}$&1.5110$_{-0.7718}^{+1.0726}$\\[1ex]
-18.0&5.8343$_{-0.8204}^{+0.8836}$&4.7361$_{-0.4413}^{+0.4823}$&3.6662$_{-0.8401}^{+1.0076}$&3.0780$_{-0.8845}^{+1.0837}$&---\\[1ex]
-17.5&6.4858$_{-0.9467}^{+1.0166}$&7.0842$_{-1.1364}^{+1.2829}$&5.9126$_{-1.2338}^{+1.4481}$&---&---
\enddata
\tablecomments{Magnitude bins with zero objects are shown as upper limits, calculated as 1/V$_{eff}$/$\Delta$M in that magnitude bin for that redshift.}
\end{deluxetable*}

Another possible issue is the source density of simulated objects.
In the smaller fields (HUDF, HUDF parallels, HFF
parallels) we input sources with twice the surface density as in the
larger fields to speed up the computing time.  As some sources (just like real galaxies) will
inevitably fall on top of real sources, and thus not be recovered, an
increased source density could result in a (slightly) lower
completeness.  This is just what is observed in these fields, as shown
in Figures~\ref{fig:probs} and \ref{fig:volumes}.  To account for this
uncertainty, we measured the spread in volume per unit area in each field at
$M_{1500} = -$21 at each redshift, which we found to be $\sim$1.5\%, 3.8\%,
6.2\%, 7.8\% and 13\% at $z =$ 4, 5, 6, 7 and 8, respectively.  At each
step in the MCMC chain, we perturbed the effective volume by this
amount to account for this systematic uncertainty in our luminosity
function results.

\subsection{Non-Parametric Approach}
 
We have also examined a non-parametric approach to studying evolution
in the luminosity function.  This is
particularly warranted at very high redshift, where the effects
responsible for suppressing the bright end of the luminosity
function and causing the exponential decline in number density (e.g., active galactic nuclei feedback, or dust attenuation) may be
less relevant.  We thus calculated the SWML
luminosity function, which is essentially the number density at a
given magnitude, free from assumptions about the functional form of number
density with magnitude.  We also calculated the SWML luminosity
function using an MCMC sampler.  In this case, as the number densities
in the magnitude bins are not linked by an overarching function, we
calculated the number density in each magnitude bin independently.

For each magnitude bin and for each field, the likelihood was
calculated (using Equations 5 and 6 above) that a given randomly drawn value of
$\phi(M)$ will give the observed number of galaxies.  The actual
recorded value of $\phi(M)$ is that which maximizes the likelihood.  While in
practice, this yields very similar results as one would get by simply
taking the observed number and dividing by the effective volume
(consistent within a few percent for bins with more than a few galaxies), our
approach has two advantages.  First, in the limits where numbers are
small, this approach is more accurate in that it properly accounts for
the Poissonian likelihood.  Secondly, this approach generates a full
probability distribution for the number densities in each magnitude
bin, allowing for the derivation of accurate asymmetric uncertainties.
Our SWML luminosity function determinations and best-fit Schechter
functions are given in Table 5 and shown in Figure~\ref{fig:lf}.

\begin{figure*}[!t]
\epsscale{1.1}
\plotone{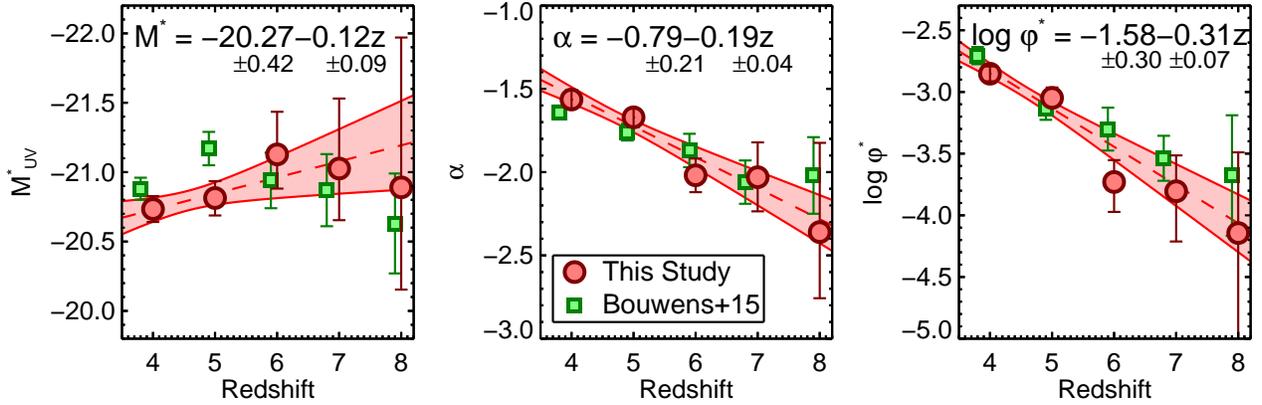}
\caption{The evolution of the Schechter function parameters.  
Red circles show results of this study, and green
squares show results from \citet{bouwens15}.  Dashed lines show
best fit evolution as a linear function of $z$ indicated in each
panel, and red shaded regions show the 68\% confidence range of the
linear form.  Contrary to previous studies, we find no significant evolution in
  $M^{\ast}_\mathrm{UV}$.  We find significant ($>$4$\sigma$) evolution in $\alpha$,
  from steeper slopes at higher redshift to shallower slopes at lower
  redshift, and in the characteristic
  number density $\phi^{\ast}$, which evolves
to higher values by a factor of 20$\times$ from $z =$ 8 to 4.}
\label{fig:schechter}
\end{figure*}  

\begin{figure*}[!t]
\epsscale{1.1}
\plotone{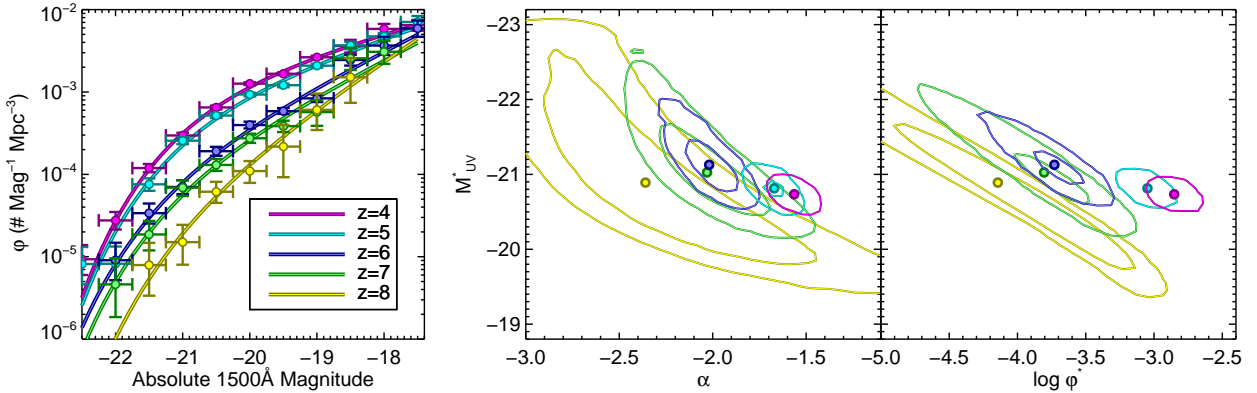}
\caption{Left) The evolution of the UV luminosity function from $z =$ 4 to
  8, where the circles and lines denote our step-wise and
  Schechter-parameterized luminosity functions, respectively.  Center) Contours of covariance between $\alpha$ and
  $M^{\ast}_\mathrm{UV}$ at $z =$ 4, 5, 6, 7 and 8.  The contours denote the 68\% and
  95\% confidence levels, while the small circles show the
  best-fitting values.  Right) Contours of covariance between $\phi^{\ast}$ and
  $M^{\ast}_\mathrm{UV}$ at $z =$ 4, 5, 6, 7 and 8.}
\label{fig:plot2}
\end{figure*} 

\begin{figure}[!t]
\epsscale{1.15}
\plotone{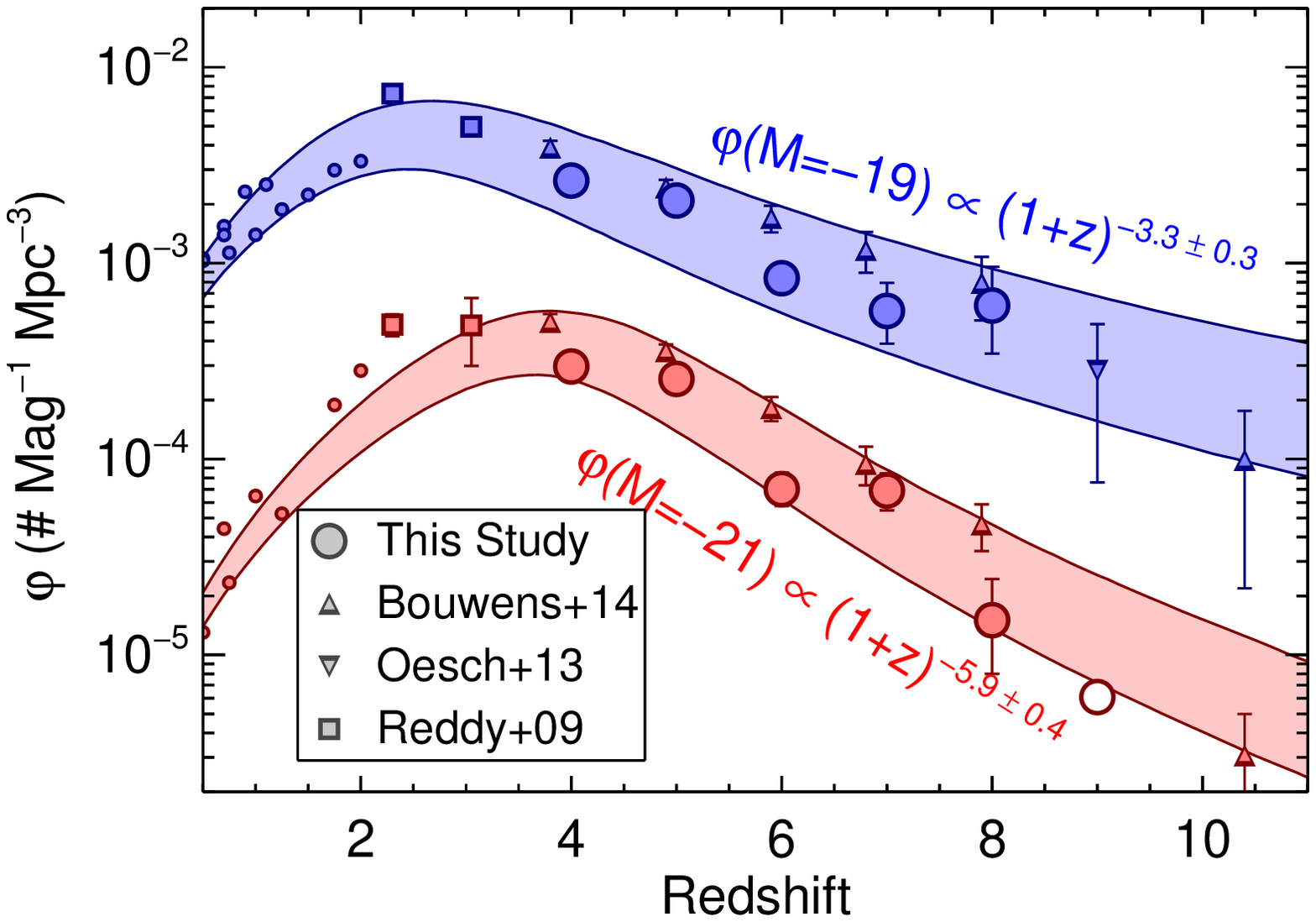}
\caption{The number densities of bright (M$_\mathrm{UV} = -$21) and faint
  (M$_\mathrm{UV} = -$19) galaxies at a variety of redshifts.  Our data are
  shown as large circles, and we also show results at high-redshift
  from \citet{bouwens15}, \citet{oesch13} and \citet{reddy09}.  At low redshift, we
  show results as small circles from \citet{arnouts05},
  \citet{oesch10} and \citet{cucciati12}.  We fit the trend of $\phi$
  with redshift, separately for our two magnitude bins, with
  the function given in Equation 7.  The shaded regions show the 68\%
  confidence ranges for each of the fits.  The value of the slope of
  this function at high-redshift is significantly steeper for bright galaxies than for
  faint galaxies, showing that from $z =$ 8 to 4, bright galaxies
  become more abundant at a faster rate than faint galaxies.  This trend reverses at $z =$ 4,
where bright galaxies stop becoming more common.  Bright galaxies peak
in number density at $z =$ 3.1--3.2, sooner than faint galaxies, which
peak at $z =$ 2.4-2.7 (68\% C. L.).  At $z <$ 2, the
abundances of both populations plummet, in line with the
evolution of the cosmic star-formation rate density.}
\label{fig:ndens}
\end{figure} 

\section{Luminosity Function Interpretation}

\subsection{Evolution}
As shown in Figure~\ref{fig:lf}, the qualitative shape of the SWML luminosity functions
at all redshifts we consider here are similar, in that bright galaxies are rare
and faint galaxies are relatively common.  Additionally, when
examining the Schechter fits (solid line), we see that they are
consistent with the SWML determinations.  The best-fit
Schechter function parameters (Table 4) surprisingly show little
evolution in $M^{\ast}_\mathrm{UV}$.  However, from $z =$ 4 to 8, the
uncertainty on $M^{\ast}_\mathrm{UV}$ gets progressively larger, to 0.4 (0.9) mag at $z
=$ 7 (8).  This is easy to understand, as at all redshifts, our dataset
contains galaxies in only 1-2 bins brightward of $M^{\ast}_\mathrm{UV}$.  Ideally,
one would prefer to have multiple bins in magnitude on either side of
$M^{\ast}_\mathrm{UV}$ to obtain robust constraints.  As shown here, that will
require a larger volume than we consider in this analysis.  In
Figure~\ref{fig:schechter}, we also fit the evolution of $M^{\ast}_\mathrm{UV}$
with redshift with a linear function, using our results at $z =$ 4--8.  We find that
$dM^{\ast}/dz =-$0.12 $\pm$0.09; thus, our data do not support a significant
evolution of $M^{\ast}_\mathrm{UV}$ with redshift.  

We also fit similar functions to see if we
detect evolution in $\alpha$ and $\phi^{\ast}$.  As shown in
Figure~\ref{fig:schechter}, we do see significant evolution in the
faint-end slope $\alpha$, with it becoming steeper at higher redshift,
as $d\alpha/dz =-$0.19 $\pm$0.04 (4.8$\sigma$
significance).  We see a similar significance in the evolution of the
characteristic number density $\phi^{\ast}$, which evolves as
$dlog\phi^{\ast}/dz =-$0.31 $\pm$0.07 (4.4$\sigma$ significance).
Thus, while $M^{\ast}_\mathrm{UV}$ does not significantly evolve with redshift
from $z =$ 4 to 8, both $\alpha$ and $\phi^{\ast}$ do, in that the
number density decreases and the faint-end slope becomes steeper with
increasing redshift.  In particular, this decline in characteristic
number density is by a factor of $\sim$20,
over a period of time of less than 1 Gyr. Although the steepening of
the faint end is
consistent with previous studies \citep[e.g.,][]{bouwens12}, the
un-evolving $M^{\ast}_\mathrm{UV}$ and strong number density evolution are the
opposite of the picture presented in the literature just one year ago
\citep[e.g.,][]{bouwens07,bouwens11,mclure13}.  This updated
evolutionary picture will be crucial when projecting number counts for
future {\it HST} and {\it James Webb Space Telescope} surveys.

In Figure~\ref{fig:plot2}, we show our determinations of the luminosity
functions together at all five redshifts, along with the joint confidence
contours on $M^{\ast}_\mathrm{UV}$, $\alpha$ and $\phi^{\ast}$. 
It is apparent that there is significant evolution in the
luminosity function, with a drop in number density from $z =$ 4
to 8, as well as a gradual steepening of the faint-end slope.  The
apparent non-evolution of the characteristic magnitude is visible as
the roughly constant magnitude of the ``knee'' of the luminosity function.

\subsection{Impact of Magnitude Uncertainties}
By definition, our method of computing the luminosity function is
dependent on magnitude binning, as we compare the observed number to
that expected based on a given model in magnitude bins of width 0.5
mag.  While galaxies close to one side
of a magnitude bin have the potential to scatter to another bin, the
typical uncertainties on the UV absolute magnitudes of
galaxies in our sample are $\sim$0.2 mag.  Additionally, galaxies can
shift both ways, thus while one galaxy moves out of a bin, another may
move in, though this effect will not be symmetric given the 
shape of the luminosity function.  In our results above, we had assumed that
magnitude scatter does not significantly impact our results.

\begin{figure*}[!t]
\epsscale{1.1}
\plotone{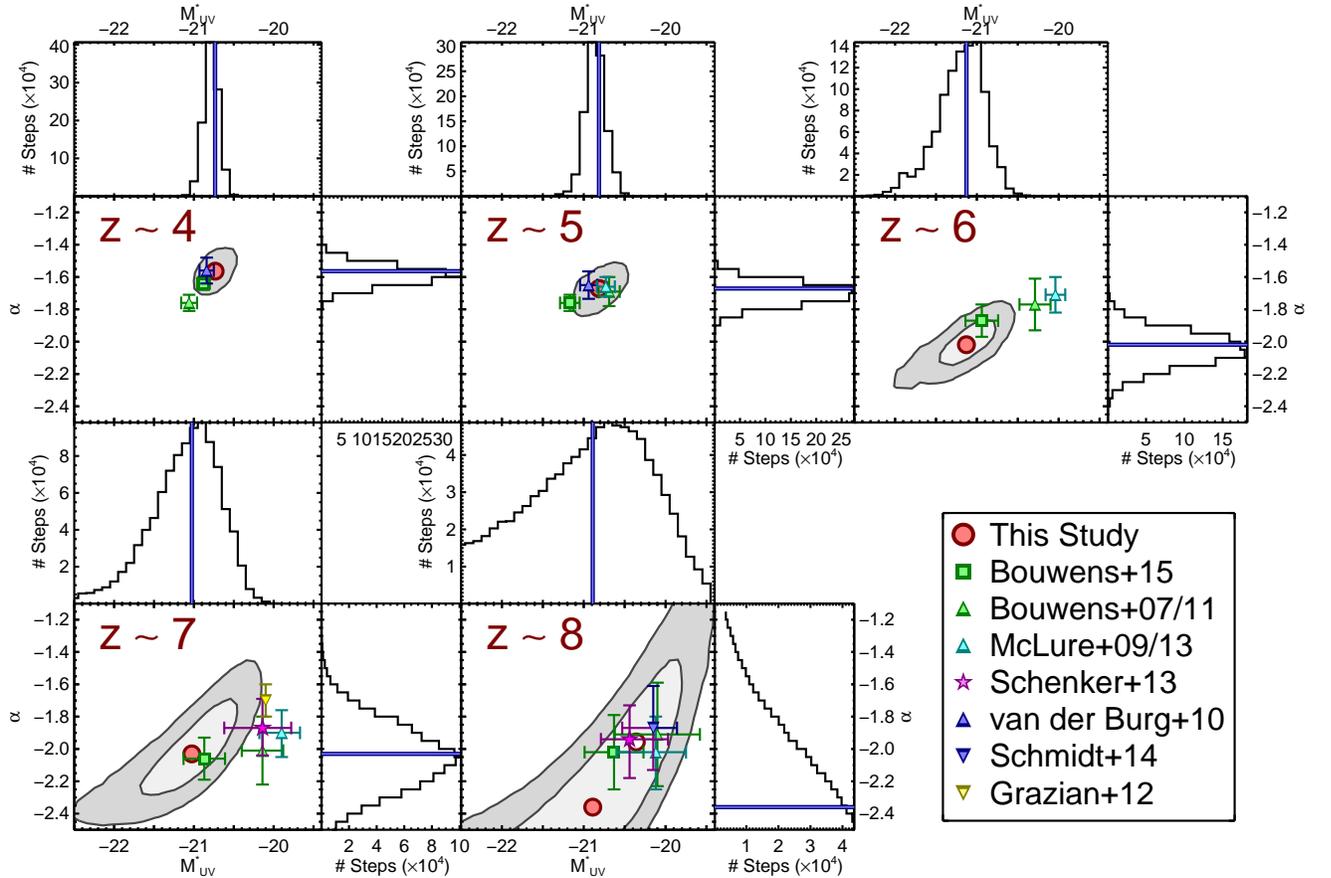}
\caption{Confidence contours on our measured value of the faint-end
  slope $\alpha$ and the characteristic magnitude $M^{\ast}_\mathrm{UV}$ at $z =$
  4, 5, 6, 7 and 8, with the light and dark shaded regions denoting 68\% and 95\%
  confidence.  The large red circles represent our fiducial best-fit luminosity
  function parameters, while the other colored symbols denote results
  from previous studies, using the same symbols as in Figure~\ref{fig:lf} (with the addition of the results from
\citet{grazian12}, shown as the yellow triangle in the $z =$ 7 panel,
who fit $\alpha$ keeping $M^{\ast}_\mathrm{UV}$ fixed to $-$20.14).  In the $z =$ 8 panel, we also show our best-fit
result when fixing $\alpha \geq -$2.3 as the white-filled red circle.
The histograms to the top and side of each contour plot show the
number of MCMC steps when a given value of $M^{\ast}_\mathrm{UV}$ or $\alpha$ was
recorded, with the median value shown by the blue line.}
\label{fig:contours}
\end{figure*}  

To investigate the impact of this assumption, we performed another
iteration of MCMC fitting to our data, here allowing galaxies to
scatter between magnitude bins.  At each step in the MCMC chain, a new
value of $M_{1500}$ was drawn for each galaxy from the 100 SED-fitting
Monte Carlo simulation results.  The spread in these values
encompassed both the photometric scatter in the observed filters and
the uncertainty in the photometric redshift (see \S 3.7).  To
compare to our fiducial luminosity function values, we recorded both
the median Schechter parameter results and the median
number density in each magnitude bin, as, unlike our fiducial MCMC
run, these varied during each step as the magnitudes changed.
At all redshifts, our fiducial values of the step-wise luminosity
functions are consistent with these ``magnitude-scatter'' values within
10\% at M $\geq -$21.5, and
typically within 2-3\%.  The sole exception is in the
brightest bin ($-$22 at $z =$ 4--6, and $-$21.5 at $z =$ 7 and 8),
where our fiducial number density values are higher by $\sim$15-20\%
(60\% at $z =$ 7, where there is only a single galaxy in this bin).
We examined the Schechter fit, to see whether this bright-end
difference affects our results.  Values of both
$M^{\ast}_\mathrm{UV}$ and $\alpha$
derived when allowing galaxies to shift between bins are consistent
with our fiducial values within 0.1 mag and $<$3\%, respectively.  We
conclude that the relatively
small ($\sim$20\%) uncertainties in the absolute magnitudes of our
galaxies do not have a significant impact on our luminosity function results.

\subsection{Non-parametric Evolution}
Given that our results show that the Schechter functional parameters
may not be a robust method of tracking galaxy evolution (e.g., a
non-evolving value of $M^{\ast}_\mathrm{UV}$ does not mean that the galaxy populations
are not evolving), we examine
the evolution in a non-parametric way.  In Figure~\ref{fig:ndens} we
show the evolution of the step-wise luminosity function, plotting the
number density corresponding to galaxies at M$_\mathrm{UV} = -$21 and $-$19
versus redshift.  From $z =$ 8 to 4 the abundance of
brighter galaxies increases faster than
faint galaxies.  This trend halts at $z =$ 4, where bright galaxies
have an approximately constant abundance down to $z =$ 2, and then turns over.  Faint
galaxies, however, continue increasing in abundance down to $z =$ 2, where
they also turn over.  This figure highlights the phenomenon of
downsizing, where bright/large galaxies grow faster at early times
\citep[e.g.,][see also Lundgren et al.\ 2014]{cowie96}.  This is different from the expectation one
would get simply from examining Schechter fits, as the luminosity
functions don't evolve much over the range 2 $< z <$ 4 \citep[e.g.,][]{reddy09}.  Given that
the trends here mimic the evolution of the cosmic SFR density, we fit
the function provided by \citet{madau14} to our data for both number
densities, given by
\begin{equation}
\phi(z)=A~\frac{(1+z)^{\alpha}}{1+[(1+z)/B]^{\gamma}}~\mathrm{mag}^{-1}~\mathrm{Mpc}^{-3} .
\end{equation}
The evolution with redshift is thus proportional to ($1+z$)$^{\alpha}$
at low redshift, and ($1+z$)$^{\alpha-\gamma}$ at high redshift, with
$A$ and $B$ as dimensionless coefficients.
Fitting the data in this way, we confirm that at $z >$ 3, bright galaxies change
in abundance faster, as ($1+z$)$^{-5.9 \pm 0.4}$, than faint galaxies,
which go as ($1+z$)$^{-3.3 \pm 0.3}$.

Another interesting aspect is to compare the trends observed to our predicted
abundance of bright $z =$ 9 galaxies (see \S 9).  The trend observed
here slightly
overestimates our predicted $z =$ 9
abundance, though if we assume the uncertainties on our $z =$ 8 number
density applies to $z =$ 9, our trend is consistent with this prediction.
In any case, this trend of abundance with redshift lends more weight to our expectation of a significant
abundance of bright $z =$ 9 galaxies.  This figure
neglects the impact of dust attenuation, as we are only looking at the
observed UV magnitudes.  The dust attenuation appears to be
luminosity dependent (bright galaxies are dustier than faint galaxies,
e.g., Bouwens et al.\ 2013), as well as being higher at lower
redshift.  Thus, correcting for dust would not only increase the
abundance of bright galaxies more than that of faint galaxies, it
would increase it by more at lower redshift, thus enhancing the
differences between faint and bright galaxies at $z >$ 4.

\subsection{Comparison to Previous Results}
Our result of a similarly bright value of $M^{\ast}_\mathrm{UV}$ at $z =$ 6, 7
and 8 is a dramatic change from previously published results.  In
Figure~\ref{fig:lf}, we show the step-wise luminosity function results
from several relevant studies from the literature.
Figure~\ref{fig:contours} shows our
uncertainty results, highlighting both the distribution of $M^{\ast}_\mathrm{UV}$
and $\alpha$ from the MCMC chains, as well as the covariance between the
two parameters, along with previous determinations of
$M^{\ast}_\mathrm{UV}$ and $\alpha$ 
\citep{bouwens07,mclure09,ouchi09,vdburg10,bouwens11,mclure13,schenker13,
willott13,bowler14}.  In this subsection we compare solely to previous
work - we reserve the comparison to the contemporaneous work by 
\citet{bouwens15} to \S 6.4.1 below.

At $z =$ 4 and $z =$ 5, both our binned luminosity function data
points as well as our Schechter function parameters are in excellent
agreement with the ground-based study of \citet{vdburg10}.  We are
also in excellent agreement with the ground-based study of
\citet{mclure09} at $z =$ 5.  We find good agreement with the
space-based study of \citet{bouwens07} at $z =$ 5, but at $z
=$ 4 the \citet{bouwens07} result lies outside our 2$\sigma$
confidence region on the Schechter function parameters, in that we
prefer a shallower faint-end slope and a fainter value for $M^{\ast}_\mathrm{UV}$.

At $z =$ 6, our binned luminosity function data points are
consistent within 1-2$\sigma$ with the \citet{bouwens07} results at
the faint end (Figure~\ref{fig:lf}).  At the bright end,
our data are \emph{higher} than those
from both ground-based studies (though again, typically only different
at the 1-2$\sigma$ level).  This is somewhat counter-intuitive,
as one may expect the ground-based studies to suffer a higher
contamination rate, particularly for relatively fainter sources at
higher redshift, due to their inability to resolve stars from
galaxies, but it may also be explained due to an aggressive sample
selection required to minimize contamination.  In any case, the
differences are not highly significant, with the exception of the
brightest data point from \citet{willott13}, which gives a number
density at $M = -$22.5 of 2.7 $\times$ 10$^{-8}$ Mpc$^{-3}$.  While
this is consistent with our upper limit at that magnitude, it is a factor of $\sim$250
lower than our number density only 0.5 mag fainter at $M = -$22 (see
Table 5).  Given the
results at similar magnitudes at lower redshifts, it is highly
unlikely that there is such a steep drop in number density over only a
0.5 magnitude interval, though future large area studies can better
investigate the difference (Bowler et al.\ in prep).  The larger
discrepancy comes when comparing the Schechter function parameters.
Specifically, \citet{bouwens07} found M$^{\ast} = -$20.29 $\pm$ 0.19,
and \citet{mclure09} found M$^{\ast} = -$20.04 $\pm$ 0.12.  Both values
are significantly (2--3$\sigma$) fainter than our derivation of M$^{\ast}
= -$21.13$^{+0.25}_{-0.31}$.  For the space-based study of \citet{bouwens07},
this is understandable, as at that time only optical data were
available, thus $z =$ 6 galaxies were selected via detections in only
one band, and a robust determination of their UV absolute magnitudes
was difficult.  For the ground-based study of \citet{mclure09}, a
cause for the difference is less clear, though certainly the different
data being used plays a role.

Comparing to previous works at $z =$ 7, we find broadly similar
results, in that our results are consistent with the derived number
densities from previous studies, yet our Schechter fit prefers a much
brighter value of $M^{\ast}_\mathrm{UV}$.  This is easier to understand, as
a number of previous studies had less data available, and thus, utilizing smaller
volumes, were unable to constrain the bright end
\citep[e.g.,][]{bouwens11,schenker13}.  The exception is the recent
work by \citet{mclure13}, which used a similar volume as our
study, though they used the CANDELS UDS field in place of our use of
the CANDELS GOODS-N field.  Examining the brightest data point from
\citet{mclure13} at $M = -$21, the number density is about a
factor of two below our data point.  However, the discrepancy is
mitigated by two factors.  First, as discussed by
\citet{bouwens15}, the use of fixed diameter circular apertures by
\citet{mclure13} systematically underestimates the fluxes for bright,
more extended, galaxies.  \citet{bouwens15} estimated the amplitude of
this effect to be $\sim$0.25 mag.  Shifting the brightest
\citet{mclure13} data point by 0.25 mag brings it into agreement
with our results.  Secondly, the CANDELS GOODS-N field appears to have
an overdensity of $z =$ 7 galaxies.  Specifically, when comparing the
number density of $z =$ 7 galaxies in the GOODS-N Deep field to the
GOODS-S Deep field in Figure~\ref{fig:fieldlfs}, GOODS-N has a higher
number density at all magnitudes.   While we have not selected galaxy
samples in the UDS, we can examine this further by
recomputing our $z =$ 7 stepwise luminosity function using only the GOODS-S and HUDF
fields.  At magnitudes fainter than $-$21 the
results do not change appreciably, as the GOODS-N Deep and Wide fields
lie on either side of our Schechter fit at those magnitudes.  However, the
results using only GOODS-S provide a number density $\sim$33\% lower
at $M = -$21.5 than our fiducial luminosity function.   This
difference is at the 1$\sigma$ level due to the
large Poisson noise contribution in this bin, and thus is
not highly significant.

We also compare to several ground-based studies at $z =$ 7.  
\citet{ouchi09} identified 22 bright $z \sim$ 7 candidate
galaxies over $\sim$0.4 deg$^2$.  Their data points based on detected
galaxies are consistent with our own, though their strict upper limits
at $M \sim -$22 push their Schechter fit to a fainter value of M$^{\ast}_\mathrm{UV}
= -$20.1, although the large uncertainty (0.76) leaves $M^{\ast}_\mathrm{UV}$ 
consistent with our fit at only slightly more than the 1$\sigma$
level.  The stepwise luminosity function from
\citet{castellano10} based on deep HAWK-I data agrees well
with our results, while the results from the zFourGE medium band
survey of \citet{tilvi13} agree at M$ = -$21.5, 
but differ by $\sim$2$\sigma$ at M $= -$20.5.

Recently, \citet{bowler14} have made a significant improvement
in search volume from the ground, discovering 34 luminous $z \sim$ 7 galaxy candidates
over 1.65 deg$^2$, from the UltraVISTA survey data over the COSMOS
field \citep{mccracken12} and the UKIDSS survey over the UDS field \citep{lawrence07}.  Broadly speaking, they are consistent with our results,
and they are highly inconsistent with the previous determinations of
M$^{\ast}_\mathrm{UV} \sim -$20 (Figure~\ref{fig:lf}).  There is a mild tension at $M = -$21.75, where
the value of our Schechter fit at that point is 2$\sigma$ higher
than their derived number density.  However, this is their faintest
magnitude bin, and is only $\sim$50\% complete, thus this data point
relies the most on the completeness correction.
In any case, the fact that \citet{bowler14} found $z \sim$ 7 candidates
out to very bright magnitudes gives us confidence that our brighter
determination of $M^{\ast}_\mathrm{UV}$ is not necessarily dominated by cosmic
variance in our fields, but is a true feature of the $z =$ 7
universe.  However, our present uncertainty on $M^{\ast}_\mathrm{UV}$ of 
$\sim$0.4 mag makes it apparent that more data are needed to constrain this
parameter further.

At $z =$ 8, we again find consistent number densities with
previous studies, though our larger volume allows us to
find more rare, bright ($M = -$21.5) galaxies than observed in some
previous surveys, pushing them to lower values of $M^{\ast}_\mathrm{UV}$ (though
again here our uncertainty on $M^{\ast}_\mathrm{UV}$ is large, so the difference
in our determination is not significantly different from previous
studies).  As noted above, in our fit of the $z =$ 8 luminosity
function, we constrained $M^{\ast}_\mathrm{UV}$ to be fainter than $-$23, to avoid
un-physically bright values, which tended to be preferred in an
unconstrained fit.  We note two important points when comparing to previous studies.
First, while our study did not utilize the pure
parallel BoRG \citep{trenti11} and HIPPIES \citep{yan11} programs, our
bright end is consistent with that from \citet{schmidt14}, based on a
determination of the $z =$ 8 luminosity function over 350 arcmin$^2$
of pure parallel data (for comparison, our search area at $z =$ 8
comprised $\sim$300 arcmin$^2$; Table 1).  The multiple sight-lines of
BoRG and HIPPIES leave their results less susceptible to cosmic
variance effects, so the agreement implies that cosmic variance may
not be strongly affecting our bright end, though we explore this in \S 6.5.

A potentially larger difference between our results and those of previous studies is also
seen at the faint end, in that our faint-end slope is possibly steeper
than previously found.  However, our uncertainty is large, such that
our result of $\alpha = -$2.36$^{+0.54}_{-0.40}$ is consistent with previous
results of $\alpha \approx -$2
\citep[e.g.,][]{bouwens11,mclure13,schenker13}.  Previous
studies use galaxies as faint as M $= -$17.5 in their determination of
the faint-end slope at $z =$ 8.  As discussed above, and shown in
Figure~\ref{fig:volumes}, we find that we fall below 50\% completeness at
$M > -$18.5, thus we do not use galaxies fainter than that in our
luminosity function determinations.  While robust estimates of the number
densities of galaxies at $-$18.5 $\leq$ M $\leq -$17.5 would certainly
improve the confidence on the faint-end slope, we use the same deep
datasets as the other referenced studies (HUDF).  We
would expect the incompleteness to be similar between all studies,
though it does depend on sample selection and the exact
details of the incompleteness simulations.  In any case, constraints
on the faint-end slope at $z =$ 8 should improve in the near future
with further data from the Hubble Frontier Fields program.  Our inclusion of
the Hubble Frontier Fields parallel imaging, even though contributing
only a small number of galaxies at $z >$ 7, did improve the
fractional error on the faint end slope ($\sigma_{\alpha}/\alpha$) by 2\% and 8\% at $z =$ 7
and $z =$ 8, respectively.

Finally, there have also been theoretical estimates of the luminosity
functions at these redshifts, most prominently from \citet{jaacks12a},
who made predictions in good agreement with our observed luminosity functions.
Specifically, their simulations also predict bright values of
$M^{\ast}_\mathrm{UV}$, of $-$21.15, $-$20.82 and $-$21.00 at $z =$ 6, 7 and 8,
respectively.  They also found quite steep faint-end slopes, of
$-$2.15$^{+0.24}_{-0.15}$, $-$2.30$^{+0.28}_{-0.18}$ and
$-$2.51$^{+0.27}_{-0.17}$ at $z =$ 6, 7 and 8, respectively.  Within
the uncertainties, these faint-end slopes are consistent with our own,
though the apparent agreement at $z =$ 8 is tantalizing (though, as
mentioned above, we cannot constrain the slope to be so steep).  Steep
faint-end slopes of $\alpha \sim -$ 2 at these redshifts were also seen by
\citet{salvaterra11} and \citet{dayal13}, though both studies also
predict a brightening in $M^{\ast}_\mathrm{UV}$ towards lower redshift
which is contrary to our observations.

\subsubsection{Comparison to Bouwens et al.\ 2015}
Recently \citet{bouwens15} published a similar study of the
evolution of the UV luminosity function at 4 $< z <$ 10.
Their sample of galaxies is larger than
ours, as in addition to the datasets we use, they selected
galaxies from the CANDELS COSMOS, EGS and UDS fields (though they did
not use the HFF parallel fields).  The data in
these other CANDELS fields have a depth similar to the GOODS-S and N Wide fields, and
thus are most useful for constraining the bright end of the luminosity
function.  Comparing our results, while
the agreement at $z =$ 5 is excellent, the \citet{bouwens15} data
points at $z =$ 4 lie at higher number densities than our own for all
but the brightest bins.  These differences result in a
slightly steeper value of $\alpha$ at $z =$ 4 ($\alpha_{Bouwens} = -$1.64 versus
$\alpha_{This Study} =-$1.56), and a slightly brighter value of
$M^{\ast}_\mathrm{UV}$ ($-$20.88 versus $-$20.73).

\begin{figure}[!t]
\epsscale{1.16}
\plotone{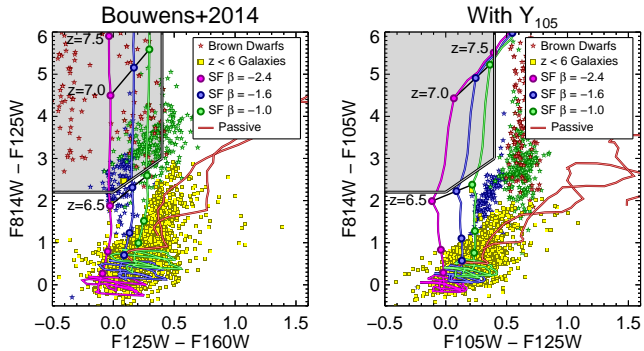}
\caption{Left) Color-selection for $z >$ 6.5 galaxies used by
  \citet{bouwens15} in the
  COSMOS, EGS and UDS fields, where {\it HST} $Y$-band data are not
  available.  Right) Improved color selection in these fields with
  the addition of hypothetical $Y$-band data.  Of particular note is
  that without $Y$-band data, the $z >$ 6.5 selection is potentially
  dominated by M, L and T dwarf stars.  These clear out of the
  selection box with the addition of $Y$-band data.  Additionally,
  galaxies with $z <$ 6 spectroscopic redshifts from the literature
  (yellow boxes) move farther from the selection box, and are less
  likely to scatter in, with the addition of the $Y$-band data.}
\label{fig:colcol}
\end{figure} 

\citet{bouwens15} also selected galaxy samples at $z =$ 6, 7 and 8.
Broadly speaking, they found similar results as we do at $z =$ 6 -- 8, in that previous
studies determined values of $M^{\ast}_\mathrm{UV}$ which were too
faint (Figure~\ref{fig:contours}).  However, investigating the actual data points in
Figure~\ref{fig:lf}, one can see that the
\citet{bouwens15} data points frequently lie above our own.  At $z =$
7 and 8, this difference is typically significant at the 1--2$\sigma$
level, though some bins at $z =$ 6 are discrepant by up to 4$\sigma$.
At $z =$ 8, the \citet{bouwens15} data points are less discrepant from our own.
However, they find both a fainter value of $M^{\ast}_\mathrm{UV}$ and a shallower
faint-end slope.  This is primarily due to their faintest data point,
which, at $M = -$17.5, is well below our 50\% completeness limit, and
lies below the extrapolation of our measured luminosity function,
pushing them to a shallower slope.  However, these differences at $z
=$ 8 are not significant, as Figure~\ref{fig:contours} shows that the
\citet{bouwens15} results lie comfortably within our 68\% confidence
contour on $\alpha$ and $M^{\ast}_\mathrm{UV}$ (in fact, at all
redshifts their Schechter parameters are consistent within the 95\%
confidence limits of our results).  If we constrained
$\alpha$ at $z =$ 8 during our fitting to be $> -$2.3, we obtain best-fit results
similar to \citet[][Figure~\ref{fig:contours}]{bouwens15}.  However, given the data at hand,
there is no robust justification for such a constraint, thus we do not
include this in our fiducial luminosity function fits.

The three CANDELS Wide fields used only by \citet{bouwens15}
lack space-based $Y$-band data, with {\it HST} data present in only
four filters ($V_{606}$, $I_{814}$, $J_{125}$ and $H_{160}$).  These
fields have deep ground-based optical data, although with much poorer angular
resolution, and occasionally shallower depth than available with {\it HST}.  
Of particular worry is contamination by stars and/or
brown dwarfs in these samples.  The left panel of
Figure~\ref{fig:colcol} shows the color-selection plane for galaxies
at $z >$ 6.5 used by \citet{bouwens15} in the CANDELS COSMOS, EGS and
UDS fields.  While the selection space used does include the likely
colors of true $z >$ 6.5 galaxies, it also contains the bulk of M, L
and T-dwarf template colors.  As shown in the right-hand panel, by
adding a single {\it HST} filter, the WFC3 $Y_{105}$-band, stellar contaminants move
out of the selection box, and lower-redshift galaxies move even
further from the selection box.  To mitigate stellar contamination,
\citet{bouwens15} used both colors including ground-based
$Y$-band data, and the Source Extractor stellarity
measurement.  However, the ground-based data are presently not very
deep, with \citet{bouwens15} typically only detecting sources with $Y
<$ 26 (M$_{UV,z=7} \leq -$21; see \S 3.4).  Additionally, the stellarity measurement
can only robustly distinguish point-sources from galaxies much
brighter than the detection limit.  Our test with the CANDELS
$H_{160}$-band imaging in the COSMOS and EGS fields show that a
robustly identified stellar sequence in the stellarity measurement is
only possible at $H_{160} <$ 25 (M$_{UV,z=7} \leq -$22).  In light
of the apparent overabundance of bright galaxies in the
\citet{bouwens15} $z =$ 6 and 7 samples compared to our results, we
conclude that the higher quality data in our fields may yield more robust
and contamination-free measurements of the number densities of bright
galaxies in the distant universe.

\subsubsection{Previously Published Measurement Uncertainties}
The differences in results, particularly on the characteristic
magnitude $M^{\ast}_\mathrm{UV}$ between our current study and previous studies in
the literature, are surprising, as in some cases the differences are
larger than what would have been expected given previously published uncertainties.  
In particular, \citet{bouwens11} initially
derived M$^{\ast} = -$20.14 $\pm$0.26 at $z =$ 7, while now
\citet{bouwens15} find M$^{\ast}
= -$20.87 $\pm$0.26 at $z =$ 7, a result which is discrepant at the
2$\sigma$ level from their previous work.  While the {\it HST} data
presented in \citet{bouwens11} (Figure~\ref{fig:lf}; green
triangles) seem insufficient to
constrain the bright end to such a relatively high precision (as they
include only the small area contained by the HUDF and ERS fields, $<$20\% of the
volume considered in our current work), that
study made use of wide-area ground-based results when deriving their
Schechter parameters to assist with constraining M$^{\ast}$.  In
particular, the dataset they used which constrained the
brightest magnitude bins was that of
\citet{ouchi09}.  Investigating the $z =$ 7 panel of
Figure~\ref{fig:lf}, one can see that the the two brightest Ouchi et al.\ data points
lie systematically below not just our results, but all published
results in that magnitude range (possibly due to an overestimate of
the contamination, see Appendix of \citet{bouwens15} for discussion).  Thus, it may be that the inclusion
of those ground-based data biased the M$^{\ast}$ result of
\citet{bouwens11} to be too faint.  This is confirmed by
\citet{bouwens15}, who perform a $z =$ 7 Schechter function fit using
only the HUDF and ERS {\it HST} data, finding M$^{\ast} = -$20.6 $\pm$0.4.
To investigate this further,
we performed a similar luminosity function fit to our data, using only data
from the HUDF09 and ERS fields, finding M$^{\ast}_\mathrm{UV} = -$20.64 $\pm$ 0.92
at $z =$ 7.  Thus, making use of only the pre-CANDELS {\it HST} data,
we find a similar value for M$^{\ast}_\mathrm{UV}$ as that found by
\citet{bouwens15} when reexamining the results from
\citet{bouwens11}, though our uncertainty is somewhat higher.
Understanding the differences in
the uncertainty computations between these studies is beyond the scope
of our work, but we note that our MCMC implementation was
designed to produce optimal uncertainties on the Schechter function
fit parameters.  As shown in Figure~\ref{fig:schechter} our current
Schechter fit uncertainties are larger than those of
\citet{bouwens15}.  While some of these differences may be due to the
fact that they used a larger volume (including all five CANDELS
fields), the different methods of computing the uncertainties likely play a role.

\begin{figure*}[!t]
\epsscale{1.15}
\plotone{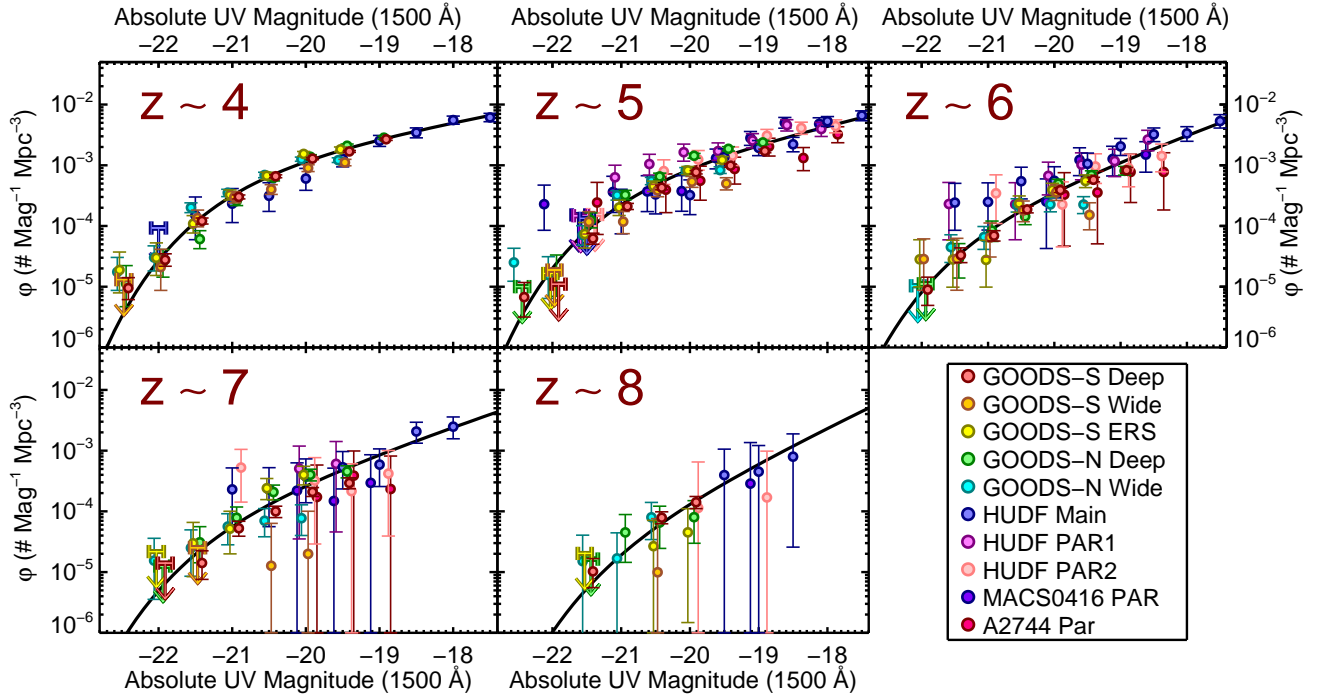}
\caption{The rest-frame UV luminosity functions at each redshift for
  each subfield.  The solid line denotes the best-fit Schechter
  function fit at each redshift.  Upper limits are shown for magnitude
bins with zero detected galaxies.}
\label{fig:fieldlfs}
\end{figure*}

\subsection{Cosmic Variance}
The impact of cosmic variance on our measurement of the luminosity
function is minimized due to our use of several fields, which are
split into four widely separated regions of the sky.  
However, as shown in Figure~\ref{fig:fieldlfs}, there is significant
variance between the different fields, particularly at $z \geq$ 5.
To estimate the effect of cosmic variance on our derived number
densities, we used two techniques: a semi-empirical technique,
combining the QUICKCV calculator provided by
\citet{newman02} \citep[using the updated version provided by
][]{moster11}, combined with the recent clustering-based bias
measurements from \citet{baronenugent14}; and a semi-analytic
technique, based on the semi-analytic models of
\citet{somerville12}. In \S 7 we discuss that these models, with a
modification to the redshift dependance of the normalization of the dust attenuation,
provide an excellent match to our measured UV luminosity functions.

\begin{figure*}[!t]
\epsscale{1.15}
\plotone{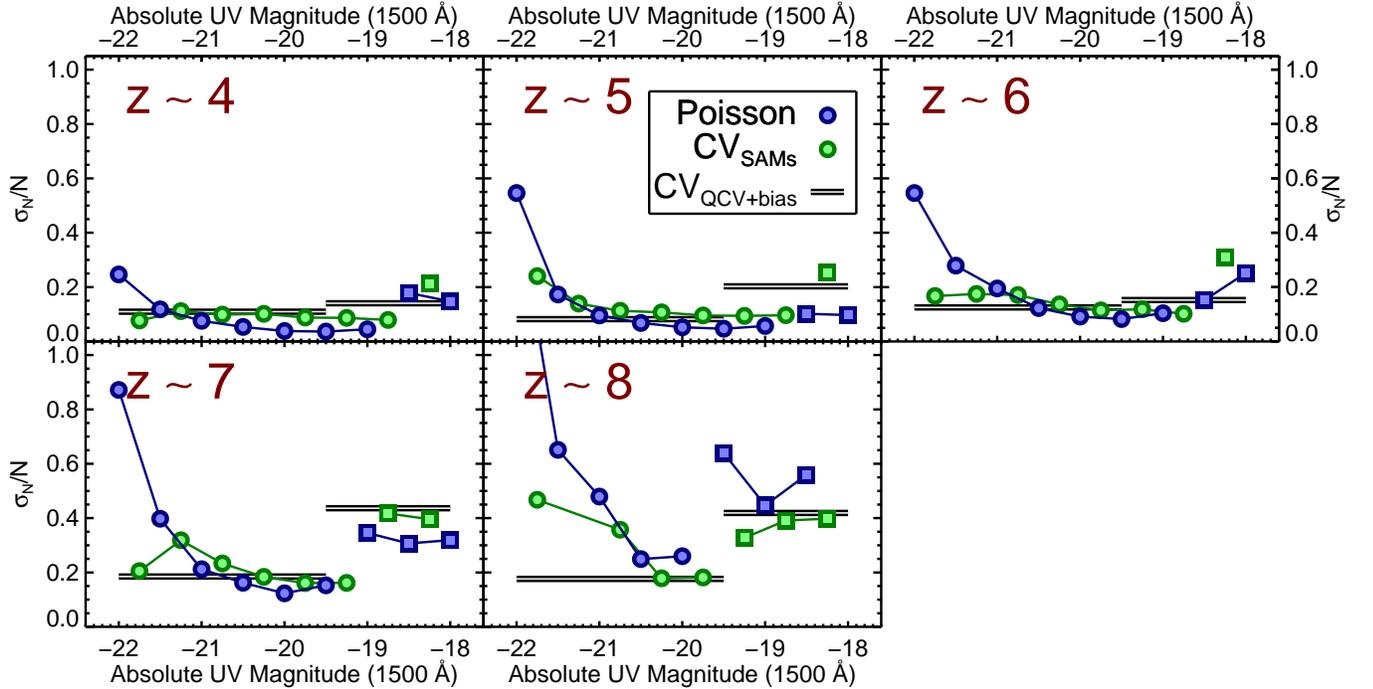}
\caption{A comparison between the fractional uncertainty due to
  Poisson noise and that due to cosmic variance.  We estimate the
  cosmic variance in two ways.  First, we use a semi-empirical method,
  combining the cosmic variance estimates for an unbiased tracer of
  mass from QUICKCV with the clustering-based bias measurements of
  \citet{baronenugent14}, as shown by the gray bars (for bright and
  faint galaxies).  Secondly, we estimate cosmic variance
  uncertainties by examining the variation in the number
  of galaxies as a function of rest-frame absolute UV magnitude from a set of
  semi-analytic models (discussed in \S 7), with the volume
  approximated as that of the two CANDELS/GOODS fields, except at
  faint magnitudes, where we use a HUDF-sized volume.  The Poisson values shown
  come from the uncertainties on the number densities shown in
  Figure~\ref{fig:lf}.  The circles and squares denote magnitudes
  where the majority of our galaxies come from the CANDELS fields and
  HUDF field, respectively.  The two estimates of the cosmic variance
  uncertainty show very good agreement.  In nearly all cases, the
  Poisson uncertainty is greater than that due to cosmic variance,
  thus cosmic variance is likely not the dominant source of
  uncertainty in our measured luminosity functions.}
\label{fig:cv}
\end{figure*}

For a given survey geometry, the QUICKCV program returns
the fractional error in a count due to cosmic variance for an unbiased
tracer of matter at a given redshift.  For this
calculation, we estimated the fractional error separately for GOODS-S, 
GOODS-N, MACS-0416 parallel, and Abell 2744 parallel
fields, adding the variances in quadrature to derive a final
value of $\sigma_{CV}$ for a given redshift bin.  In the GOODS-S
field, we included the area from the three HUDF09 fields, as even the
parallel fields are separated by only a few arcmin from the GOODS-S
proper.  For the input survey geometries, we estimate rectangular regions of the approximate
shape of the GOODS fields, with an enclosed area equal to the GOODS-S
Deep+Wide+ERS+HUDF09 fields for GOODS-S, and the GOODS-N Deep+Wide for
GOODS-N.  The field geometries were thus 10.2$^{\prime}$ $\times$
15.03$^{\prime}$ for GOODS-S, 9.51$^{\prime}$ $\times$
14.65$^{\prime}$ for GOODS-N, and 2.1$^{\prime}$ $\times$
2.1$^{\prime}$ for each of the HFF parallel fields.  However, at faint
magnitudes, our galaxy sample primarily comes from the HUDF, thus we
also estimated the QUICKCV-derived cosmic variance uncertainty with
the HUDF area only, with a geometry of 2.26$^{\prime}$ $\times$
2.26$^{\prime}$.  To convert these unbiased estimates of cosmic
variance to values appropriate to our galaxy sample, we use the
recently published clustering-based bias measurements from
\citet{baronenugent14}, who used the galaxy sample of
\citet{bouwens15} for their calculation (they did not measure the
clustering at $z =$ 8, thus we use the $z =$ 7 bias values for our $z
=$ 8 cosmic variance estimate).  They estimated the bias for
both bright and faint galaxies, splitting their sample at
$M_\mathrm{UV} = -$19.4 at each redshift.  Our estimates of the
fractional uncertainty on galaxy counts due to cosmic variance from
this method is shown as the gray bars in Figure~\ref{fig:cv}, where we
show values of this quantity for both bright and faint galaxies.

For our semi-analytic cosmic variance estimate, we used mock catalogs
of the \citet{somerville12} SAMs, which cover an area $\sim$40$\times$
larger than that of the combined CANDELS/GOODS
fields.  We thus extract independant, GOODS-sided volumes from these catalogs, exploring
the variation in number counts in the independant volumes as a
function of UV absolute magnitude.  At magnitudes brighter than
$-$18.5 (at $z =$ 4--6; $-$19 at $z =$ 7; $-$19.5 at $z =$ 8)
we estimated our survey as being
two 10$^{\prime} \times$ 16$^{\prime}$ fields, representing a
combination of the CANDELS/GOODS fields with the five single-WFC3
pointing fields (HUDF09 and HFF).  At fainter magnitudes, where the
majority of our objects come from the HUDF, we assume a single
2.26$^{\prime} \times$ 2.26$^{\prime}$ field.  We calculate the
1$\sigma$ fractional uncertainty on the number density, $\sigma_{\rm
  cv}/N$, by bootstrap resampling
galaxies in a given M$_\mathrm{UV}$ bin at each redshift.
This 1-sigma fractional uncertainty includes the Poisson noise, thus
we subtract the Poisson errors using the recipe of Gehrels (1986) to
calculate the fractional uncertainty on the number density due to
cosmic variance only.
The uncertainty for the total survey volume is calculated by
adding the variance for two GOODS-sized fields in quadrature for
bright bins (and the HUDF-only for faint bins), and
is shown in Figure~\ref{fig:cv}.

Comparing the SAM-derived cosmic variance values to those from
QUICKCV, we find generally excellent agreement.  The SAM method
predicts, as expected, a larger cosmic variance uncertainty for the brightest
galaxies, though this is understood as the \citet{baronenugent14}
``bright'' sample encompassed galaxies down to M$_\mathrm{UV} = -$19.4, and thus
likely has a relatively faint median magnitude.  Future measurements
of the bias in finer-resolution bins of bright galaxies can probe this
effect further.  The agreement at the faint end is generally good as
well, with the exception of $z =$ 6,
where the semi-empirical method predicts a somewhat low uncertainty,
due to the \citet{baronenugent14} bias measure at $z =$ 6 which is
slightly lower than at $z =$ 5.  While both of these methods are
estimates, and thus may not be extremely accurate, the general
agreement between these two different techniques imply that our
estimates of the cosmic variance uncertainty are not highly inaccurate.

To assess the impact of cosmic variance
on our measured luminosity functions, we compared both of our
estimates of the cosmic variance uncertainty
to the Poisson noise from our step-wise luminosity functions, all
shown in Figure~\ref{fig:cv}.  
At all redshifts, the data at M $< -$21 are dominated by
Poisson noise (M $< -$ 20.5 at $z \geq$ 6), thus we do not expect
cosmic variance to be dominating the uncertainties on the bright end
of the luminosity functions derived here.  
However, cosmic variance may play some role at the
faint end, where we are restricted to small fields.  At $z =$ 7, there
does appear to be a step in the stepwise luminosity function at $M
\geq -$18.5, where the $M = -$19 point is below our best-fit Schechter
function, and the $M = -$18.5 point is above.  At M $\geq -$18.5 our data come
from only the HUDF main field, thus this break represents the point
where we become reliant on a single small field.  As shown in
Figure~\ref{fig:cv}, the cosmic variance uncertainty at faint
magnitudes at $z =$ 7 is $\sim$40\%, somewhat larger than the Poisson
uncertainty, thus, cosmic variance may
bear some responsibility for this discontinuity in the luminosity
function at the faint end.  Future measures of the luminosity function
at $M \geq -$18.5 from the Hubble Frontier Field lensing program may
improve these constraints, but while faint galaxies may be found, the
volumes will still be incredibly small \citep[e.g.,][]{robertson14}.  Thus, robust constraints on
the number densities at this faint level at $z \geq$ 7 may need to
wait until the {\it James Webb Space Telescope}.  We conclude that
while the effects of cosmic variance are
not negligible, they are not the dominant source of uncertainty on the
abundances of the bright objects we have discovered at very high
redshift.

\subsection{Do the Data Support a Schechter Function?}
When allowed to choose any value of $M^{\ast}_\mathrm{UV}$, our $z
=$ 8 Schechter function fit preferred very bright values of
$M^{\ast}_\mathrm{UV}$, such that all observed data points lay on the faint-end
slope part of the function.  This implies that the $z =$ 8 luminosity
function is consistent with a single power law.  Such a functional
form is what one might expect when the feedback effects which govern
the bright end at lower redshift (mainly feedback due to accreting
supermassive black holes) disappear, or if dust attenuation ceases to
be a factor.
\citet{bowler14} recently postulated that the $z =$ 7 luminosity function is better fit
by a double-power law, rather than a Schechter form.  
At $z =$ 7, our step-wise data appear consistent with the Schechter fit out to the
brightest magnitudes we cover.  To see whether our data show a
preference for a Schechter functional form at all redshifts, we performed three fits
to the data -- a Schechter fit, a single power law,
and a double power law.  
To place these fits on equal ground, we found the best-fit parameters
for each function using a simple maximum likelihood routine.  For
the Schechter fit, we used the function shown in Equation 4,
investigating a range of $M^{\ast}_\mathrm{UV}$ with $\Delta M =$ 0.1 mag, and
$\alpha$ with $\Delta\alpha =$ 0.02.  We approximated a single power
law using the Schechter functional form with $M^{\ast}_\mathrm{UV}$ fixed at
$-$30.  For the double power-law, we used the form given in Equation 2
of \citet{bowler14}, which is similar to the Schechter function at the
faint end, but replaces the bright end with a second power law with
slope $\beta$.  In all cases, $\phi^{\ast}$ is found as the normalization such that
the total number of expected objects for a given function is equal to
the number observed.  The likelihood that a given functional form
represents our data was calculated in an identical manner as in \S5.1,
using Equations 5 and 6.

To compare the results from these fits at each redshift, we used the
Bayesian information criterion (BIC).  This is similar to a $\chi^2$
statistic, except that it takes into account both the number of data
points and the number of free parameters.  For a model to be
preferred over a competing model, it must have a BIC lower by at least
2.  This is sensible, as adding a free parameter must yield a better
fit for that model to be preferred.  The BIC is calculated as
\begin{equation}
BIC = -2\ \ln (\mathcal{L}) + k\ \ln(N)
\end{equation}
where $N$ is the number of data points, and $k$ is the number of free
parameters \citep{liddle04}.  For the Schechter, double
power law, and single power law
fits, the number of free parameters are 2 ($M^{\ast}_\mathrm{UV}$, $\alpha$), 3
($M^{\ast}_\mathrm{UV}$, $\alpha$, $\beta$) and 1 ($\alpha$), respectively (we
do not count $\phi^{\ast}$ as a free parameter as it is a
normalization).  The number of data points is the number of galaxies
in our sample used in the fit, which is restricted to those brighter
than the 50\% completeness limits discussed above.  This gives N $=$
2788, 1812, 605, 221 and 47 galaxies at $z =$ 4, 5, 6, 7 and 8, respectively.

\begin{deluxetable}{cccccc}
\tabletypesize{\small}
\tablecaption{Comparison of Luminosity Function Fits}
\tablewidth{0pt}
\tablehead{
\colhead{Redshift} & \colhead{BIC} &
\colhead{BIC} & \colhead{BIC} &
\colhead{$\Delta$BIC} & \colhead{$\Delta$BIC}\\
\colhead{$ $} & \colhead{Schechter} &
\colhead{Double} & \colhead{Power} &
\colhead{Sch-Dou} & \colhead{Sch-Pow}
}
\startdata
4&358&377&641&$-$18.5&$-$283\\
5&540&562&694&$-$21.7&$-$153\\
6&350&361&376&$-$11.1&$-$25.7\\
7&225&234&235&$-$9.28&$-$9.83\\
8&86.9&92.3&86.7&$-$5.36&\phantom{$-$}0.26
\enddata
\tablecomments{The comparison of the Bayesian information criterion
  statistic for fits to our $z =$ 4, 5, 6, 7 and 8 luminosity functions
  using a Schechter, double power-law and single power law functional
  form.  A difference in the absolute value of BIC between two models of $\geq$2 (6) is
  positive (strong) evidence for the preference of one model over
  another.  A Schechter function is strongly preferred over a single
  power law at all redshifts except $z =$ 8, where our data cannot
  distinguish between the two models.}
\end{deluxetable}

\begin{figure*}[!t]
\epsscale{0.95}
\plotone{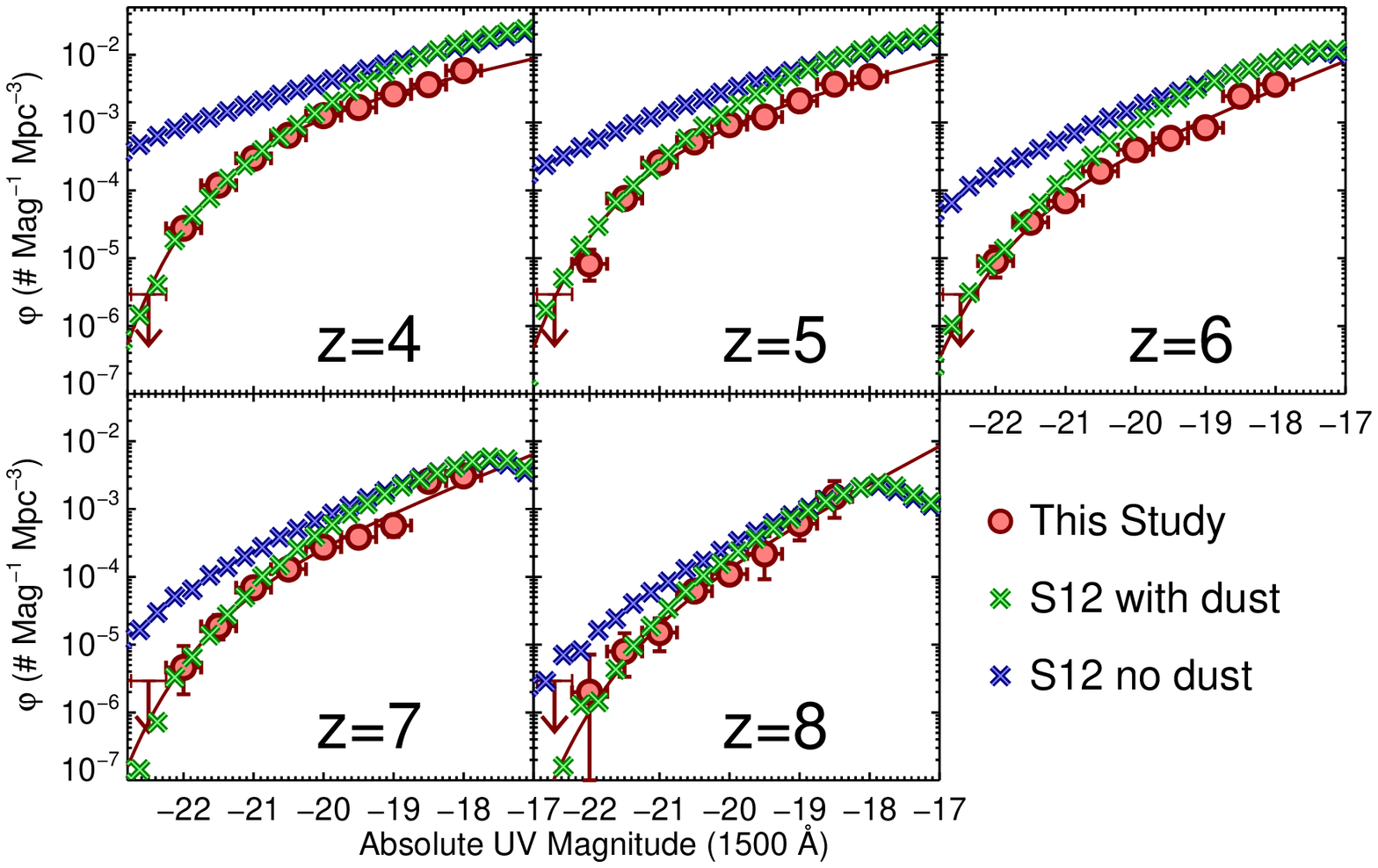}
\caption{Rest-frame ultraviolet luminosity functions at $z =$ 4--8,
  comparing our observations to the semi-analytic models (SAMs) of
  \citet{somerville12} (S12).  The crosses mark the models of
  S12 both with and without dust. In order for the models to be
  consistent with the observations, it is clear that some dust is
  needed at all redshifts (except perhaps at $z=8$), particularly at
  the bright end. Using a simple model for dust attenuation, our
  results suggest that the dust-to-metal ratio or dust geometry must
  change as a function of redshift, a continuation of a trend already
  pointed out in previous studies.  The turnover at very faint
  luminosities in the SAM predictions is due to the resolution limit of the {\it
    Bolshoi} simulation.}
\label{fig:sams}
\end{figure*}

The results from this analysis are shown in Table 6.  A difference in
the absolute value of the BIC of 2 is interpreted as ``positive''
evidence, while a difference
of 6 or higher is ``strong'' evidence, both in favor of the model with the
smaller value.  In Table 6, in addition to the value of the BIC, we
show the difference between the BIC values for the Schechter versus
double power-law, and Schechter versus single power-law.  In this
formalism, a negative difference is in favor of the Schechter
function.  Comparing the Schechter function versus the double
power-law, we find that a Schechter form is strongly preferred to either a
double or single power law at $z =$ 4--7.
This is not surprising, as there is clearly a deficit of observed galaxies
at the bright end when compared to the best-fit power law (Figure~\ref{fig:lf}).  
However, no such deficit is visible at $z =$ 8, and this is confirmed as both
the Schechter fit and
the single-power law fit have effectively identical values of the
BIC.  We conclude that our data support an exponential decline at the bright end of
the luminosity function at $z =$ 4, 5, 6 and 7.  At $z =$ 8, we do not see any evidence for a decline
in the bright end, at least out to $M = -$21.5.  Further data are
needed to show whether one can either detect,
or rule out a decline at the bright end at $z =$ 8.  If
the latter ends up the case, it could indicate
either a significant change in the halo masses of bright galaxies, a
drop in dust attenuation in bright galaxies, or 
a change in the physics governing the feedback in bright galaxies in
the distant universe.

\section{Comparison to Semi-Analytic Model Predictions}

In this section we compare our observations with predictions from
theoretical models set within the predominant $\Lambda$ Cold Dark
Matter ($\Lambda$CDM) paradigm. All such models, whether based on
numerical hydrodynamics or semi-analytic techniques, currently rely
upon phenomenological ``sub-grid'' recipes to treat the physics on
scales smaller than those that can be directly resolved. These
processes include star formation and feedback from massive stars,
supernovae, and supermassive black holes. The phenomenological recipes
are parameterized and must be empirically calibrated. Here, we compare
our new observations at $z =$ 4 -- 8 with predictions from the models
presented by \citet[][hereafter S12]{somerville12}. The sub-grid recipes in these
models have been calibrated using a set of observations at $z\sim 0$,
and \citet{somerville12} presented a comparison with available
observations from $z\sim 0$--5. It is therefore very interesting to
test these model predictions --- with no re-tuning of the free
parameters controlling physical processes --- in the higher redshift
regime probed in this work.

Figure~\ref{fig:sams} shows our estimates of the rest-UV luminosity
function compared with the S12 semi-analytic model (SAM) predictions with and without
dust.  It is already interesting that the dust-free model predictions
are even in plausible agreement with the observations; i.e., the model
predictions lie above the observations at all luminosities and
redshifts.  Next we can ask the question: what characteristics must
the dust extinction have in order to be consistent with the
observations? One can immediately see that the dust extinction must be
differential with both luminosity (more luminous galaxies are more
extinguished) and redshift (galaxies are less dusty at higher
redshift). We use a simple approach to model the dust extinction: as
in S12, we assume that the face-on dust optical depth in the $V$-band
is given by $\tau_{0, V} = \tau_{\rm dust} m_{\rm cold} Z_{\rm cold}
/{r_{\rm gas}}^2$, where $m_{\rm cold}$ is the mass of cold gas in the
disk, $Z_{\rm cold}$ is the metallicity of the cold gas, $r_{\rm gas}$
is the exponential scale radius of the gaseous disk, and $\tau_{\rm
  dust}$ is a normalization parameter. The values of $m_{\rm cold}$,
$Z_{\rm cold}$, and $r_{\rm gas}$ are predicted by the SAM. We treat
$\tau_{\rm dust}$ as a free parameter. We then assign random
inclinations to our galaxies and use a ``slab'' model to compute the
inclination-dependent extinction (see S12 for details). We used a
Calzetti attenuation curve to compute the attenuation at 1500 \AA\ at
each redshift.

In S12, we showed that if we normalize $\tau_{\rm dust}$ to match
observations at $z\sim 0$ and use a fixed value, our model
overpredicts the dust extinction at higher redshift. Similarly here,
we find that the empirical redshift-dependent function for $\tau_{\rm
  dust}$ adopted in S12 based on observations at $z\la 5$ overpredicts
the extinction at $z\ga 5$. We empirically adjust $\tau_{\rm dust}$ to
obtain a good fit to the bright-end of the observed LF in the five redshift bins shown,
and find that $\tau_{\rm dust} \propto \exp(-z/2)$ produces a
reasonably good fit over this redshift range, where $z$ is
redshift. This may be physically interpreted as either a changing
dust-to-metal ratio, or a systematic evolution in the dust geometry
relative to our simple slab model. The required luminosity and
redshift dependence of the dust extinction is in qualitative agreement
with observational conclusions drawn based on the UV colors
\citep{finkelstein12a,bouwens13}.  While the agreement at the bright
end is excellent, the prediced faint-end is too steep, particularly at
lower redshift.  This could imply that the predicted luminosity
dependence of the dust attenuation results in too little attenuation
at faint magnitudes, or it could reflect on the impact of feedback on
the star-formation in the simulations.  Performing a similar analysis
with stellar mass functions and UV luminosity functions in tandem can
break this degeneracy.

In future work, we plan to investigate whether the dust extinction
parameters derived from SED fitting on the observations are consistent
with the empirical SAM requirements. In addition, we plan to use these
models, which plausibly match the observed UV luminosity functions, to
make predictions for the clustering, stellar fractions, and other
properties of high redshift galaxies. We will also show the results of
varying the sub-grid recipes for star formation and feedback, to
illustrate what physical insights can be gained from these
observations. For the moment, however, it is intriguing that the
models that were developed to explain galaxies at a very different
epoch are plausibly consistent with these new observations.

\section{Evolution of the Cosmic Star-Formation Rate Density}
While the evolution of the shape of the luminosity function can
provide interesting constraints on the physics of galaxy evolution, the
integral of the luminosity function provides a key measure of the
total number of UV photons produced at a given redshift.  This is a
key constraint in two ways. First, the integral of the total SFR density provides a
key check against the measured stellar mass density \citep[e.g.,][]{bouwens13}.  Secondly,
assuming a conversion between UV and ionizing photons, this measure
can determine whether galaxies are producing enough ionizing photons to
reionize the Universe at a given redshift
\citep[e.g.,][]{madau96,finkelstein12b,robertson13}.

\begin{deluxetable}{cccc}
\tabletypesize{\small}
\tablecaption{Rest-Frame UV Luminosity Densities and SFR Densities}
\tablewidth{0pt}
\tablehead{
\colhead{Redshift} & \colhead{log $\rho_{UV}$} &
\multicolumn{2}{c}{log SFR Density}\\
\colhead{$ $} & \colhead{(ergs s$^{-1}$ Hz$^{-1}$ Mpc$^{-3}$)} &
\multicolumn{2}{c}{(M\sol\ yr$^{-1}$ Mpc$^{-3}$)}\\
\colhead{$ $} & \colhead{Observed} &
\colhead{Observed} & \colhead{Dust-corrected}
}
\startdata
4&26.26$_{-0.01}^{+0.01}$&-1.59$_{-0.01}^{+0.01}$&-1.03$_{-0.21}^{+0.23}$\\[1ex]
5&26.17$_{-0.01}^{+0.01}$&-1.69$_{-0.01}^{+0.01}$&-1.20$_{-0.25}^{+0.20}$\\[1ex]
6&25.88$_{-0.02}^{+0.02}$&-1.97$_{-0.02}^{+0.02}$&-1.68$_{-0.18}^{+0.24}$\\[1ex]
7&25.77$_{-0.06}^{+0.06}$&-2.09$_{-0.06}^{+0.06}$&-1.85$_{-0.16}^{+0.22}$\\[1ex]
8&25.65$_{-0.19}^{+0.19}$&-2.20$_{-0.19}^{+0.19}$&-2.20$_{-0.19}^{+0.19}$
\enddata
\tablecomments{All values have been computed down to M$_{UV} = -$17.
  The dust correction was derived based on the values of E(B-V)
  derived from SED fitting, with the dust-corrected SFR densities
  including an uncertainty term from the spread of extinction values
  at a given absolute magnitude.  The SFRs were computed assuming the
  \citet{kennicutt98} conversion from the UV luminosity density
  ($\rho_{UV}$), assuming a Salpeter IMF, and a constant
  star-forming population with age $\geq$ 100 Myr.}
\end{deluxetable}

We calculated the luminosity density at each redshift, integrating
down to M$_\mathrm{UV} = -$ 17.  This is approximately the magnitude of the
faintest galaxy in our $z =$ 8 sample, and also facilitates comparison
with recent works which use a similar magnitude limit.  Galaxies
likely exist beyond this magnitude limit
\citep[e.g.,][]{trenti12,alavi14}, which we will consider in the next
subsection.  We utilized the results of our MCMC luminosity function
fitting chain to derive a robust estimate of both the rest-frame
UV specific luminosity density ($\rho_\mathrm{UV}$, in units of erg s$^{-1}$
Hz$^{-1}$ Mpc$^{-3}$) and its uncertainties.  In each step of the
chain, we calculated $\rho_\mathrm{UV}$ by taking the luminosity function from
the best-fit Schechter
function parameters for that step, and integrating it from $-$23 $<
M_{1500} < -$17.  To convert this number to a SFR density, we use the
relation adapted from \citet[][$\rho_{SFR} = 1.25 \times 10^{-28} \rho_\mathrm{UV}$]{kennicutt98}, which converts the specific UV
luminosity density to a SFR density ($\rho_{SFR}$), assuming a Salpeter IMF and a
constant star-formation history over $\geq$ 100 Myr.  The original
coefficient from \citet{kennicutt98} was 1.4; however, updated
stellar population models \citep[e.g.,][]{bruzual03} imply a somewhat
smaller value.  We chose a value of 1.25 to be consistent with the
assumptions used in \citet{bouwens15}, though we note an even lower
coefficient of 1.15 was used by \citet{madau14}.  The quoted value
of $\rho_\mathrm{UV}$ or $\rho_{SFR}$ is the median of the values recorded
from all of the MCMC steps, while the 68\% confidence range is taken
to be the central 68\% of values.

\begin{figure*}[!t]
\epsscale{0.9}
\plotone{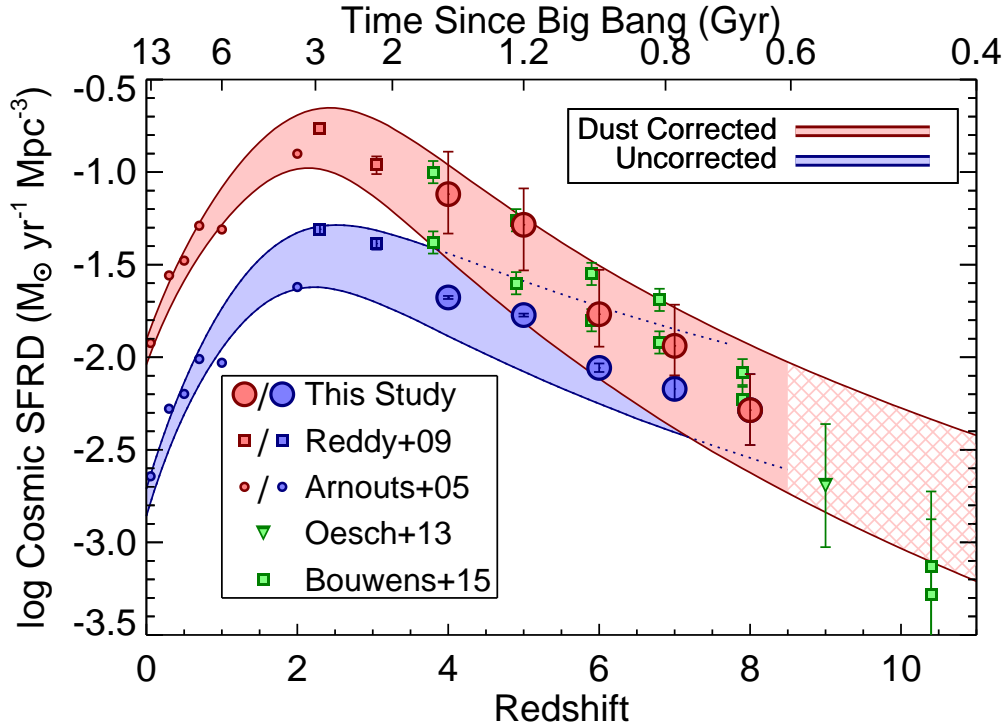}
\caption{The evolution of the cosmic star-formation rate density,
  derived by integrating the best-fit Schechter function at all
  redshifts to $M_\mathrm{UV} < -$ 17.  Our data are shown as large circles.
To extend this analysis to lower redshifts, we also show the values at $z
\sim$ 2--3 from \citet{reddy09}, and from $z =$ 0--2 from
\citet{arnouts05}.  For both studies, we integrated the published
best-fit Schechter function parameters to $-$17 to derive the
uncorrected values of $\rho_\mathrm{UV}$.  We used the published ratio of the
dust-corrected -- to -- unobscured values of $\rho_\mathrm{UV}$ from
\citet{reddy09} to calculated the dust-corrected values at $z \sim$
2--3.  At $z \leq$ 2, we used the dust correction from
\citet{schiminovich05}, which assumes a constant value of A$_\mathrm{UV} =$
1.8 at all redshifts.  We used Equation 7 to fit
  the observed trends, deriving the uncertainties on this fit via 10$^4$ Monte Carlo
  simulations, showing as the shaded region the 68\% confidence range
  from these fits.  The total (dust-corrected) SFR
  density evolves with (1+$z$)$^{-4.3 \pm 0.5}$ from $z =$ 3--8.  The green
  symbols show the high-redshift results from \citet{bouwens15} and
  \citet{oesch13}, which were not included in the fit, but are
  consistent with the observed trends at the $\sim$1$\sigma$ level.}
\label{fig:madau}
\end{figure*}

Although the UV luminosity is a relatively easy observable in this
epoch, the major drawback in its use as a SFR indicator is its
susceptibility to attenuation by dust.  As a bevy of recent work has
shown, this dust correction is important even out to $z \sim$ 7--8
\citep[e.g.,][see also \S 7]{finkelstein12a,dunlop13,bouwens13}.  To calculate the
total SFR density, we corrected the observed SFR density for
extinction using a new iteration of SED fitting, including the deep {\it Spitzer}/IRAC
data (\S 3.6), which is a
crucial probe of the rest-frame optical light, providing better
constraints on the dust attenuation.  Using these updated extinction results, we
calculated a sigma-clipped median and standard deviation for the
best-fit extinction values at a given redshift in four
magnitude bins: $< -$ 21, $-$21 to $-$20, $-$20 to $-$19, and $-$19 to
$-$17.  We recover previously observed trends that dust extinction
lessens with both increasing redshift and decreasing UV luminosity \citep[e.g.,][]{finkelstein12a,bouwens13}.
The values of $E(B-V)$ for bright galaxies decreases from 0.15 at $z =$ 4
to 0.02 at $z =$ 7, and for faint galaxies from 0.06 at $z =$ 4, to
0.0 at $z =$ 7.  The small numbers, limited wavelength coverage, and
faint magnitudes of $z =$ 8
galaxies make it difficult to measure their extinction, therefore we assumed
$E(B-V)$ $=$ 0 for all $z =$ 8 galaxies.  The spread in $E(B-V)$ values at
all redshifts and luminosities is $\sim$0.1, thus we assume this value
in all cases (with the exception of $z =$ 8, where we fix $E(B-V)$
to zero).  To include this uncertainty in $E(B-V)$ in our derived
dust-corrected SFR density, in each step of the chain we draw a new
value of $E(B-V)$ for a given redshift and magnitude bin, modifying the
fiducial value by a number drawn from a Gaussian 
distribution with a standard deviation equal to the $E(B-V)$ spread of
0.1.  The values of $\rho_\mathrm{UV}$ and the observed and dust-corrected
values of $\rho_{SFR}$ are given in Table 7.  These
values do not include potential sub-millimeter galaxies which lie
below our rest-frame UV detection limits.  However, as we are
observing at $z >$ 4, we expect their impact on the total cosmic SFR
density to be minimal (see Table 7 from \citet{bouwens12}).

In Figure~\ref{fig:madau}, we show our derived values
  of the cosmic SFR density.  Our results are for the
most part consistent with those of \citet{bouwens15}, although our
observed values are lower by a few $\sigma$ at $z =$ 4 and 5,
likely due to our shallower faint-end slopes at these redshifts.  
To study the evolution of $\rho_\mathrm{UV}$ with redshift,
  we fit the function provided by \citet{madau14}, given in Equation
  7.  Although we have
  included lower-redshift data in our fit, we do not discuss here our
  results for the low-redshift slope or the peak redshift, as these
  are better obtained from \citet{madau14}, who use a
  compilation of several sources, including far-infrared observations.
 As we are adding data at high redshift, it is interesting to examine
 the trends there.  We find that  at $z >$ 3 the uncorrected values of $\rho_\mathrm{UV}$
evolve as ($1+z$)$^{-2.4 \pm 0.3}$, while the dust corrected values
evolve as ($1+z$)$^{-4.3 \pm 0.5}$.  Most interesting, 
the observed trend of the evolution of the total SFR density is
consistent within 1$\sigma$ of the published results at $z =$ 9 from
\citet{oesch13} and at $z =$ 10 from \citet{bouwens15}.  
Thus, the \citet{oesch13} and \citet{bouwens15} are consistent with a smooth
extrapolation of our derived cosmic SFR density at $4 < z < 8$ to higher
redshift, with no break.

\begin{figure*}[!t]
\epsscale{0.58}
\plotone{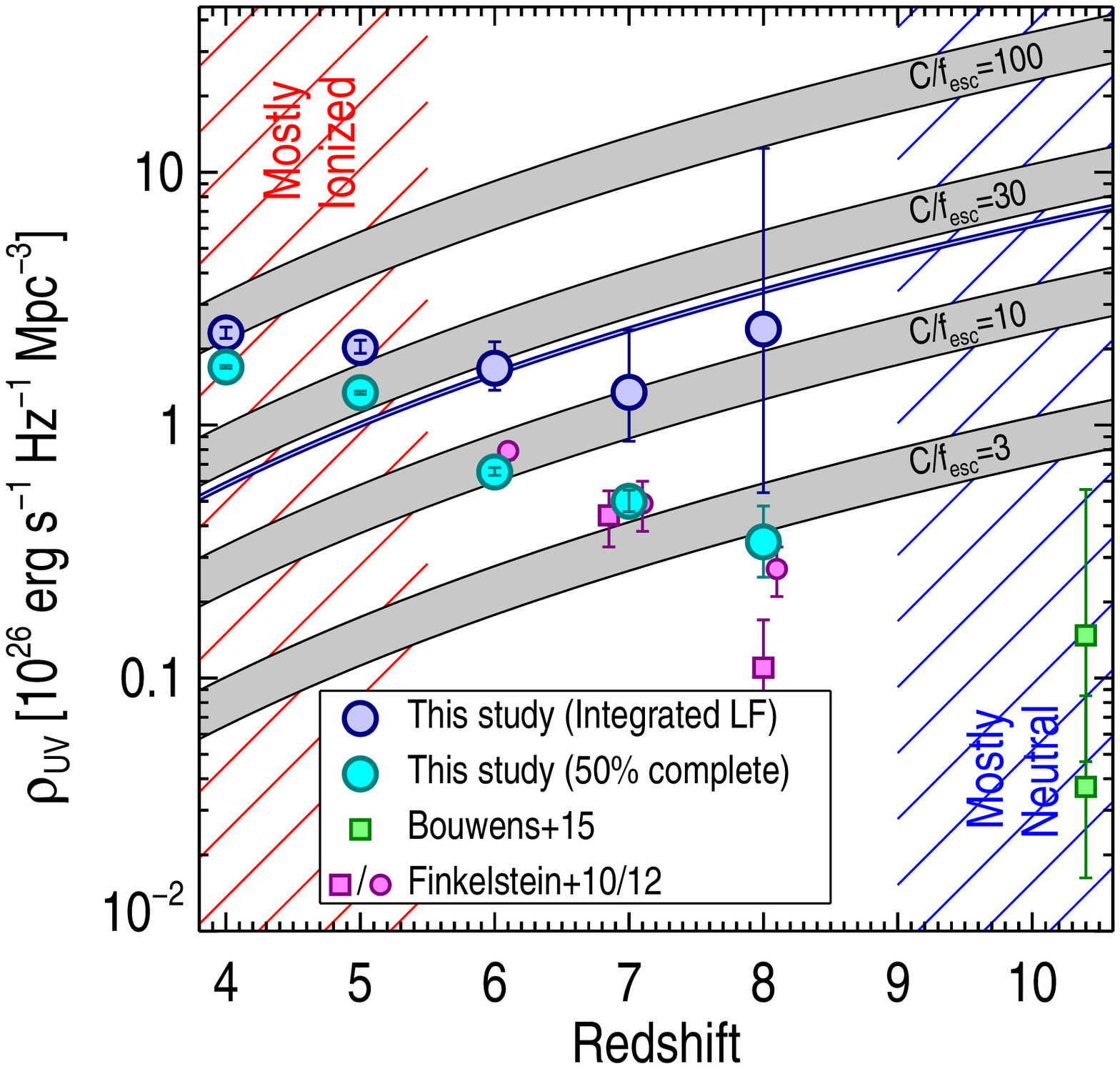}
\plotone{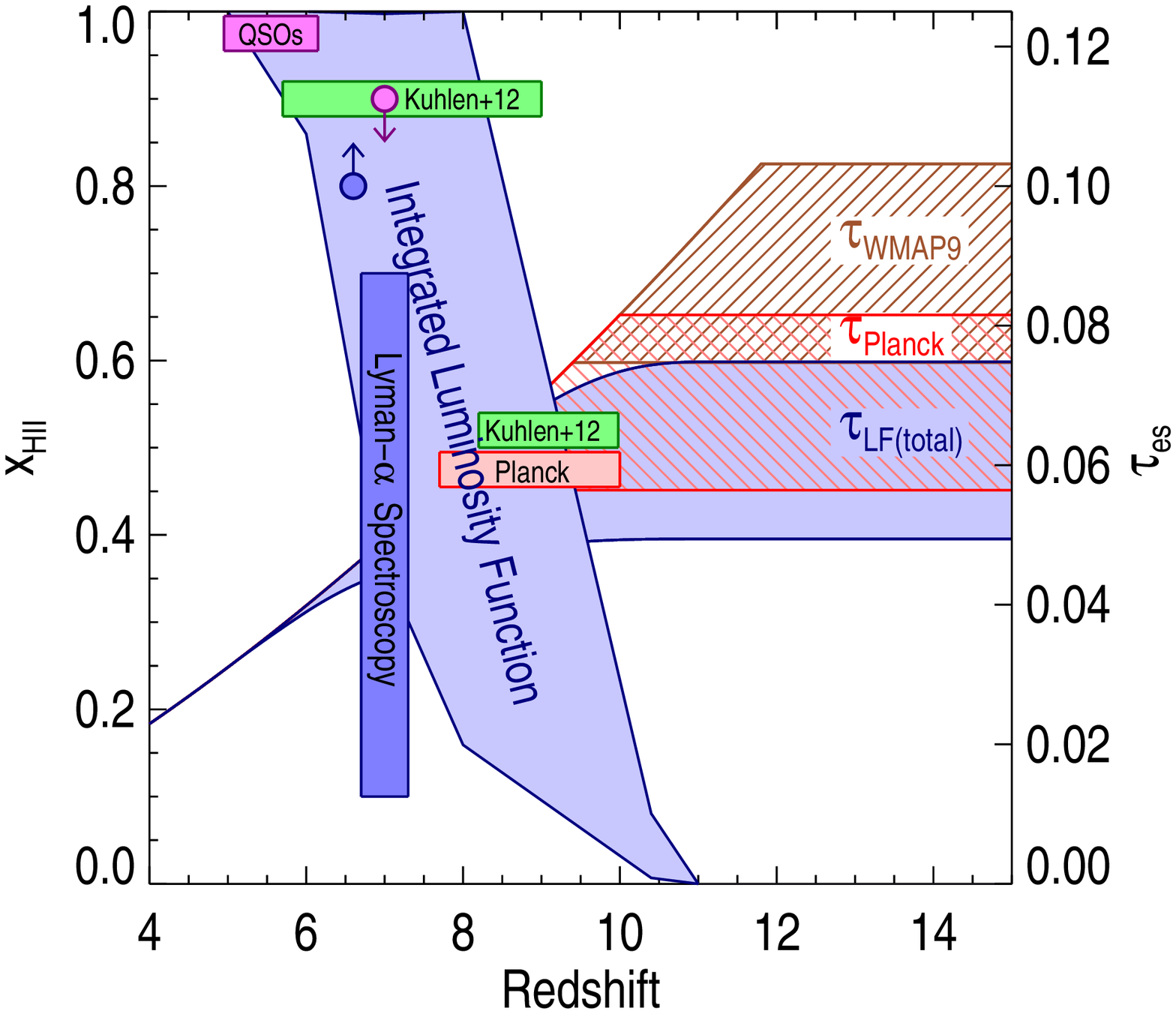}
\caption{Left) The specific luminosity density
    ($\rho_\mathrm{UV}$) versus redshift (similar to Figure 3 from
    \citet{finkelstein12b}).  Here we show our luminosity functions
    integrated down to M$<-$13 as blue circles.  The cyan circles
    denote the value of $\rho_\mathrm{UV}$ when we integrate down to our
    50\% completeness limit ($-$18 at $z =$ 7).  Recent results from
    \citet{bouwens15} at $z \approx$ 10 are shown in green, with the
    lower and upper squares representing limiting magnitudes of $-$17
    and $-$13, respectively.  The wide gray curves denote the value of $\rho_\mathrm{UV}$ needed to sustain a fully
reionized IGM at a given redshift, for a given ratio of the clumping
factor C over the escape fraction of ionizing photons fesc (Madau et
al. 1999).  The thin blue curve shows our fiducial value of $C=$3 and
$f_{esc} =$ 13\%.  Right) The volume ionized fraction, $x_{HII}$, of the IGM which can be
sustained given the integral of our luminosity functions at $z =$ 4--8
(as well as that at $z =$ 10.4 from \citet{bouwens15}).  We assume the
luminosity function extends to M$_\mathrm{UV} = -$13, $C=$3 and
$f_{esc} =$ 13\% (this escape fraction is the highest that does not
violate constraints set by the Ly$\alpha$ forest at $z =$ 6;
Finkelstein et al.\ 2012b).
We plot constraints on $x_{HII}$ from spectroscopy of
quasars at $z <$ 6 from \citet{fan06} and at $z =$ 7 from
\citet{bolton11}.
The blue circle denotes constraints on $x_{HII}$ from
the evolution in the Ly$\alpha$ luminosity function from $z =$ 5.7 to 6.6 from
Ouchi et al. (2010), while the blue bar denotes the range of $x_{HII}$
values inferred from $z \approx$ 7 follow-up Ly$\alpha$ spectroscopic
studies \citep[e.g.,][]{pentericci14,tilvi14,faisst14}.  The
instantaneous redshift for reionization from {\it Planck}
(8.8$_{-1.1}^{+1.2}$) is indicated by the red rectangle. The derived
50\% and 90\% $x_{HII}$ redshifts from the study
of \citet{kuhlen12} are shown in green.  The right-hand axis
corresponds to the hatched regions, which show the Thomson optical
depth to electron scattering ($\tau_{es}$) as predicted by our integrated luminosity
functions (blue) compared to {\it Planck} (red) and WMAP9 (brown).
Compared to previous results, the improved constraints on the
luminosity functions yield a tighter range of possible reionization
histories.  Broadly speaking, we find a
picture where the Universe is fully ionized by $z =$ 6, with the neutral
fraction becoming non-negligible at $z \geq$ 7, with $\tau_{es}$ highly
consistent with the {\it Planck} value.}
\label{fig:reion}
\end{figure*}  

\subsection{Constraints on Reionization}
Although it is presently generally assumed that galaxies dominated the
ionizing photon budget for the reionization of the IGM, 
it has been difficult for observations to obtain robust proof.
Analyses of the IGM via line-of-sight quasar observations have been
able to show that reionization was likely complete by $z \sim$ 6
(e.g., \citet{fan06}, though see \citet{mesinger10} and
\citet{becker14}).  Additionally, observations of the cosmic microwave background
(CMB) radiation constrain the total optical depth due to electron scattering,
which, while it cannot directly inform the duration of reionization,
it can give an estimate of the reionization redshift ($z_{reion}$) if
reionization was instantaneous.  The results from the {\it Wilkinson
  Microwave Anisotropy Probe} (WMAP) 9-year dataset give $\tau_{es} =$ 0.088
$\pm$ 0.014, which corresponds to $z_{reion} =$ 10.6 $\pm$ 1.2
\citep{hinshaw13}, while the recent {\it Planck} results suggest
$z_{reion} =$ 8.8$_{-1.1}^{+1.2}$ \citep{planck15}.

The primary reason for the current uncertainties in the contribution
of galaxies to reionization lies in the uncertainty in the faint-end
slope measurements, and also in the assumptions of the escape fraction
of ionizing photons ($f_{esc}$) and the clumping factor in the IGM
($C$).  The clumping factor is primarily constrained theoretically, but
most studies agree that it is low ($<$6) at high redshift
\citep[e.g.,][]{faucher08,pawlik09,mcquinn11,finlator12}.
To infer from observations of galaxies a number of escaping
ionizing photons, one first needs to take the observed UV light, and
assume an IMF and a metallicity.  Then, to calculate the number of
these ionizing photons available for reionization, one then needs to
multiply by an assumed value of $f_{esc}$.  It is difficult to
constrain $f_{esc}$ directly at high-redshift, as the correction for intervening IGM absorption systems is
extremely high at $z >$ 4.  Significant effort is being expended on
observationally constraining $f_{esc}$ at $z <$ 4.  Although bright
galaxies at $z \sim$ 1 have very low escape fractions (relative to the
UV emission) of $f_{esc,rel} <$ 2\% \citep{siana10}, escaping ionizing
emission has been observed from small fractions of galaxies probed by
studies at $z \sim$ 3--4 \citep[e.g.,][]{steidel01,shapley06,iwata09,vanzella10,nestor11},
though some ground-based studies may suffer from contamination by
intervening sources \citep[e.g.,][]{vanzella12}.  
Recent results imply that escape fractions from star-forming galaxies
at $z \sim$2--3 range from 5--20\%, with lower-mass galaxies,
especially those with Ly$\alpha$ in emission, having a greater
likelihood of having detectable escaping ionizing emission \citep[e.g.,][]{nestor13,mostardi13}.

\citet{finkelstein12b} used measurements of the emission rate of
ionizing photons from observations of the Ly$\alpha$ forest in quasar
spectra to place an upper limit on $f_{esc}$ from galaxies.  Assuming
that the rest-frame UV luminosity function extended down to $M_\mathrm{UV} =
-$13, the escape fraction must be f$_{esc} <$
13\% to avoid violating the Ly$\alpha$ forest measurements of
\citet{bolton07}.  Using this value, and assuming $C =$ 3, 
the luminosity functions available at the time were consistent
with a wide range of reionization histories, including an end redshift
as late as $z \lesssim$ 5, and an ionized fraction at $z \sim$ 7 from
30-100\%.  \citet{kuhlen12} and \citet{robertson13} did similar
analyses, folding in additional observables (e.g., the Ly$\alpha$
forest and CMB), found that in order to complete
reionization by $z \sim$ 6, the luminosity function must extend much
deeper than can presently be observed, and/or the average escape fraction must
be higher at higher redshift.

Here, we use our updated luminosity functions to reexamine the
contribution of galaxies to reionization.  Figure~\ref{fig:reion} shows
both the observable specific UV luminosity density ($\rho_\mathrm{UV}$), which we define to
be that above our 50\% completeness limit, as well as the total
$\rho_\mathrm{UV}$, which we define as the integrated luminosity function
down to $M_{1500} = -$13.  We then compare these values to the
critical number of UV photons necessary to sustain an ionized IGM at a
given redshift, taken from \citet{madau99}.  This figure is similar to
Figure 3 from \citet{finkelstein12b}, thus we refer the reader there
for more details.  Effectively, these critical curves depend on
assumptions about the stellar IMF, metallicity, $f_{esc}$ and clumping
factor.  The first two are responsible for the conversion from
observed UV photons to intrinsic ionizing photons.
We assumed a Salpeter IMF, and the width of the curves denote the
impact of changing the metallicity from 0.2 $\leq Z$/$Z$\sol\ $\leq$
1.0.  We show several curves for the reader's choice of the ratio of
$C/f_{esc}$.  Here, we use a fiducial value of $C =$ 3 and $f_{esc} =$
13\%, consistent with \citet{finkelstein12b}.

The right panel of Figure~\ref{fig:reion} shows the ionization history
of the IGM, comparing our derived value for the total specific UV
luminosity density to our fiducial model of C $=$ 3 and f$_{esc} =$
13\%, folding in the values at $z =$ 10.4 from \citet{bouwens15} to
extend our analysis beyond $z =$ 8.  Our luminosity
functions are consistent with a reionization history that starts at $z
\sim$ 11, and ends by $z >$ 5.  Although the exact value of the volume
ionized fraction in the
IGM is uncertain between these redshifts, due to the persistent uncertainty in the faint-end
slope, our results imply the following constraints (given the caveat of
our assumptions).  At $z =$ 6, we can constrain x$_{HII} >$ 0.85 (1$\sigma$), while
out to the limit of our observations at $z =$ 8 the data are
still consistent with a fully ionized IGM (68\% C.L.\ of 0.15 $<
x_{HII} = <$ 1.0).  We find a midpoint of reionization (x$_{HII} =$ 0.5) of 6.7 $< z <$ 9.4 (68\% C.L.).

Broadly speaking, measurements from quasar spectra as well as from
Ly$\alpha$ emission
from galaxies support a reionization scenario consistent with
what we derive (Figure~\ref{fig:reion}).  The constraints from Ly$\alpha$ emission
are heavily model dependent, and studies claiming a very low value
of $x_{HII}$ may be assuming a velocity offset of Ly$\alpha$ from
systemic which is too high \citep[e.g.,][]{stark14}.  
While our results are in slight tension (1.3$\sigma$) with the results from WMAP9,
they are in excellent agreement with the more recent results from {\it
  Planck}.  From our
fiducial reionization history, we find $\tau_{es} =$ 0.063 $\pm$
0.013, highly consistent with the measurement from {\it Planck}
of $\tau_{es} =$ 0.066 $\pm$ 0.012.  We remind the reader that our
results did not use the {\it Planck} results as a constraint; rather,
the inferred reionization history from our observed luminosity
functions (with our assumptions on $C$ and $f_{esc}$) are in remarkable
agreement with the {\it Planck} observations.

Future observations are necessary to improve the constraints on
reionization from galaxies.  Specifically, more robust measurements of
the faint-end slope $\alpha$ at $z =$ 6--8 can dramatically shrink the
uncertainties on $\rho_\mathrm{UV}$, subsequently reducing the width of our
plausible values of $x_{HII}$.  Likewise, improving the measurements
at $z \geq$ 9 will inform us on whether the ionization fraction of the
IGM at that early time was significantly non-zero.  Even a small
contribution ($\sim$10\%) to $x_{HII}$ at early times will erase any
discrepancy between our current observations and those from WMAP.  The
Hubble Frontier Fields program will improve both of these
areas, though definitive results will likely not be obtained until the
{\it James Webb Space Telescope} ({\it JWST}) era.
\vspace{2mm}

\section{Evolution of Galaxies at $z \geq$ 9}
Studies of galaxies at $z \geq$ 9 are now only in their nascent phase,
but {\it HST} surveys such as CANDELS and UDF12 are beginning to probe
this early epoch.  The first robust results on galaxies in this epoch
were published by \citet{ellis13}, who used the new F140W data in the
HUDF from the UDF12 program to discover galaxies at $z \sim$ 9.  This
filter allows $z \sim$ 9 galaxies to be detected in two bands (F140W and
F160W), dramatically reducing the contamination due to noise from
F160W-only studies alone \citep[\S3.8.1; c.f.,][]{bouwens11c}.  \citet{ellis13}
discovered the first robust sample at $z >$ 8.5, finding seven candidate galaxies.  \citet{mclure13} followed
this up with an analysis of the $z =$ 9 luminosity function, finding
number densities at the faint end (M$_\mathrm{UV} \sim -$18) only slightly
lower than at $z =$ 8.  \citet{oesch13} also analyzed the $z =$ 9
luminosity function, also finding seven $z \sim$ 9 candidate galaxies
in the HUDF.  Although the GOODS-S
field lacks the F140W data necessary to detect potential $z =$ 9
galaxies in two-bands, \citet{oesch13} added the full CANDELS/ERS
GOODS-S field to improve their constraints at the bright end.
However, they found no $z =$ 9 candidates in this larger field.
Their published luminosity function is consistent with that from
\citet{mclure13} at the faint end.  Bolstered with their additional
constraints due to the inclusion of
the non-detections from the larger GOODS-S field, \citet{oesch13} fit
a luminosity function (keeping the faint-end slope and normalization
fixed), finding a surprisingly faint value for
$M^{\ast}_\mathrm{UV}$ of $-$18.8 $\pm$ 0.3.  
However, this derivation was based
on the assumption that the luminosity function shows luminosity
evolution at $z \geq$ 6 --- a trend which we have now shown to be
unlikely.  Given this new insight, as well as the presence of a
plethora of bright galaxies at $z =$ 7 and 8, we consider it likely that
the \citet{oesch13} estimate of the bright end of the $z =$ 9
luminosity function is underestimated.  

A number of recent papers have described empirical evidence that
galaxies at high redshift have star-formation histories that increase
with time \citep[e.g.,][]{papovich11,salmon15,finlator11,jaacks12b,lundgren14}.
Most recently this has been examined by \citet{salmon15}, who found
that the star-formation rates of galaxies from $z =$ 3 to 6 are
consistent with a power-law of the form $\Psi(t) = (t/\tau)^{\gamma}$
  (with $\gamma$ = 1.4 $\pm$ 0.1 and $\tau =$ 92 $\pm$ 14 Myr).  This
  analysis assumed that studying galaxies at a constant number density
  allows one to trace the progenitors and descendants of a galaxy
  population \citep[e.g.,][]{vandokkum10,leja13}, and their
  star-formation history was measured for a constant cumulative number density of
  2 $\times$ 10$^{-4}$ Mpc$^{-3}$.  Although
  the accuracy of this constant number density technique was initially studied at $z
  <$ 3, recent evidence shows that it likely works out to $z \sim$ 8
  \citep[albeit with a possible slight evolution of number density
  with redshift;][Jaacks et al.\ in prep]{behroozi13c}.
 
\begin{figure*}[!t]
\epsscale{1.0}
\plotone{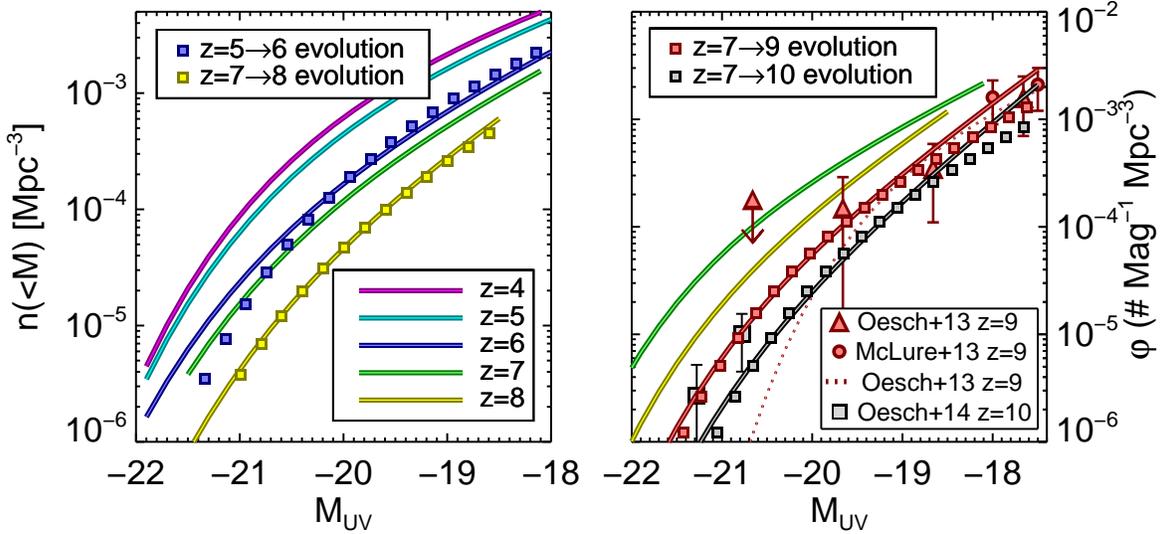}
\caption{Left) The solid lines denote cumulative luminosity functions at $z =$ 4 to 8
  from our study.  We evolve the observed luminosity functions
  to higher redshift, assuming that the M$_\mathrm{UV} \propto$ SFR, that
  the SFR rises with time as $\Psi \propto t^{1.4}$ \citep{salmon15},
  and that galaxy progenitors and descendants share a common number
  density \citep[e.g.,][]{leja13}.  We show two results: the predicted
  $z =$ 6 luminosity function, evolved from $z =$ 5 (blue squares),
  and the predicted $z =$ 8 luminosity function evolved from $z =$ 7
  (yellow squares).  Though there are small discrepancies at $z =$ 6,
  the match between predicted and observed is
  excellent at $z =$ 8.
Right)  Our differential luminosity functions at $z =$ 7 and 8,
with $z =$ 9 data from the literature (triangles and circles).  The
red (gray) squares show our predicted $z =$ 9 (10) luminosity function, continuing
the evolution from $z =$ 7 as shown in the left panel.  The red line
shows the $z =$ 7 best-fit Schechter function, dimming $M^{\ast}_\mathrm{UV}$ to
$-$19.95, providing an excellent match for our predicted $z =$ 9
luminosity function.  The dashed ride line shows this same function,
but with $\alpha$ equal to the $z =$ 8 value of $-$2.3.
This predicted luminosity function shows excellent agreement
with the observed faint end at $z =$ 9 when assuming the $z =$ 8 value
of $\alpha$, but is significantly higher at the bright
end compared to the published luminosity function of \citet{oesch13}.
However, the recent discovery of bright $z =$ 10 candidate galaxies by
\citet{oesch14} (large gray squares) imply that bright galaxies are
indeed present at this early epoch.  However, we note that the $z =$
10 number densities from \citet{oesch14} are actually more consistent
with our $z =$ 9 predictions than $z =$ 10, thus clearly more work is
needed to sort out the bright end at such early times.
We conclude that when
sufficient data exists for a large-volume survey for $z =$ 9 galaxies,
large numbers of bright galaxies will be discovered.
}
\label{fig:z9}
\end{figure*}  

Using our updated luminosity functions, we examine whether the
star-formation histories at this
earlier epoch are consistent with a similar functional form.  Figure~\ref{fig:z9} shows the cumulative luminosity functions at $z =$ 4 to 8 from our
analysis.  Using the Salmon et al.\ rising SFH, we can
evolve our $z =$ 7 cumulative luminosity function back in time to $z
=$ 8 via:
\begin{equation}
\Psi_{z=8} = \Psi_{z=7} \left(\frac{t_{z=8}}{t_{z=7}}\right)^\gamma
\end{equation}
where $\Psi$ is the SFR, $t_z$ is the cosmic time elapsed since formation to a given
redshift, and using the \citet{kennicutt98} conversion between
M$_\mathrm{UV}$ and SFR (with the updated coefficient of 1.25).  The available data cannot constrain the formation redshift ($z_f$),
as it is degenerate with the star-formation history exponent, thus we
assume a value of $z_f =$ 18, which gives a close match between
predicted and observed $z =$ 8
cumulative luminosity functions.
Figure~\ref{fig:z9} shows this predicted $z =$ 8 cumulative luminosity
function alongside our observed one.  A very close match is seen at nearly all
magnitudes.  Our predicted $z =$ 8 luminosity function
slightly under-predict the UV luminosity at $M_\mathrm{UV} > -$19.
However, as discussed above, our constraints on the faint end of the
luminosity function at $z =$ 8 are tenuous at best.  The agreement at
the bright end is
excellent.  While we did not correct for dust in this
analysis, dust is highly unlikely to change these results
(particularly at the bright end where we are interested), as
bright/massive galaxies at 4 $< z <$ 7 all have similar UV slopes
\citep{finkelstein12a,bouwens13}.

It is apparent when examining our cumulative luminosity
functions in Figure~\ref{fig:z9} that this type of evolution will not
work at all redshifts, as our luminosity functions are not uniformly
spaced in magnitude (e.g., the $z =$ 4 and 5, and $z =$ 6 and 7
cumulative luminosity functions are very close together).  We examined
one other redshift, evolving the observed $z =$ 5 luminosity function
to $z =$ 6.  We find a decent match, though
this under-predicts the bright end, and over-predicts the faint end.
In any case, as we are most interested in extrapolating to $z >$ 8,
the fact that the predicted evolution works extremely well from $z =$
7 to 8 gives us confidence that extrapolating to higher redshifts is
reasonable.  This assumed evolution is stronger than that
observed from $z =$ 6 to $z =$ 7.  Had we
assumed a SFH which matched the evolution from $z =$ 6 to $z =$ 7, we would have
\emph{over-predicted} the $z =$ 8 LF.  Our use of a SFH which matches
the observed $z =$ 7 to $z =$ 8 evolution thus yields a conservatively
low $z =$ 9 predicted luminosity function.

Given the relative paucity of observational information at $z >$ 8,
the fact that our assumed SFH matches the evolution from $z =$ 7 to 8
makes it interesting to continue our study out to $z =$
9 (continuing to evolve the $z =$ 7 luminosity function due to its
smaller uncertainties compared to our observed $z =$ 8 luminosity
function).  The red squares in Figure~\ref{fig:z9} show the expected
$z =$ 9 luminosity function from our model, alongside our
observed luminosity functions at $z =$ 7 and 8.  We calculated
the expected $z =$ 9 luminosity function by again taking the $z =$ 7
luminosity function and evolving it out to $z =$ 9 assuming the
star-formation history discussed above (and $z_f =$ 18 for all number
densities/magnitudes).  As shown in this figure, our predicted $z
=$ 9 luminosity function is consistent at
the $\sim$1$\sigma$ level with all published data points
from \citet{mclure13} and \citet{oesch13}.  The insignificant
under-prediction at the faint end is due to the fact that our
analysis effectively keeps the
faint-end slope fixed to the $z =$ 7 value, while in reality it may
become steeper.  Given that our assumed SFH was derived from brighter
galaxies \citep{salmon15}, the slight underprediction at the faint end is not surprising.

Figure~\ref{fig:z9} shows
a Schechter function which matches our predicted
$z =$ 9 luminosity function at all but the faintest magnitudes.  
To derive this Schechter function we used the measured significant evolution in $\phi^{\ast}$
and $\alpha$ with redshift shown in Figure~\ref{fig:schechter} to find 
$\phi^{\ast}_{z=9} =$ 4.3 $\times$ 10$^{-5}$ and $\alpha _{z=9} = -$2.50.  With these two parameters, we find a reasonable
agreement with our predicted $z =$ 9 luminosity function with
M$^{\ast}_{z=9} = -$20.65.  This implies a slight dimming in the
characteristic magnitude at $z =$ 9, though much less than that
implied by \citet{oesch13}, and within the uncertainties of our
observed $z =$ 7 and 8 values.
Regardless of the Schechter parameterization, our predicted luminosity
function shows a much higher number density of bright galaxies than that reported by
\citet{oesch13}, yet still moderately consistent with the observed data points
from both \citet{oesch13} and \citet{mclure13} at the faint end.  Using our predicted
$z =$ 9 Schecter function, we would expect to see 4.2 $z =$
9 galaxies at M$_\mathrm{UV} < -$20.3 ($H <$ 27) in a GOODS-sized field.  Based on
Poisson statistics alone, this is in mild tension with the zero
galaxies at these magnitudes reported by \citet{oesch13}.  We also
show results from this analysis when evolving $z =$ 7 to $z =$ 10, which predict
M$^{*}_{z=10} = -$20.55 when using our assumed evolution of
$\phi^{\ast}$ and $\alpha$ with redshift (or $\sim$1 $M_\mathrm{UV} < -$20.5
galaxy per GOODS field).

Recently, \citet{oesch14} have performed a new search for
extremely distant galaxies, finding four bright $z =$ 10 candidates in
the GOODS-N field and two new candidates from a re-analysis of the
GOODS-S dataset.  Although as mentioned above, these fields do not
have deep F140W data, \citet{oesch14} used the $< 3\sigma$ detections
of these galaxies in the extremely shallow 800s 3D-HST (PI van Dokkum)
F140W pre-imaging data to place these galaxies at $z =$ 10.  Even
though these galaxies are only detected in one band with {\it
  HST}, three of them are detected in IRAC (though at least one may be
affected by blending from a nearby bright sources), thus
their presence is intriguing.  
Figure~\ref{fig:z9} shows the number densities of these sources
from \citet{oesch14}; there is
excellent agreement with our predicted $z =$ 9 evolution, though these
data are much
higher in abundance than our predicted $z =$ 10 luminosity function.
 Although these sources may be at $z =$ 10 rather than $z =$ 9, if real, their
presence confirms that bright galaxies are relatively abundant at $z
>$ 8.5.

Finally, we examine the change in the integrated luminosity density at
$z =$ 9 with our proposed luminosity function compared to that from
\citet{oesch13}.  Here we use our derived $z =$ 9 Schechter function,
which matches our predictions at bright magnitudes, and matches the
observed faint galaxy counts from the HUDF.  The integrated luminosity
density (from $-$23 to $-$17) is 
$\sim$50\% higher than that published
in \citet{oesch13}.  Thus, the precipitous decline in the luminosity density
\citep{oesch13,oesch14} may be
less than previously thought \citep[e.g.,][]{ellis13,coe13,behroozi14}.
While these results are intriguing, we
conclude that in order to robustly probe
the bright end of the $z =$ 9 luminosity
function, we require a significantly increased searchable area
with the correct filter set (allowing more than single-band
detections) to discover these distant galaxies. 
Constraints on the full shape of the luminosity function in this
distant epoch are crucial to design the most efficient surveys with
{\it JWST}.

\section{Conclusions}

Combining the extremely deep data available in the HUDF with the
still deep yet much wider data available from CANDELS in the
GOODS-South and GOODS-North fields allows robust samples of galaxies
to be discovered across a large dynamic range of UV luminosity at $z
=$ 4, 5, 6, 7 and 8.  Using a robust photometric redshift selection
technique, we have discovered a sample of nearly 7500 galaxies at 3.5 $<
z <$ 8.5 over five orders of UV magnitude, and
over a volume of 0.6--1.2 $\times$ 10$^6$ Mpc$^3$.  We discovered a
large number of bright ($M_\mathrm{UV} < -$21)
  galaxies at these redshifts, in excess of predictions based on
  previous estimates of the luminosity functions at $z \geq$ 6.  

\begin{itemize}

\item Our sample selection performs very well when comparing to
  available spectroscopic redshifts.  We perform various
  tests to estimate the contamination rate, which we find at worst to
  be $\leq$15\%, and more likely to be $\leq$5--10\%.  This is
  consistent with contamination estimates
  based on the colors of the most likely contaminants, dusty
  star-forming galaxies at $z \sim$ 2.  Although the GOODS
fields are only two of five CANDELS fields, the remaining three fields
contain relatively shallow $Y$-band data, which can result in increased
sample contamination, as well as a reduced ability to separate
galaxies into $z =$ 6, 7 and 8 samples. 

\item Our large
  volume probed allows us to make a
  robust determination of the amplitude and shape of the
  bright end of the luminosity function, which can be used as a
  crucial probe of the physics dominating galaxy evolution.  We used a MCMC technique to
  estimate the luminosity function, to better characterize the
  uncertainties, both on the step-wise luminosity function, as well as
  on the parameters of the Schechter functional form.  Our results
  agree with previous studies at the faint end, but deviate from some
  previous studies at the
  bright end, where our data allow us to better constrain
  the abundance of rare, bright galaxies.  We find results consistent
  with a non-evolving characteristic magnitude ($M^{\ast}_\mathrm{UV} \approx
  -$21), with our values of $M^{\ast}_\mathrm{UV}$ at $z =$ 6 and 7
  brighter than the previous values of \citet{bouwens07,bouwens11} at
  $\sim$2$\sigma$ significance.  Both the faint-end slope ($\alpha$) and the normalization
  ($\phi^{\ast}$) do significantly evolve with increasing redshift, to steeper
  and lower values, respectively.  This is in contrast to previous
  results, which determined that the evolution of the luminosity
  function was primarily in luminosity.

\item We explored whether a Schechter functional form is required by
  the data, or whether a single (or double) power-law is a better fit
  for our luminosity functions;
  a single power-law form of the luminosity function may be expected
  at very high-redshift, when dust may not be present, and/or feedback due to AGN activity is no
  longer sufficient to suppress star-formation in the most massive
  galaxies.  At $z =$ 6 and 7, a Schechter (or
  double power-law) is required to fit the bright end.  However, at $z
  =$ 8, a single power-law provides an equally good fit to the data.
  Although larger volumes will need to be probed to improve the
  estimates of the abundances of bright $z =$ 8 galaxies, if a power
  law is preferred, it could imply that we
 may be observing the era when feedback stops affecting
  massive galaxies.  Comparing to semi-analytical models, we find that the evolution
  in our luminosity function can be explained by a changing impact of
  dust attenuation with redshift.  In a future work we will explore
  whether this is a unique constraint, or whether a combination of
  feedback and dust changes can reproduce the observations.

\item We measure the evolution of the cosmic star-formation rate
  density by integrating our observed luminosity functions to the
  observational limit of M$_\mathrm{UV} = -$17, and
  correcting for dust attenuation.  The cosmic
  SFR density evolves as (1$+z$)$^{-4.3 \pm 0.5}$ at $z \geq$ 4.  This
  smoothly declining function with increasing redshift is consistent
  with published estimates of the SFR density at $z \geq$ 9.

\item We investigate the constraints on the contribution of galaxies to
  reionization by integrating our luminosity functions down to
  $M_\mathrm{UV} = -$13.  Our fiducial results (assuming
  $C/f_{esc} =$ 23, which does not violate Ly$\alpha$ forest
  constraints at $z \leq$ 6) are consistent
  with a reionization history that begins at $z >$ 10, and completes
  at $z \approx$ 6, with a midpoint at 6.7 $< z <$ 9.4.  However, the
  uncertainties, particularly at $z \geq$ 7 are high, due to the
  relatively high uncertainty in the faint-end slope, such that our
  observations are consistent with an IGM at $z =$ 8 that is anywhere
  from completely ionized, to 85\% neutral.

\item The presence of bright galaxies at $z =$ 6 -- 8 has
  interesting implications for the luminosity functions at higher
  redshift.  We used empirically derived star-formation histories to
  evolve our $z =$ 7 luminosity function back to $z =$ 9, and predict
  that $\sim$4 bright (M$_\mathrm{UV} < -$20.3) galaxies should be detectable
  per GOODS-sized field.  
This is contrary to initial observational results, which,
  using single-band detections found no bright $z =$ 9 galaxies,
  though consistent with emerging results that some bright galaxies
  may exist at $z =$ 10.  Future wider-area studies with two-band
  detections will provide a more robust estimate of the bright end of
  the $z =$ 9 luminosity function.

\end{itemize}

This study highlights the power of combining deep and wide-area
studies to probe galaxy populations at very high redshifts, a topic
that will remain highly active through the advent of {\it JWST}.  
These results leave us with a variety of
questions.  What is responsible for the apparent abundance of bright
galaxies at $z >$ 6?  Is this tied in with a reduction of feedback, or
is some other physical process in play?  Does this trend continue out
to higher redshifts, or does the luminosity density fall off
dramatically at $z >$ 8, as has been proposed?  Although these issues are
inherently intertwined, we can make progress on these
issues with future wide-area {\it HST} surveys.  This will allow
us the most complete view of the high-redshift universe by the end of
this decade, allowing us to make full use of {\it JWST}.

\acknowledgements
We thank Kristian Finlator, Brian Siana, Rychard Bouwens, Pascal
Oesch, Dan Jaffe and Jon Trump for useful
conversations.  SLF acknowledges support from the University of Texas
at Austin College of Natural Sciences.  MS was supported by a NASA
Astrophysics and Data Analysis Program
award issued by JPL/Caltech.
RJM acknowledges ERC funding via the award of a consolidator grant.
This work is based on observations made with the NASA/ESA Hubble Space Telescope,
obtained at the Space Telescope Science
Institute, which is operated by the Association of Universities for
Research in Astronomy, Inc., under NASA contract NAS 5-26555. These
observations are associated with program \#12060.
This work is also based in part on
observations made with the Spitzer Space Telescope, which is operated
by the Jet Propulsion Laboratory, California Institute of Technology
under a contract with NASA. Support for this work was provided by NASA
through an award issued by JPL/Caltech.


\end{document}